\documentclass[aps,
english,preprintnumbers,nofootinbib,
twocolumn]{revtex4-1}

\usepackage{amsfonts,amsmath,amssymb}
\usepackage{graphicx}
\usepackage[utf8]{inputenc}
\usepackage{hyperref}
\usepackage{babel}

\newcommand\e{{\rm e}}

\newcommand\be{\begin{equation}}
\newcommand\ee{\end{equation}}
\newcommand\bea{\begin{eqnarray}}
\newcommand\eea{\end{eqnarray}}

\begin{document}

\def\rhoo{\rho_{_0}\!} 
\def\rhooo{\rho_{_{0,0}}\!} 

\begin{flushright}
\phantom{
{\tt arXiv:2006.$\_\_\_\_$}
}
\end{flushright}

{\flushleft\vskip-1.4cm\vbox{\includegraphics[width=1.15in]{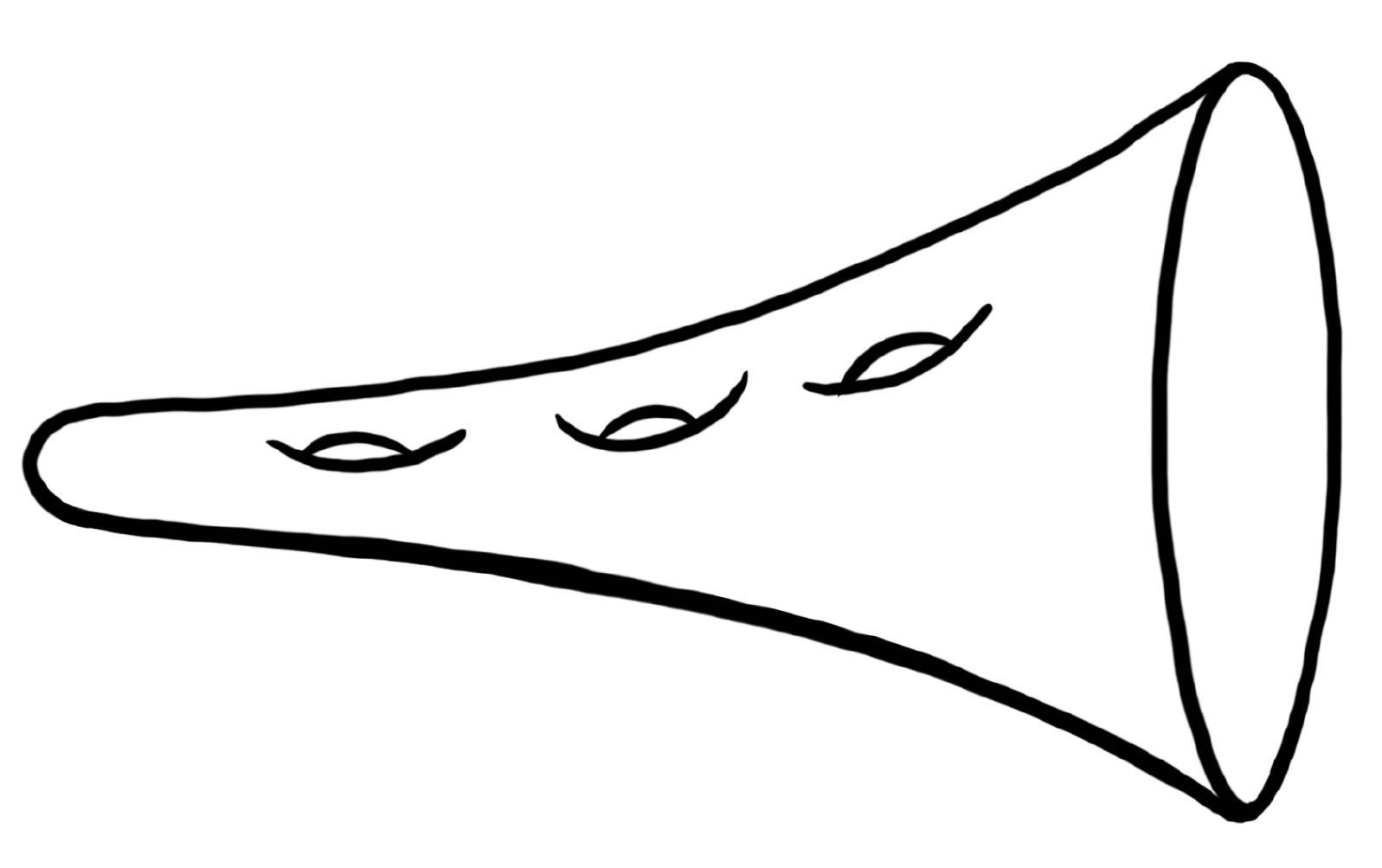}}}

\title{On the Quenched Free Energy of JT Gravity and Supergravity}
\author{Clifford V. Johnson}
\email{johnson1@usc.edu}
\affiliation{Department of Physics and Astronomy\\ University of
Southern California \\
 Los Angeles, CA 90089-0484, U.S.A.}


\begin{abstract}
The quenched free energy, $F_Q(T){=}{-}T\langle \ln Z(T)\rangle$, of various JT gravity and supergravity theories is explored, taking into account the key non-perturbative physics that is accessible using their matrix model formulations.  The leading low energy physics of these systems can be modelled by the Airy and (a family of) Bessel models, which arise from scaling limits of matrix ensembles. The $F_Q(T)$s of  these models are directly computed  by explicit sampling of  the matrix ensembles, and how their properties are connected to the statistical mechanics of the underlying discrete spectrum of the ensembles is elucidated. Some of the low temperature ($T$)  features of the results confirm recent observations by Janssen and Mirbabayi. 
The results are then used as benchmarks for exploring an intriguing formula proposed  by Okuyama for computing $F_Q(T)$  in terms of  the connected correlators  of its partition function,  the wormholes of the gravity theory. A low $T$ truncation of the correlators helps render the formula practical, but it is shown that this is at the expense of much of its accuracy. The significance of the statistical interpretation of $F_Q(T)$  for black hole microphysics is discussed. 
\end{abstract}


\maketitle

\section{Introduction}

\label{sec:introduction}

The thermodynamics of Jackiw--Teitelboim (JT) gravity~\cite{Jackiw:1984je,Teitelboim:1983ux}, treated as a fully quantum gravity theory by summing over all possible geometries and topologies,  has been of keen interest in recent years. While there has been a great deal of progress, a most important quantity that has remained elusive is the complete free energy of the model. This is  the ``quenched'' free energy in the sense used most frequently in the condensed matter and statistical physics literature, $F_Q(\beta){=}-\beta^{-1}\langle \ln Z(\beta)\rangle$. This quantity  is distinct from the ``annealed'' free energy $F_A(\beta){=}{-}\beta^{-1}\ln \langle Z(\beta)\rangle$ (where $\beta{=}1/T$ is the inverse  temperature). The use of $\langle\cdots\rangle$ is a reminder to think in terms of  disorder and thermodynamic averages,  over configurations of constituents or of couplings, or both.  

In condensed matter and statistical physics the distinction between $F_Q$ and~$F_A$ is especially important when studying systems with random  fluctuating properties, as might be encoded in the strength of the couplings between constituents. The annealed average   treats random configurations and random couplings on the same footing, allowing the system to explore them both. The quenched average instead freezes the couplings first, and lets the system explore its thermodynamical configurations. Only  afterwards is there an averaging over couplings. It is this latter that is the more meaningful thermodynamic quantity arising from taking the average over an ensemble of systems, at some~$T$. 

There is a distinction between quenched and annealed in a theory of gravity too, and  statistical interpretations analogous to those made in the condensed matter context  can in principle be made, especially when there is a holographic dual~\cite{Maldacena:1997re,Witten:1998qj,Gubser:1998bc,Witten:1998zw} description of the dynamics in terms of non-gravitational physics. Traditionally, however, the free energy is usually discussed in regimes of high temperature, where  the system is dominated by some set of typical high energy (essentially classical) configurations, and the difference between $F_A$ and $F_Q$ goes away.  For example, in  a  traditional semi-classical quantum gravity treatment of black hole thermodynamics, the black hole saddle point solution (with infinitesimally small fluctuations about it) is such a configuration set, and it is indeed the (easier to compute) annealed free energy that is used to calculate  thermodynamic quantities such as the internal energy, $U{=}{-}\partial_\beta \ln Z(\beta)$, and the entropy $S{=}\beta(U{-}F)$.

Generally speaking, to explore fully the thermodynamics in a theory of quantum gravity, it is desirable (perhaps even essential) to compute the quenched free energy. It should be expected that $F_Q(T)$, if it can be computed, should have a statistical interpretation. This will naturally emerge in the approach studied in this paper, which builds the result out of quantities computed (non-perturbatively) using  matrix models. Moreover, given the dictionary between random matrix model quantities and JT gravity quantities on the one hand~\cite{Saad:2019lba}, and JT gravity and the near-horizon geometry of higher dimensional black holes on the other~\cite{Achucarro:1993fd,Nayak:2018qej,Kolekar:2018sba,Ghosh:2019rcj},  a natural interpretation in terms of the statistical mechanics and thermodynamics of those black holes suggests itself, and it potentially opens up an avenue to revisit old questions~\cite{Preskill:1991tb,Maldacena:1998uz,Page:2000dk} about the thermodynamics of extremal black holes at the very lowest temperatures. This will be unpacked below.

\subsection{Background}

\begin{figure}[h]
\centering
\includegraphics[width=0.40\textwidth]{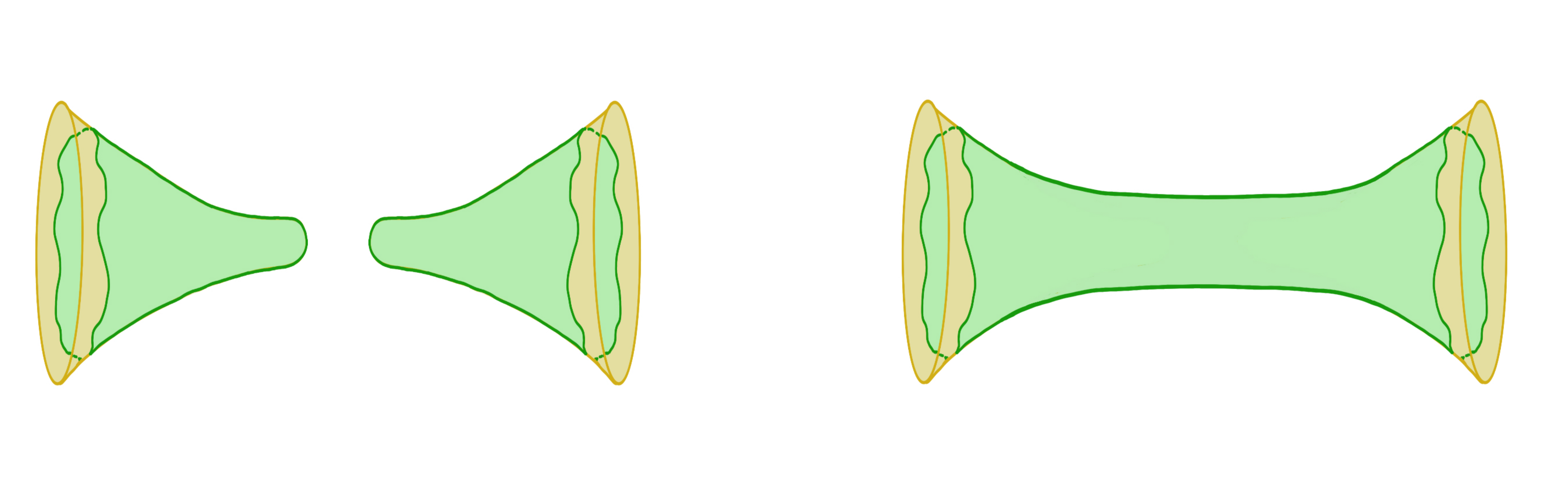}
\caption{\label{fig:blackholes_vs_wormholes} Connected and disconnected contributions to the two-point correlator $\langle Z(\beta) Z(\beta)\rangle$. For JT gravity the loops are  in an asymptotic spacetime that is ``almost'' AdS$_2$  in the sense that the boundaries (edges of the green shaded region) have finite length $\beta$.}
\end{figure}

On the gravity side of things,  there is a natural averaging associated with having to do  the gravity path integral over various geometries and topologies.  In two Euclidean dimensions, the basic quantity of interest is the (boundary) loop of fixed length $\beta$. Attached to the loop are all the possible bulk geometries allowed by the dynamics. $Z(\beta)$ will mean the partition function for this arrangement.  It is extremely natural  to incorporate both connected and disconnected diagrams in computations  such as {\it e.g.} multiple correlators $\langle Z(\beta_1) Z(\beta_2)\cdots\rangle$. These simply represent the connected and disconnected geometries with the loops $\beta_1$, $\beta_2$, {\it etc.,} as boundaries. (See {\it e.g.} the contributions to the two-point correlator in figure~\ref{fig:blackholes_vs_wormholes}.)
In fact, if the loops are small ({\it i.e.,}~high~$T$) then disconnected diagrams are dominant, while for large loops the connected diagrams are favoured.

Focusing on  JT gravity (and variants thereof), when  the  dynamics are understood to be dual~\cite{Maldacena:2016hyu,Jensen:2016pah,Maldacena:2016upp,Engelsoy:2016xyb,Almheiri:2014cka} to some non-gravitational theory such as the Sachdev-Ye-Kitaev (SYK) model~\cite{Sachdev:1992fk,Kitaev:talks}, incorporation of the connected or ``wormhole'' geometries imply a non-factorization that invites an interpretation~\cite{Maldacena:2004rf,Cotler:2016fpe,Harlow:2018tqv,Saad:2019lba} of gravity as being an ensemble average of the dual system (at least for the right kinds of question), an interesting  issue in its own right.\footnote{For  recent work addressing this matter, see {\it e.g.,} refs.\cite{Afkhami-Jeddi:2020ezh,Maloney:2020nni,Cotler:2020ugk,Bousso:2020kmy,Belin:2020hea,Saad:2021rcu,Cotler:2021cqa,Pollack:2020gfa,Blommaert:2019wfy,Penington:2019kki,Goel:2021wim,Benjamin:2021wzr,Janssen:2021stl,Verlinde:2021kgt,Verlinde:2021jwu}.} An example is in the computation of the spectral form factor for the SYK model, which requires an explicit ensemble average on the SYK side to get a smooth quantity representing the ``typical'' behaviour, but a simple inclusion of a wormhole diagram  on the JT gravity side in order to capture the same behaviour~\cite{Saad:2018bqo}.

A natural ``dual'' formulation of 2D gravity, and particularly JT gravity and supergravity, is the (double-scaled) random matrix model. It can be intuitively thought of as relevant along a number of  different routes. One way goes back to explicitly performing the path integral over all possible geometries by writing it as an $N{\times}N$ matrix-valued field theory whose Feynman diagrams have, upon counting powers of $N$ in t'Hooft's large $N$ expansion, a topological interpretation in terms of random 2D surfaces~\cite{'tHooft:1973jz}. The diagrams yield tesselations/discretizations of the surfaces and the toy field theory performs the path integral. The continuum limit is achieved by taking the double-scaling limit~\cite{Brezin:1990rb,Douglas:1990ve,Gross:1990vs}. The other way goes back to Wigner~\cite{10.2307/1970079}, where the matrix model is, by construction, an ensemble average over a class of matrices. This averaging aspect of matrix models was less of a focus in the gravity applications of  old, but clearly it is relevant now in view of the discussion of SYK, more general connections to quantum chaos~\cite{Maldacena:2015waa}, and related topics.  The various ensembles are each naturally associated with an SYK-like structure that motivates a JT-like gravity~\cite{Stanford:2019vob}: Hermitian for perturbative JT gravity~\cite{Saad:2019lba}, $\boldsymbol{\beta}{=}2$ in the Dyson-Wigner classification, along with $\boldsymbol{\beta}{=}1,4$  from that class and the seven $(\boldsymbol{\alpha},\boldsymbol{\beta})$ Altland-Zirnbauer classes defining a rich variety of other JT gravity and supergravity theories. 

In a sense, the  matrix model picture of the gravity theory therefore naturally connects the built-in averaging of quantum gravity (summing over fluctuating surfaces with some fixed length loops) with the more familiar averaging procedures in statistical mechanics ({\it e.g.,} determining the typical behaviour in the ensemble of traces of large powers of the matrix). So naturally the discussion now returns to the computation of $F_Q(T){=}{-}\beta^{-1}\langle\log Z(\beta)\rangle$ in the gravity theory. Unfortunately, it is notoriously hard to compute the average  of the logarithm $\langle\ln Z(\beta)\rangle$. In condensed matter, a common way to proceed is to use the ``replica trick''~\cite{Edwards_1975}, defining $\langle\ln Z(\beta)\rangle$ in terms of averages over multiple insertions of the partition function into the averaging process:
\begin{equation}
\langle \ln Z(\beta) \rangle = \lim_{n\to0} \frac{\langle Z(\beta)^n\rangle -1}{n}\ ,
\end{equation}
Notably,  it is often hard or ambiguous (or both) as to how to do the $n\to0$ continuation.

\begin{figure}[h]
\centering
\includegraphics[width=0.25\textwidth]{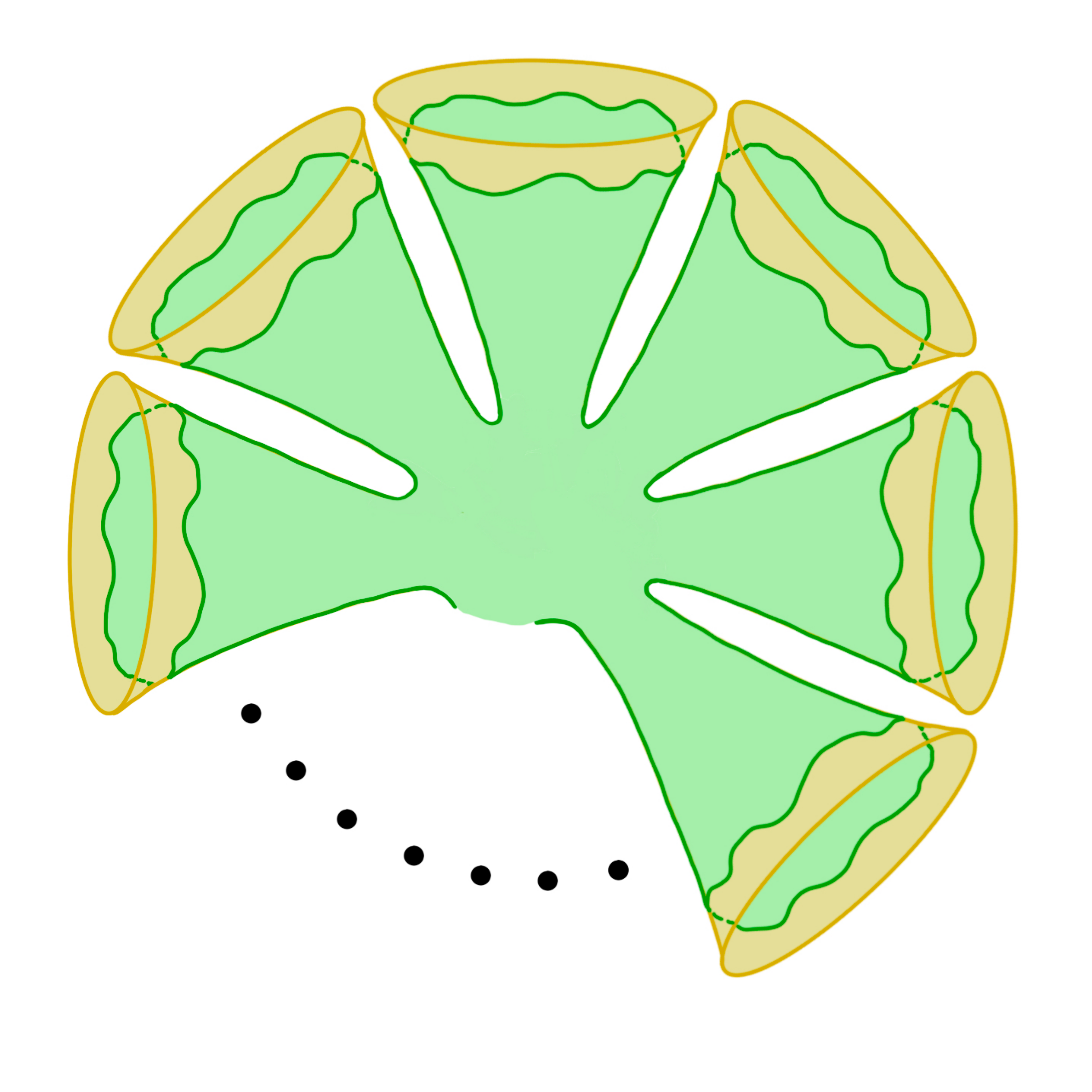}
\caption{\label{fig:n-point-wormhole} $\langle Z(\beta)^n\rangle_c$ is a wormhole with $n$ boundaries. }
\end{figure}

In a significant step in finding the quenched free energy of JT gravity,  Engelhardt, Fischetti, and Maloney~\cite{Engelhardt:2020qpv} focussed on  the role of  wormholes in a gravitational replica computation.  Naturally, 
the wormhole geometries the connected contributions to the correlations/averaging  (see {\it e.g.,} figure~\ref{fig:n-point-wormhole}),  played a natural role in their replica computation.
While some valiant computations were performed, there was a struggle to fully resolve the computation since there arose the  issue of finding an unambiguous  way of  determining and handling the $n$-dependence. There was a suggestion that the phenomenon of replica symmetry breaking~\cite{Parisi:1979mn,Parisi:1979ad,PhysRevLett.35.1792}, familiar from systems such as spin glasses, might be relevant  in order to resolve the matter. Notably, the incorporation of non-perturbative (in the topological expansion) contributions to the physics was not manifest in an approach that is perturbative at the outset.  

Later, ref.~\cite{Johnson:2020mwi} tried to use the matrix model approach to JT gravity to understand features of the replica approach. There were broadly  two key motivations there. The first is that the matrix model approach  allows for a very efficient and succinct method for computing correlators of $Z(\beta)$, including non-perturbative effects, which in fact are known in some examples to be highly significant at low temperature, and hence relevant to the final form of $F_Q(T)$.  The second is that in the very low temperature limit, the leading contribution to the  ``wormhole''  connected correlator $\langle Z(\beta)^n\rangle_c $  has a very simple  $n$--dependence and $\beta$--dependence, making them potentially ideal probes of the low-temperature phase. This simplicity suggested that perhaps in that limit there was no ambiguity leading to complicated phenomena such as replica-symmetry breaking. 

However, working with a simple toy model and taking the $n\to0$ limit led to a divergent coefficient of the $\beta$--dependence,  for which only an heuristic resolution was provided. Nevertheless, the work suggested that the leading non-trivial low temperature dependence of $F_Q$ should be (minus) quadratic in $T$, with a curvature scale determined by non-perturbative physics. This will find support in the results presented here.

Recently,  Okuyama  proposed~\cite{Okuyama:2021pkf} a formula~(\ref{eq:okuyama-formula}) for how to build $\langle \ln Z(\beta)\rangle $ (and hence~$F_Q$) from connected correlators of $Z(\beta)$. It follows from a remarkably simple and robust derivation that requires no special limits (in~$\beta$, or couplings, {\it etc.}). Exploring the properties and consequences of the formula  will be carried out in this paper. Simply put,  assuming that the formula is correct, when it is combined with  {\it non-perturbative} matrix model results, it in principle completes the journey of refs.~\cite{Engelhardt:2020qpv,Johnson:2020mwi}  by successfully combining the connected $\langle Z(\beta)^n\rangle$, but  in a manner that sidesteps the replica trick issues. 
Moreover, since the formula applies to all temperatures, it in principle solves the problem of replica ambiguities everywhere, since although it is more involved to compute the correlators at subleading orders in $\beta$, they are can be given an unambiguous  matrix model definition that is free of pathologies, and hence $F_Q(\beta)$ is unambiguously defined.

However, for the formula to be fully useful, closed forms for the $\langle Z(\beta)^n\rangle_c$ need to be input, and such results are only available in very special cases (some discussed below). Using the aforementioned low temparature simplification of $\langle Z(\beta)^n\rangle_c$ renders the formula manageable, and aspects of this truncations are thoroughly explored here and applied to a number of examples. Unfortunately, the effectiveness of the formula (using the truncation) is considerably reduced, since (as discussed in detail later) the truncation effectively omits key information about the  lowest energy microstates.\footnote{The original version of this manuscript missed this issue, and took the results obtained from the formula too seriously.}

The importance for $F_Q(T)$ of the underlying microstate physics, particularly at low energy (and hence non-perturbative in character) is the main lesson of the detailed studies presented in  this paper (see also the precursor discussion in ref.~\cite{Johnson:2020mwi}). Several results will be directly computed for many matrix models, and a clear picture  emerges. It is summarized next.

\subsection{Outline of New Results}
\label{sec:summary-of-results}

In the low energy regime, the end of the tail of the spectral density $\rho(E)$ can take a number of different forms as $E{\to}0$, depending upon non-perturbative effects peculiar to each JT gravity model. However, there really is only one non-perturbative feature  that is universal to all the different kinds of model, and it is the appearance of undulations in~$\rho(E)$, increasing in amplitude as~$E$ is reduced. The peaks and troughs in $\rho(E)$ are the signature of the underlying discrete spectrum (or class of spectra) that is being averaged over. It is natural to suppose that this is also to be associated with the underlying discrete black hole spectrum for which JT supplies the near-extremal physics, and hence to interpret results in terms of the statistics of the black holes. The separation between states is $O(\e^{-\#S_0})$ where~$S_0$ is the extremal entropy. Since $S_0{\sim}1/G$, where~$G$ is  Newton's constant,  this is non-perturbative in the gravity theory. Rather than a sum of $\delta$-functions, the peaks broaden out and merge into the smooth function~$\rho(E)$.  The undulations are invisible at  any order in perturbation theory in small $\hbar{=}\e^{-S_0}$, but emerge in a non-perturbative treatment where $O(\e^{-1/\hbar})$ effects can be incorporated.

\begin{figure}[h]
\centering
\includegraphics[width=0.48\textwidth]{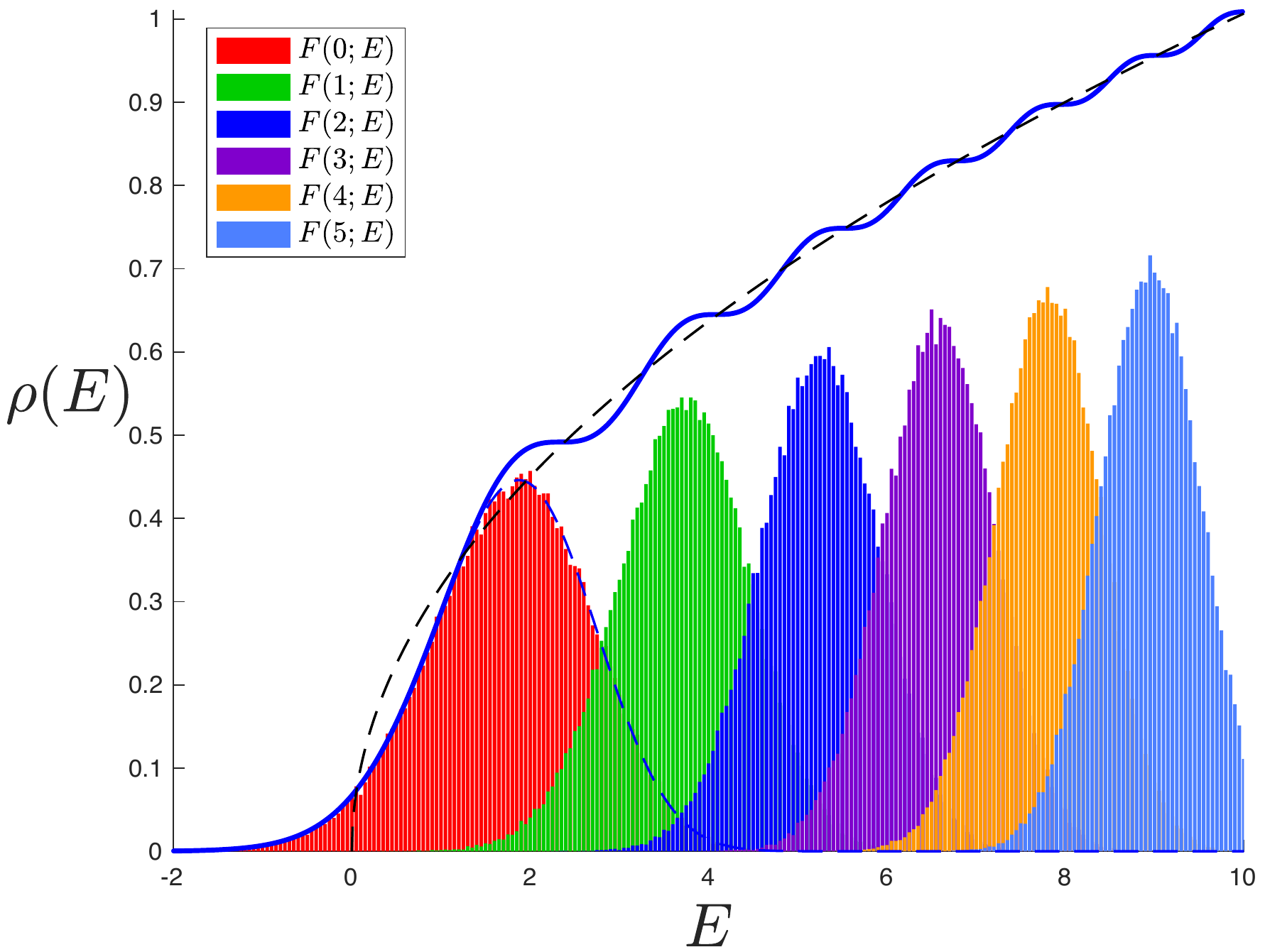}
\caption{\label{fig:toy-densities-2} The spectral density $\rho(E)$ for the Airy model (solid line). The rising dashed line is the leading result $\rhoo(E){=}\sqrt{E}/(\hbar\pi)$. The  histograms   $F(n;E)$ are frequencies of the $n$th energy level, extracted numerically from a  Gaussian random system of $100{\times}100$  Hermitian matrices, for $100K$ samples. Note the  correspondence with the undulations in~$\rho(E)$. The blue dashed peak  is the exact Tracy-Widom distribution~\cite{Tracy:1992rf} $F(0;E)$  for the ground state $E_0$, and  $\langle E_0\rangle\simeq1.77$.}
\end{figure}

The precise description of the states that the peaks hint at can be given by the double-scaled matrix model to which the JT gravity is equivalent.  An instructive  toy example is the popular Airy model, with spectral density shown as  the solid curve in figure~\ref{fig:toy-densities-2}. The dashed curve is the classical (disc order) result~$\rhoo(E)=\sqrt{E}/(\pi\hbar)$. 
Beneath the curve are the individual peaks of the spectrum that were obtained, as an explicit illustration, by numerically sampling (100K times) an ensemble of Gaussian distributed random $N{\times}N$ Hermitian matrices for $N{=}100$. The statistics of the neighbourhood of the endpoints can be readily studied in this way, zooming in on the scaled region of the endpoint as appropriate for the double-scaling limit. (Section~\ref{sec:free-energy-direct} will describe this useful procedure in detail.) Histograms can be generated for the frequency $F(0;E)$ of the location of the $0$th excited state (the ground state), the first ($F(1;E)$), and so on.\footnote{These peaks are familiar from the work of Forrester\cite{FORRESTER1993709}, and Tracy and Widom\cite{Tracy:1992rf}, (and many works since then) on the statistics of the endpoints of various random matrix distributions. For example, ref.~\cite{Tracy:1992rf} showed that the function $F(n;E)$   can be obtained by solving  ordinary differential equations of Painlev\'e type. It will emerge later in this paper that analogous results (some exact) by Edelman~\cite{EDELMAN199155}, and by Forrester and Hughes~\cite{doi:10.1063/1.530639}, will be relevant for a related class of models, of Bessel type.} The first six $F(n;E)$ are  shown in figure~\ref{fig:toy-densities-2}, showing the peaks that develop. Indeed, they  align with the undulations of the smooth Airy model density, $\rho(E)$. Their sum coincides with it precisely.

\begin{figure}[t]
\centering
\includegraphics[width=0.48\textwidth]{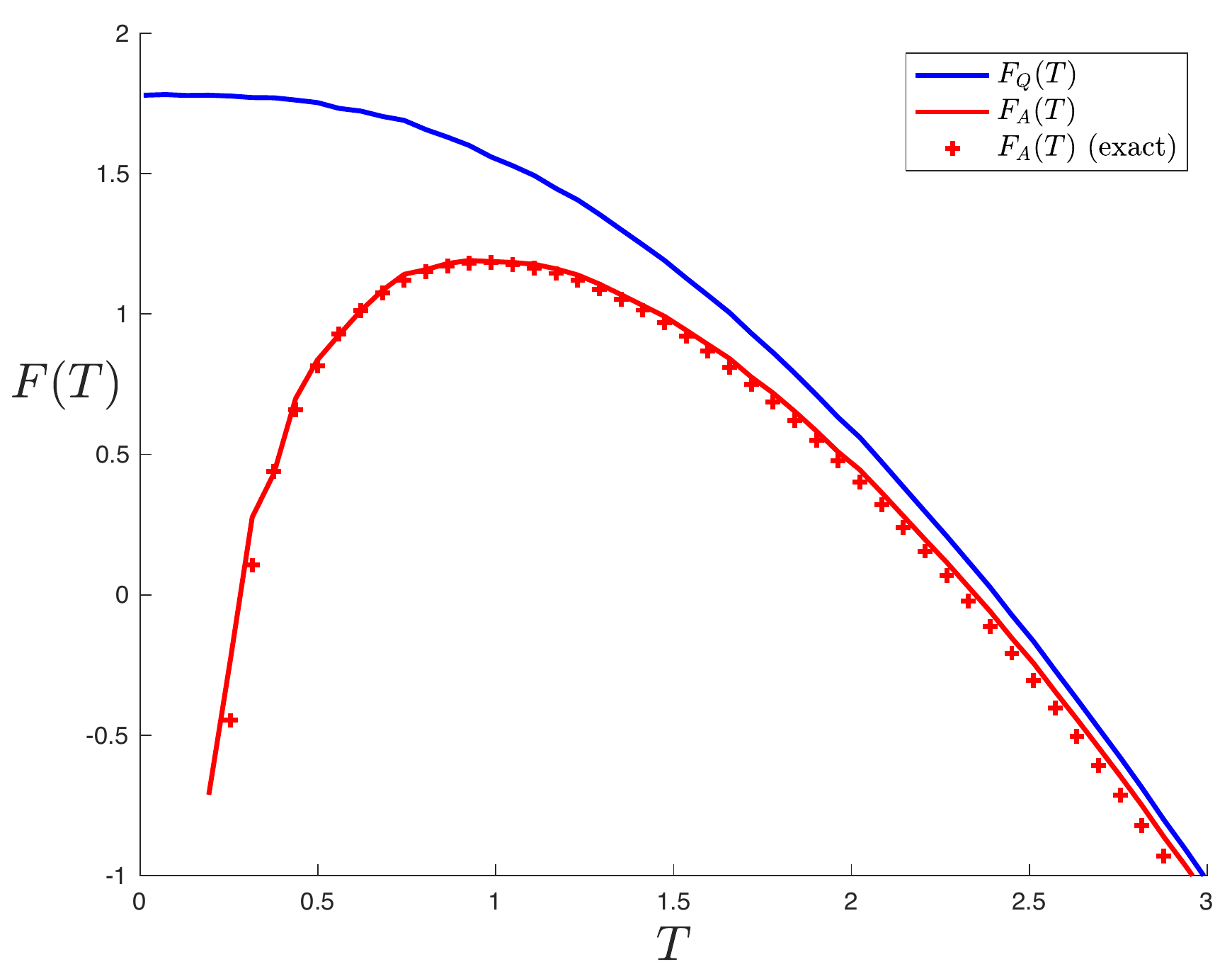}
\caption{\label{fig:OMFG} The directly computed free energy $F_Q$ (upper, blue) for the Airy model constructed explicitly from $100{\times}100$ Gaussian random Hermitian matrices.   The annealed result,~$F_A$ is in red (lower) and the cross marks are the result of computing $F_A$ from the $Z(\beta)$ obtained by  Laplace transforming the Airy spectral density. Note that $F_Q(0){\simeq}1.77$ the average ground state $\langle E_0\rangle$.}
\end{figure}

This paper will present studies of the quenched free energy for a number of double-scaled matrix models, establishing (and confirming)  some general features by first (Section~\ref{sec:free-energy-direct}) working on matrix models for which much can be computed explicitly by sampling, and casting the results in terms of known analytic results about the matrix ensembles where available. Several models will be studied in this way, with many new and illuminating results for $F_Q(T)$. Some of the results will confirm a recent elegant paper by Janssen and Mirbabayi~\cite{Janssen:2021mek} on the low energy behaviour of $F_Q(T)$, obtained by certain scaling arguments, but the results here are for more general $T$ and for a wider class of models. The output of the studies is that:
\be
F_Q(T)=\langle E_0\rangle+f(T)\ ,
\ee
where $\langle E_0\rangle$ is the average ground state of the system. See for example figure~\ref{fig:OMFG} for the Airy model. 

In all cases studied,   $f(T)$ is a monotonically decreasing function. This makes sense since its form is naturally determined by the details of the available states that successively turn on at higher energies beyond the ground state, and these are captured  at leading order by quantities such as the entropy ($S{=}{-}\partial_T F$) and the specific heat $C{=}T\partial_T S$, which are (in a stable model) positive. These are inherently statistical quantities, and this is reflected in  the results. A simple linear result for $F_Q(T)$ arises  when the energies are in fixed definite states (since the entropy is simply just the logarithm of their number; Matrix models where this happens will be discussed), and more general behaviour comes from richer underlying statistics.  At low $T$, $f(T){=}{-}A T^p+\cdots$, where $A$ and the power $p$ are determined by the statistical distribution of the gap $r{=}E_1{-}E_0$ between the first excited state and the ground state.  (This was first shown in ref.~\cite{Janssen:2021mek} and is confirmed here in several new cases.) For  a class of Bessel-type models presented, the gap is even tunable, and  the leading dependence for the statistics of $r$ can be determined analytically. The result is that the leading form of $F_Q(T)$ can be written in closed form, and directly tested (successfully) by sampling many members of the ensemble.

After establishing several such  results in the context of the exact models, the attention  turns (Section~\ref{sec:the-formula}) to the study of  formula~(\ref{eq:okuyama-formula}) as a tool for working out $F_Q(T)$ in cases when direct sampling of the underlying matrices is either not practical or not desirable. Such a tool is needed for the full models of JT gravity (and variants). Several general features of the formula are unpacked  for the first time (see Section~\ref{sec:general-properties}). A particularly interesting feature is how the low temperature physics is entangled with the behaviour of the $x{\to}\infty$ regime of the integral at the heart of the formula. It has features reminiscent of an RG flow, and deserves further study.

 The question of the effectiveness of the low temperature truncation (obtained by using a simplified form for the correlators; see Section~\ref{sec:truncation}) is a focus, and it can be answered by studying the same  models discussed exactly earlier, and comparing results. The overall observation is that the truncated formula only gives a rough approximation to the correct low energy results. This is attributable to the fact that the full form of the connected correlators contain information that is equivalent to knowing all the statistical properties of individual peaks  that was observed earlier to emerge in $F_Q(T)$.  The truncation of parts of the correlators throws most of that information away, by essentially using only the peaks' sum, $\rho(E)$, in the computation.  For example, applying the truncated formula to the Airy example gives $F_Q(0){\simeq}2.51$ as shown in  figure~\ref{fig:free-airy-approx}. (Note that ref.~\cite{Okuyama:2021pkf} also presented this result). This value is considerably higher than the correct value of $\langle E_0\rangle{\simeq}1.77$ (seen in figure~\ref{fig:OMFG}), and moreover the fall--off from there is quadratic (instead of quartic), reflecting the absence of the information about the statistics of $E_1{-}E_0$.  This is one of several examples presented in Section~\ref{sec:FQ-for-Bessel-Airy} where the known results from the exact models are used to benchmark the truncated formula. In Section~\ref{sec:full-JT-studies} the truncated formula is deployed on more complete JT gravity and supergravity models, where enough intuition has now been developed (from the toy models) to assess the results' correctness. 

Generally speaking, it  emerges that the truncated formula seems to settle on a $T{=}0$ value that corresponds to the energy at which the density $\rho(E)$ changes most slowly,  in the low energy regime. For some models, this is in the rough neighbourhood of the  average ground state, but in some cases the truncated formula gets it considerably wrong. The full formula~(\ref{eq:okuyama-formula}) remains potentially quite powerful, it must be said, but generally less practical since it requires the complete form of the correlators to be input. As mentioned in the closing remarks in Section~\ref{sec:closing-remarks}, it would clearly be of interest to seek  a different simplification scheme (perhaps an alternative low energy truncation scheme on the formula itself, and not the individual correlators)  that is more accurate or well-controlled.

{\it Note added:} Since this manuscript appeared, further research by the Author has shown how to compute the statistics of the individual energy  states of the full JT gravity spectrum, allowing a computation of the JT gravity $F_Q(T)$. Since the methods used are somewhat different from this paper, the results are reported elsewhere~\cite{Johnson:2021zuo}.

\section{Free energy by Direct \,\,\, Enumeration}
\label{sec:free-energy-direct}

The random matrix model of $100{\times}100$ Hermitian matrices used to generate data for illustration purposes (figure~\ref{fig:toy-densities-2}) can also be used for direct evaluation of the quenched  (and annealed) free energy. The results can then be used later to benchmark other methods for  computing $F_Q(T)$, since, as the saying goes, the numbers don't lie. In fact, it is very worthwhile carefully carrying out this procedure for both classes of matrix model that underlie the most well-studied JT gravity and JT supergravity examples. They are ({\bf A}) Hermitian random $N{\times}N$ matrices $H$ where there is no restriction and  ({\rm B}) Hermitian random matrices in ``Wishart''~\cite{10.2307/2331939,doi:10.1063/1.1704274} form $H{=}MM^\dagger$ where $M$ is a Gaussian random complex matrix. Case~{\bf A} leads, in the Gaussian case at large $N$, to the Airy model. Case~{\bf B} leads to (a family of) Bessel models in the large $N$ limit. 

\subsection{Random Matrix  Testbed: Airy Model}
\label{sec:testbeds-A}

It is straightforward (with a few lines of code in most off the shelf computational suites) to simply generate randomly (with a Gaussian probability)  an $N{\times}N$ Hermitian matrix. Here $N$ will be chosen to be reasonably large. Another line or two can produce a list (ordered, if desired) of the eigenvalues, $e_i$, of that matrix. Storing the results and doing it repeatedly  is all one needs in order to do some illustrative experimental work on the statistics of this class of matrix. To be concrete, {\tt MATLAB} was used and~$N$ was chosen to be 100, and even after a few thousand random samples, a histogram of the frequency of the eigenvalues shows the emergence of  the famous Wigner semi-circle law~\cite{10.2307/1970079} obeyed by the energy eigenvalues. The endpoints are at $\pm2\sqrt{N}$, and so rescaling so that $\lambda{=}e/\sqrt{N}$, the unit-normalised density approaches  ${\tilde\rho}(\lambda){=}\sqrt{4-\lambda^2}/2\pi$.  This is so famous a result as to not warrant a figure. 
More interesting is a focus on the endpoints at large~$N$, which is what the double-scaling limit involves. (Up to orientation they are identical, so the left one will be chosen.)  A consultation of refs.~\cite{Bowick:1991ky,FORRESTER1993709} shows that  working in the $e$ units where the inter-energy spacing is $O(1)$, the double-scaled physics results from blowing up the neighbourhood of endpoint energies by a factor  $N^\frac16$. Any pure numerical factor will do, but the simple choice $E=N^\frac16(e+2\sqrt{N})$ turns out to be a correct normalization to match the conventions of this paper when $\hbar=1$. The same data set that was used to generate the semi-circle law can be used now (assuming one was prudent to keep track of the ordering of the spectrum in each sample) to generate histograms for the smallest energy $E_0$, the next smallest, $E_1$, and so forth. The {\it sum} of these histograms is the famous Airy spectral density so beloved of discussions of the leading tail of JT gravity. The equation for this density is:
\be
\label{eq:airy-density}
\rho(E) =  \hbar^{-2/3}\left[({\rm Ai}(\zeta)^{\prime})^2 -\zeta{\rm Ai}(\zeta)^2 \right]
\ee
where $\zeta{=}{-}E/\hbar^{2/3}$. (The leading (disc) contribution in this model is $\rhoo(E){=}\sqrt{E}/(\pi\hbar)$.)  When $\hbar{=}1$, $\rho(E)$ matches the scaling chosen above for the zooming-in.\footnote{There is a potential confusion here about this correspondence, since readers will also be thinking of $\hbar$ as something to do with $1/N$ from other discussions of the double-scaling limit. This confusion goes away upon realizing that $\hbar$ is $1/N$ times a scaling parameter $\delta$ to some negative power, where $\delta\to0$ as $N\to\infty$, leaving a finite constant $\hbar$, whose value can then be chosen at will in the scaled theory.}  

The individual energy histograms and the overall Airy envelope were plotted in the Introduction as figure~\ref{fig:toy-densities-2}. (For $N{=}100$ and 100K samples, this takes less than a minute to generate on a typical desktop computer.) This exact alignment between the peaks and the Airy density, and hence the precise meaning of the non-perturbative undulations of the matrix model spectrum is perhaps less well-appreciated (and certainly less frequently explicitly demostrated) to people working on the gravity side of things (although it is apparently second-nature to the statistical mechanics community) and so a figure was worthwhile. 

It would be  neglectful to not use these same datasets to examine the free energy of the matrix model.\footnote{The Author is  embarrassed that doing this did not occur to him until recently.} What this means pragmatically here is simply pick a value of $\beta$ and compute  $Z(\beta){=}{\rm Tr}[{\rm e}^{-\beta H}]$ for a given sample, which in the scaling limit  is just: $Z(\beta){=}\sum_i^N{\rm e}^{-\beta E_i}$.  Now simply run over all the samples, as before,  doing this evaluation every time, storing the results. To form the quenched free energy for the ensemble average, take the logarithm of each $Z(\beta)$ thus computed and then compute the average: $F_Q(\beta){=}{-}\beta^{-1}\langle\log Z(\beta)\rangle$. Taking the average and then the logarithm gives the annealed free energy $F_A(\beta){=}{-}\beta^{-1}\log\langle Z(\beta)\rangle$. Looping over a range of~$\beta$ values completes the job, and the results are presented in figure~\ref{fig:OMFG}, also already displayed in the Introduction.

Several features are worth noting. The first is that in contrast to $F_A(T)$ this result gives, as it should, a manifestly positive entropy (above the extremal value),  $S{=}{-}\partial F_Q/\partial T$, all the way down to $T=0$ where it vanishes. The second feature that will occupy lots of discussion to come, is the value~$F_Q(0)$. It approaches~$1.77$, which is a good approximation to the value of the average ground state energy, whose distribution is the first (red) peak in figure~\ref{fig:toy-densities-2}. This distribution is often referred to as the Tracy-Widom distribution, famous for its appearance in a range of statistical physics contexts. It is amusing  to see it in this black hole/JT gravity/quantum chaos context. This  robust  result (and others like it) will  later be used as a benchmark for  computations  to follow.

\begin{figure}[h]
\centering
\includegraphics[width=0.4\textwidth]{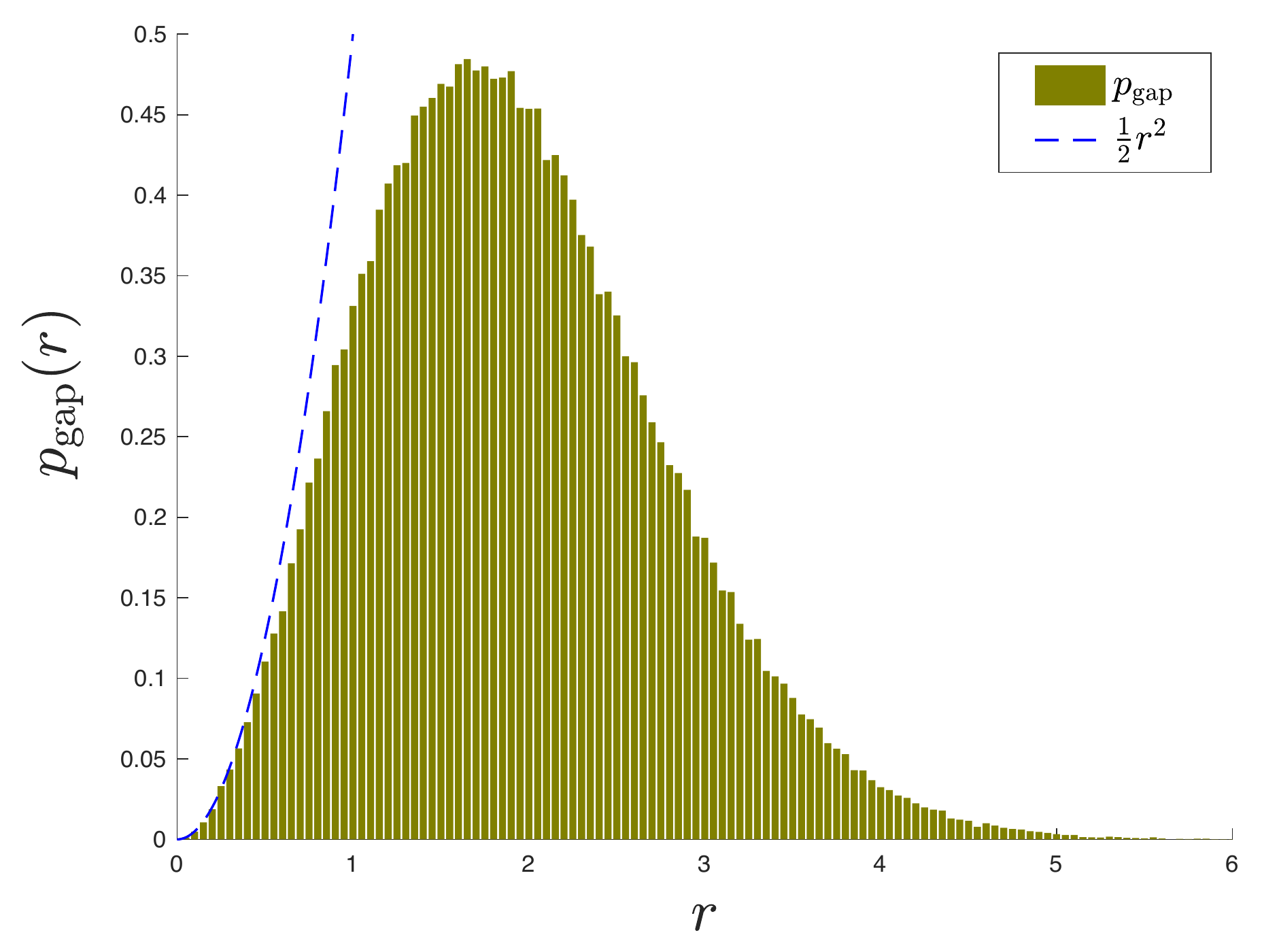}
\caption{\label{fig:gap_distribution_airy} The distribution of the gap between the ground state and the first excited state for the Airy case, compared to the leading dependence. }
\end{figure}

It should also be pointed out that this result was conjectured in ref.~\cite{Okuyama:2020mhl}, and (up to a scaling of conventions) explicitly argued in ref.~\cite{Janssen:2021mek} using scaling arguments on the matrix integral itself. Moreover,  ref.~\cite{Janssen:2021mek} computed the dependence for the leading low $T$ dependence to be $-7\pi^4T^4/360$, which seems to match this result nicely from $T{=}0$ out to about $T\sim0.25$.  Ref.~\cite{Janssen:2021mek} shows that this term is controlled by the leading dependence  of the probability distribution $p_{\rm gap}(r)$ of the  spacing $r{=}E_1{-}E_0$ between the ground state and the first excited state, according to:
\be
\label{eq:leading_FQ}
F_Q(T)=\langle E_0\rangle-T\!\int_0^\infty \!\! dr \,\,p_{\rm gap}(r)\log(1+{\rm e}^{-r/T}) +\cdots\ ,
\ee
and for the Airy case, $p_{\rm gap}(r)=\frac12 r^2+\cdots$, the leading dependence found by Perret and Schehr~\cite{Perret_2014} (see also refs.~\cite{FORRESTER2007,Witte_2013}), gives the numerical coefficient. It is straightforward to extract this distribution   numerically from the data already gathered. See figure~\ref{fig:gap_distribution_airy} with the leading dependence superimposed. The distribution for the gap for the spectrum of other examples should be expected to play similar roles when understanding the quenched free energy, and several cases will be studied below. A straightforward rule of thumb will be that a leading power law $p_{\rm gap}\sim r^n$ is to be expected (since the distribution starts out from zero), and so the integral will yield an overall $-T^{n+2}$ leading dependence.

\subsection{Random Matrix  Testbed: Bessel Models}
\label{sec:testbeds-B}
Given the success of the previous subsection, it makes sense to carry out the exercise for the case where $H{=}MM^\dagger$, where $M$ is a Gaussian randomly generated complex matrix. The eigenvalues of $H$ are manifestly positive. The system can be usefully thought of as the previous Hermitian system, but with a ``wall''   placed at $E{=}0$, stopping them from flowing to negative values.  A simple modification of the lines of code used last time can be used to generate $H$.  This will give a model which will be labelled as~$\Gamma{=}0$. Other integer~$\Gamma$ can easily be obtained by a further modification: The complex matrix~$M$ need not be square. So, start instead with a random~$M$ that is $(N+\Gamma){\times}(N+\Gamma)$, and then simply delete either~$\Gamma$ rows or~$\Gamma$ columns before forming $H$. An important note is that that in the case where rows are chosen, $H$ will have $\Gamma$ repeated zero eigenvalues. In the Altland-Zirnbauer~\cite{Altland:1997zz} classification of random matrix ensembles, this will be the  $(\boldsymbol{\alpha},\boldsymbol{\beta}){=}(2\Gamma+1,2)$ system, where the case with rows deleted has $\Gamma$ negative and the case with columns has~$\Gamma$ positive. In fact, $\Gamma$ can also be half-integer in this classification scheme, but that isn't accessible here in terms of counting rows and columns of these matrices. Nevertheless much will be learned  with just integer~$\Gamma$. 

As before, large numbers of samples  of the case $N{=}100$ can be readily generated, and a pattern emerges. The analogue of Wigner's semi-circle law  in this case is the Marchenko-Pastur law~\cite{zbMATH03259601}, a shape with the same~$e^\frac12$ behaviour at one end (away from the wall), but a~$e^{-\frac12}$ divergence at the wall end. The end is at $4N$ this time ($H$ being instead the product of two matrices), and so the inter-energy spacing is of order 1 here. Defining $\lambda{=}e/4N$, the unit-normalized density is ${\tilde\rho}(\lambda){=}\sqrt{2-\lambda}/(\sqrt{\lambda}\pi)$.

The random matrix model literature on the wall endpoint (now our interest) of such distributions often refers to them as ``hard-edge'' (in contrast to the unconstrained ``soft-edge'' of the semi-circle distribution that gives rise to Airy after scaling). In this case, a Bessel model instead arises~\cite{FORRESTER1993709,Tracy:1993xj}. The analogue of the zooming-in change of variables done in the previous section turns out to be $E=2Ne$, following from the fact that having a hard edge for the  eigenvalues to bump into is similar to having neighbours on both sides, unlike for the soft edge.

\begin{figure}[t]
\centering
\includegraphics[width=0.48\textwidth]{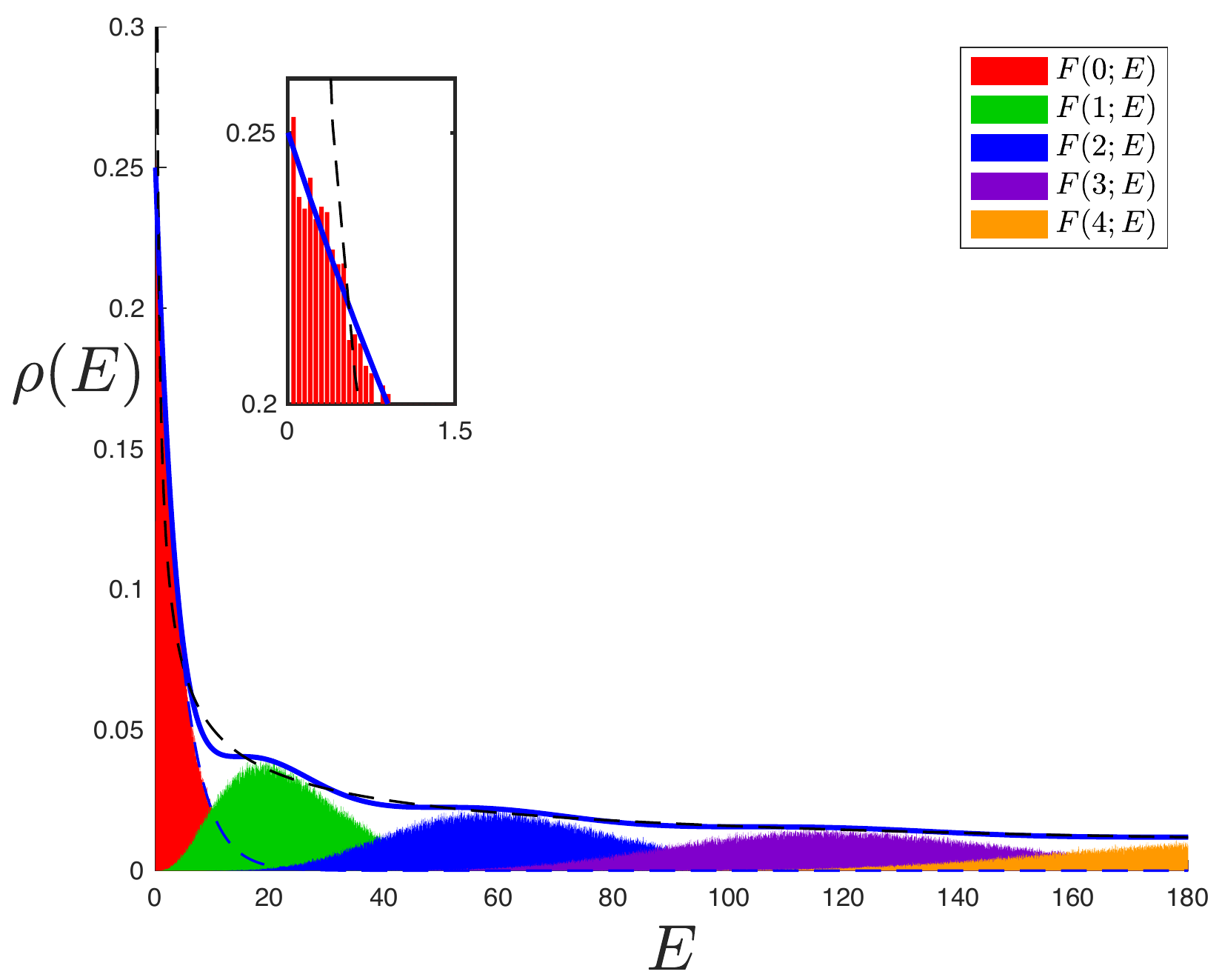}
\caption{\label{fig:toy-densities-3} The spectral density $\rho(E)$ for the $\Gamma{=}0$ Bessel model (solid line). The  dashed line is the leading result $\rhoo(E){=}1/(2\pi\hbar\sqrt{E})$. The  histograms  of $F(n;E)$ are frequencies of the $n$th energy level, extracted numerically from a  Gaussian random system of $100{\times}100$  Hermitian matrices $H{=}MM^\dagger$, for $200K$ samples. Note the  correspondence with the undulations in~$\rho(E)$.   $F(0;E){=}\frac14{\rm e}^{-E/4}$  is the exact form  for the distribution of ground states, with mean   $\langle E_{\rm 0}\rangle{=}4$.}
\end{figure}

Once again, histograms can be made of the statistics of the lowest energy (ground state), next lowest, and so forth, giving distributions called $F(0;E)$, $F(1;E)$, and so on, as before. Again, it is illuminating to see how (with the scaling of the previous paragraph) the peaks line up precisely with the familiar undulations  of the  Bessel models' spectral density  (with $\hbar{=}1$):
\be
\label{eq:bessel-density}
 \rho(E)\! = \frac{1}{4\hbar^2}\!\left[J_\Gamma^2(\xi)\!+\!J_{\Gamma+1}^2(\xi)\!-\!\frac{2\Gamma}{\xi}J_\Gamma(\xi)J_{\Gamma+1}(\xi)\right]\ ,
\ee
 (where $ \xi\equiv{\sqrt{E}}/{\hbar}$) and as noted above, when $\Gamma$ is a negative integer there are $\Gamma$ zero eigenstates, and so  $|\Gamma|\delta(E)$  should be added to the spectral density. The leading form is $\rhoo(E){=}1/(2\pi\hbar\sqrt{E})$.  The case of $\Gamma=0$ is distinguished by having a non-zero and finite value of $\rho(0)=1/(4\hbar^2)$, a non-trivial feature to be discussed later as it is important in the $(1,2)$ JT supergravity and also the stable non-perturbative definition of JT gravity provided in ref.~\cite{Johnson:2019eik,Johnson:2020exp}. This feature is matched by the histogram of the ground state values. See~figure~\ref{fig:toy-densities-3} for the $\Gamma{=}0$ case.

\begin{figure}[t]
\centering
\includegraphics[width=0.48\textwidth]{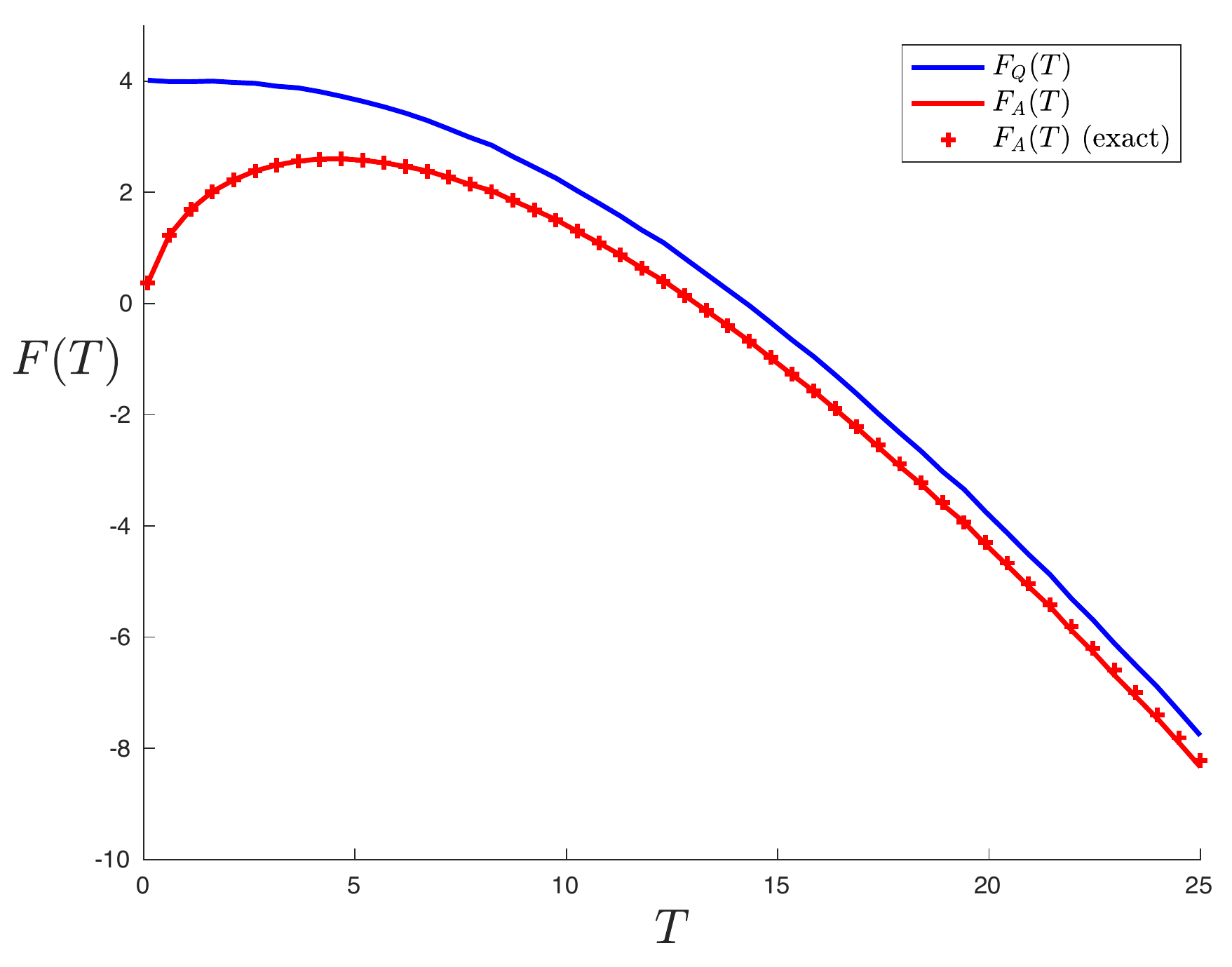}
\caption{\label{fig:bessel-free-1} The directly computed free energy $F_Q$ (upper, blue) for the Bessel model constructed explicitly from $100{\times}100$ Gaussian random  matrices $H{=}MM^\dagger$.   It lands at $\langle E_0\rangle{=}4$, the exactly known value of the averaged ground state. The annealed result,~$F_A$ is in red (lower) and the cross marks are the result of computing $F_A$ from the $Z(\beta)$ obtained by  Laplace transforming the  spectral density.}
\end{figure}

Similarly to the previous case, $F_Q(T)$ and $F_A(T)$ can be readily computed, with the result given in figure~\ref{fig:bessel-free-1}. As with the Airy case, as $T{\to}0$, $F_Q(T)$ approaches (with very good accuracy) the mean value of the ground state, which in fact is exactly $\langle E_0\rangle{=}4$, a satisfying result. The exact result follows from the fact that~\cite{10.1137/0609045,EDELMAN199155,FORRESTER1993709}   the distribution of ground states is known exactly as:
\be
\label{eq:exact-ground-1}
F(0;E){=}\frac14{\rm e}^{-\frac{E}{4}}\ . 
\ee
Folding in a factor of $E$ and integrating gives the result.

\begin{figure}[t]
\centering
\includegraphics[width=0.4\textwidth]{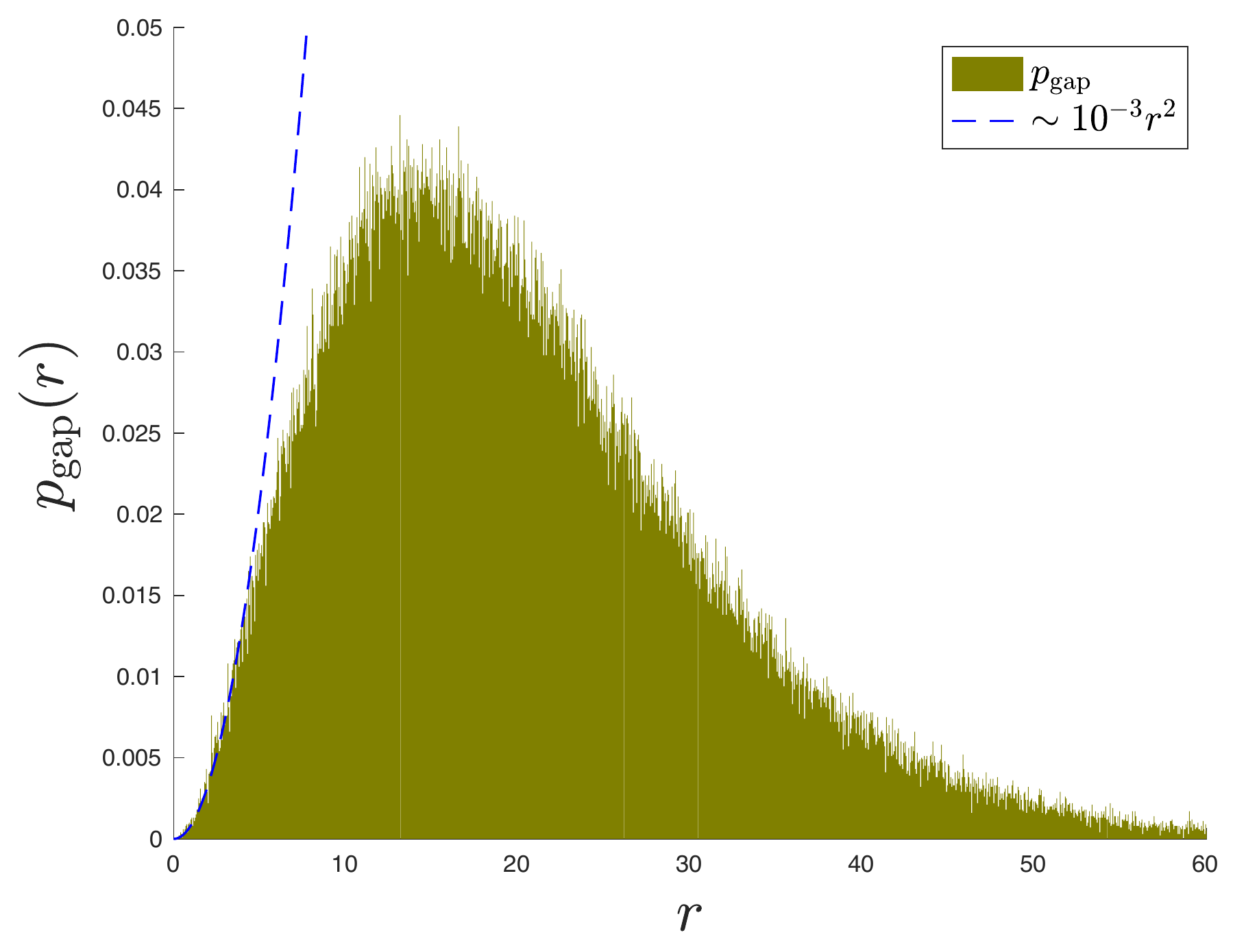}
\caption{\label{fig:gap_distribution_bessel} The distribution of the gap between the ground state and the first excited state for the $\Gamma{=}0$ case, compared to an estimate of the leading dependence. }
\end{figure}

The relatively flat approach to the the $T{=}0$ axis  has a natural explanation too. As with the previous case, its leading form should be obtained by inserting the leading form of the distribution $p_{\rm gap}(r)$ of the difference between the ground state and the first excited state. It is a different function for this matrix model, and the functional form does not appear to be in the literature. The behaviour can be extracted from the data numerically however, and it is given in figure~\ref{fig:gap_distribution_bessel}.  The leading part seems likely to be well fit by quadratic bahaviour again, but with a much smaller coefficient of order $10^{-3}$ (what is sketched there is $r^2/1200$, but it would be useful to determine the precise coefficient). This  results in a correspondingly smaller~$-T^4$ behaviour at small $T$ for this case as compared to the Airy case.

\begin{figure}[h]
\centering
\includegraphics[width=0.48\textwidth]{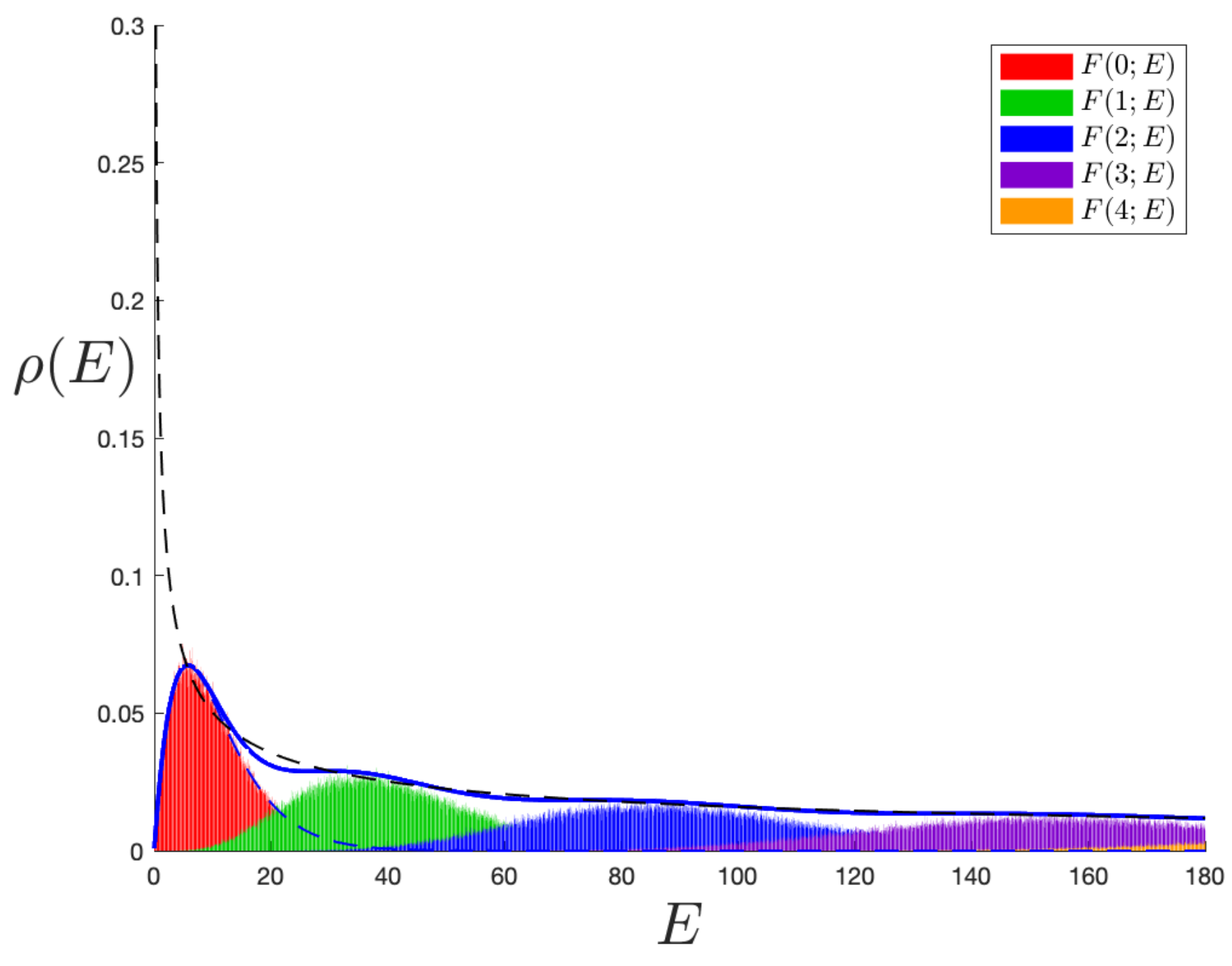}
\caption{\label{fig:toy-densities-4} The spectral density $\rho(E)$ for the $\Gamma{=}1$ Bessel model (solid line). See figure~\ref{fig:toy-densities-3}'s caption for more details.  $F(0;E)$  is the distribution for the ground state, and the mean value is $\langle E_{\rm 0}\rangle{=}4\e\simeq10.9$. See text.} 
\end{figure}

Exploring both negative and positive integer $\Gamma$ is  very instructive, giving new classes of behaviour. Taking $\Gamma{=}1$ first, the analogous results for the spectrum are given in figure~\ref{fig:toy-densities-4}. 

In this case, $\rho(0){=}0$ and the ground states are bunched away from the wall. It turns out that their distribution can be worked out from expressions in ref.~\cite{FORRESTER1993709} to be:
\bea
\label{eq:exact-ground-2}
F(0;E)&=&-\frac{d}{dE}\left(\e^{-\frac{E}{4}}I_0(\sqrt{E})\right)\nonumber\\
&=&\frac14 \e^{-\frac{E}{4}}\left(I_0(\sqrt{E})-\frac{2}{\sqrt{E}}I_1(\sqrt{E})\right)\ ,
\eea
where $I_n(x)$ is the $n$th modified Bessel function in $x$, and  computing the mean energy from this gives exactly $\langle E_0\rangle{=}4\e\simeq10.9$.

\begin{figure}[t]
\centering
\includegraphics[width=0.48\textwidth]{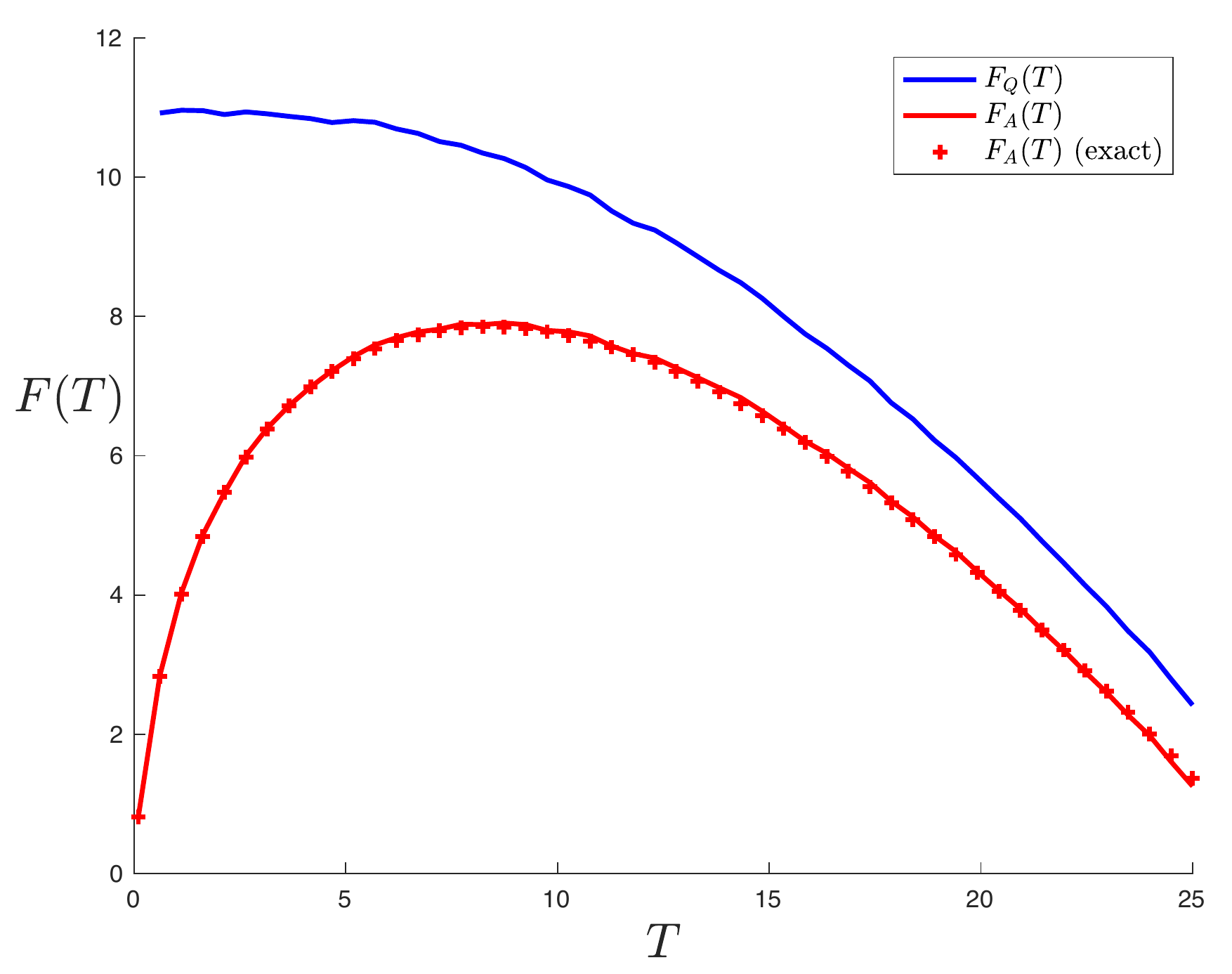}
\caption{\label{fig:bessel-free-2} The directly computed free energy $F_Q$ (upper, blue) for the $\Gamma{=}1$ Bessel model.   The annealed result,~$F_A$ is in red (lower) and the cross marks are the result of computing $F_A$ from the $Z(\beta)$ obtained by  Laplace transforming the  spectral density.  $F_Q(T)$ lands on 4$\e\simeq10.9$ at $T=0$, the exact average value of the ground state. }
\end{figure}

The directly computed free energies, $F_Q(T)$ and $F_A(T)$ are  given in figure~\ref{fig:bessel-free-2}. Like before, as $T{\to}0$, $F_Q(T)$ approaches $\langle E_0\rangle{=}4\e$. The flatness of the approach is intermediate between the two previous cases, and indeed, an examination of the statistics of the gap  $r$ yields a distribution curve qualitatively similar to the previous two cases, where again the leading behaviour  is again quadratic in~$r$, leading to $-T^4$ behaviour.

This all generalizes rather nicely. Higher $\Gamma$ can be readily computed, with values of $F_Q(0)$ read off, and indeed they correspond to $\langle E_0\rangle$ for the smallest energy. Happily, the general distribution   has been written in closed form in ref.~\cite{doi:10.1063/1.530639}:
\bea
\label{eq:general-peak}
F(0;E)&=&-\frac{d}{dE}\left(\e^{-\frac{E}{4}}{\rm det}[I_{j-k}(\sqrt{E})]_{j,k=1\cdots\Gamma}\right)\ .
\eea
For example, for $\Gamma{=}2$, 
\be
F(0;E)=\frac14\e^{-\frac{E}{4}}\left[I_0^2(\sqrt{E})-\left(1+\frac{4}{E}\right)I_1^2(\sqrt{E})\right]\ ,
\ee
and from this can be computed:
\be
\langle E_0\rangle=4\e^2[I_0(2)-I_1(2)]\simeq20.36\ .
\ee
 The pattern from these first few values of $\Gamma$ suggests that $\langle E_0\rangle$ (and hence $F_Q(0)$ for these models) for general $\Gamma{\geq}0$ is $4e^\Gamma{\times}$ multiplied by a constant built out of evaluations of $\{I_0,\hdots I_\Gamma\}$. It would be interesting to work out the exact expression.

\begin{figure}[t]
\centering
\includegraphics[width=0.48\textwidth]{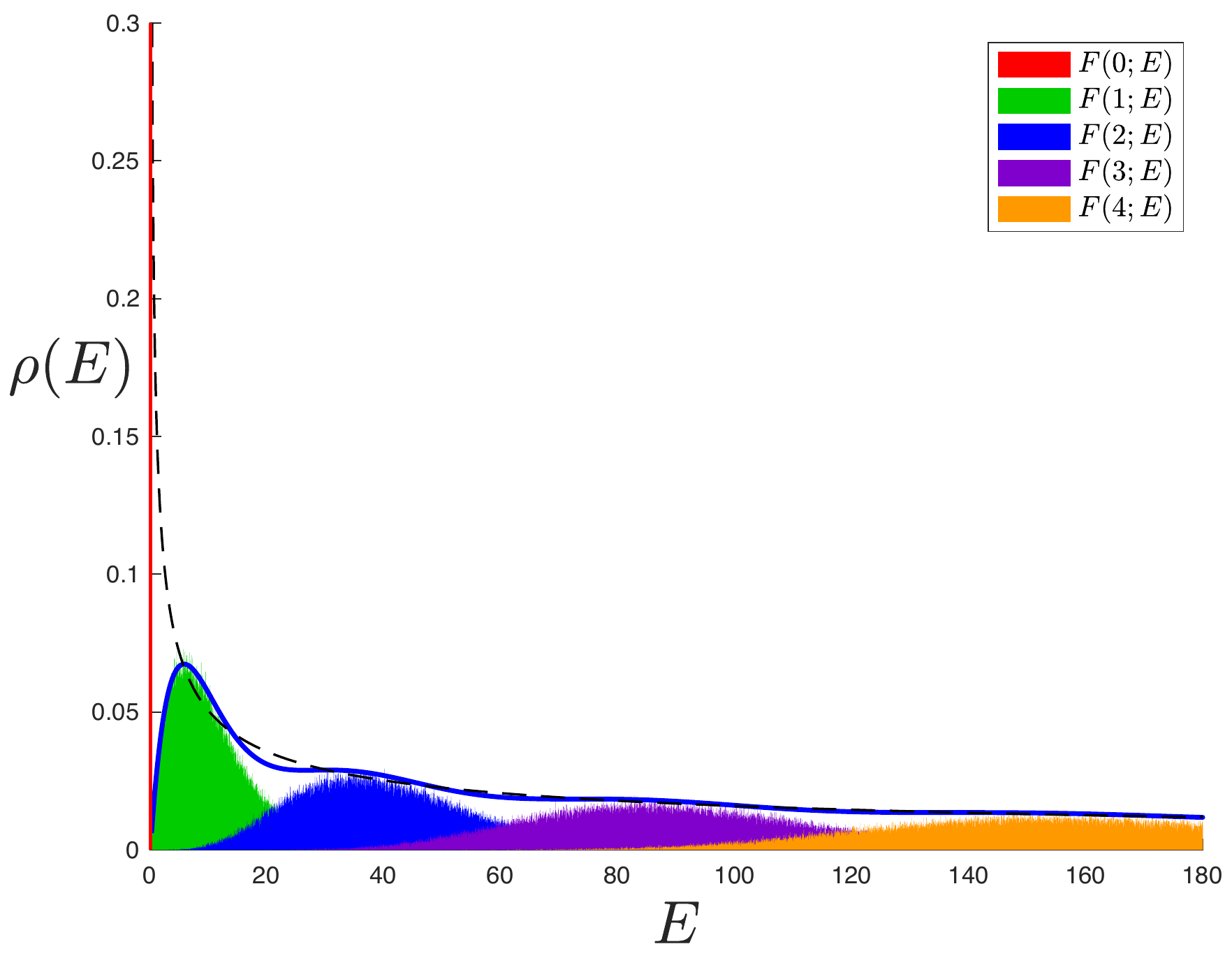}
\caption{\label{fig:toy-densities-5} The spectral density $\rho(E)$ for the $\Gamma{=}-1$ Bessel model (solid line). See figure~\ref{fig:toy-densities-3}'s caption for more details.  $F(0;E)$  is the distribution for the ground state, and here it is a delta function at $E=0$. This should be compared to figure~\ref{fig:toy-densities-3},  the $\Gamma{=}1$ case.}
\end{figure}

\begin{figure}[t]
\centering
\includegraphics[width=0.48\textwidth]{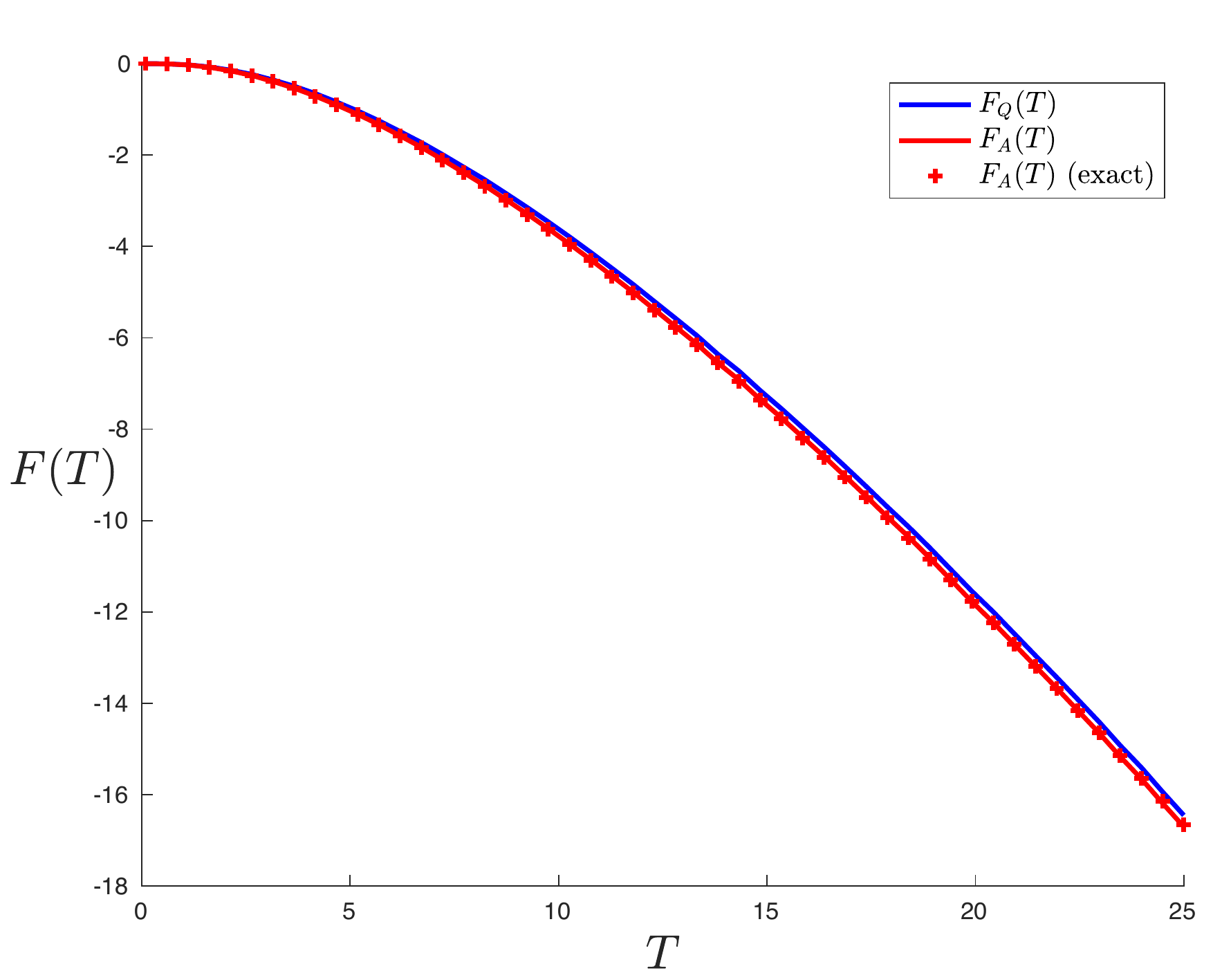}
\caption{\label{fig:bessel-free-3} The directly computed free energy $F_Q$ (upper, blue) for the $\Gamma{=}-1$ Bessel model.   The annealed result,~$F_A$ is in red (lower) and the cross marks are the result of computing $F_A$ from the $Z(\beta)$ obtained by  Laplace transforming the  spectral density.}
\end{figure}

It is very interesting to compare these results to the case of $\Gamma{<}0$, starting with $\Gamma{=}{-}1$. Notice that, due to identities satisfied by the Bessel functions, the smooth part of the spectral densities are identical. This generalizes to other cases of non-zero integer $|\Gamma|$. This is why it  is crucial that there is the addition of $|\Gamma|\delta(E)$ to the spectral density for negative $\Gamma$.\footnote{In fact, this reflects observations made in refs.~\cite{Carlisle:2005mk,Carlisle:2005wa} about the nature of solutions of the underlying string equations for integer~$\Gamma$. Moreover the additional zero energy states for negative $\Gamma$ are in accord with the idea presented there that $\Gamma$ (with a sign convention switch) counts threshold bound state solitons of the associated quantum mechanics. Notice also that $|\Gamma|$ corresponds to the integer parameter $|\nu|$ in ref.~\cite{Stanford:2019vob}, which has an interpretation as counting ``Ramond punctures'' in JT supergravity.} This  results in the following relation between the partition functions of the models:
\be
Z(\beta)_{\Gamma>0}=Z(\beta)_{-\Gamma}+|\Gamma|\ ,
\ee
where the $\beta$ dependent part comes from Laplace transforming the smooth function $\rho(E)$ given in equation~(\ref{eq:bessel-density}.) This should make a difference to both the annealed and quenched free energies of the model, since both the partition function and the ground state are significantly different.  This is all borne out in the directly constructed models, as shown (for $\Gamma{=}{-}1$) in figure~\ref{fig:toy-densities-5} for the spectrum, and figure~\ref{fig:bessel-free-3} for the free energy. Indeed,  $F_Q(T)$ goes to zero as $T{=}0$, reflecting the fact that the ground state is zero (not on average, but in all members of the ensemble).

\begin{figure}[t]
\centering
\includegraphics[width=0.48\textwidth]{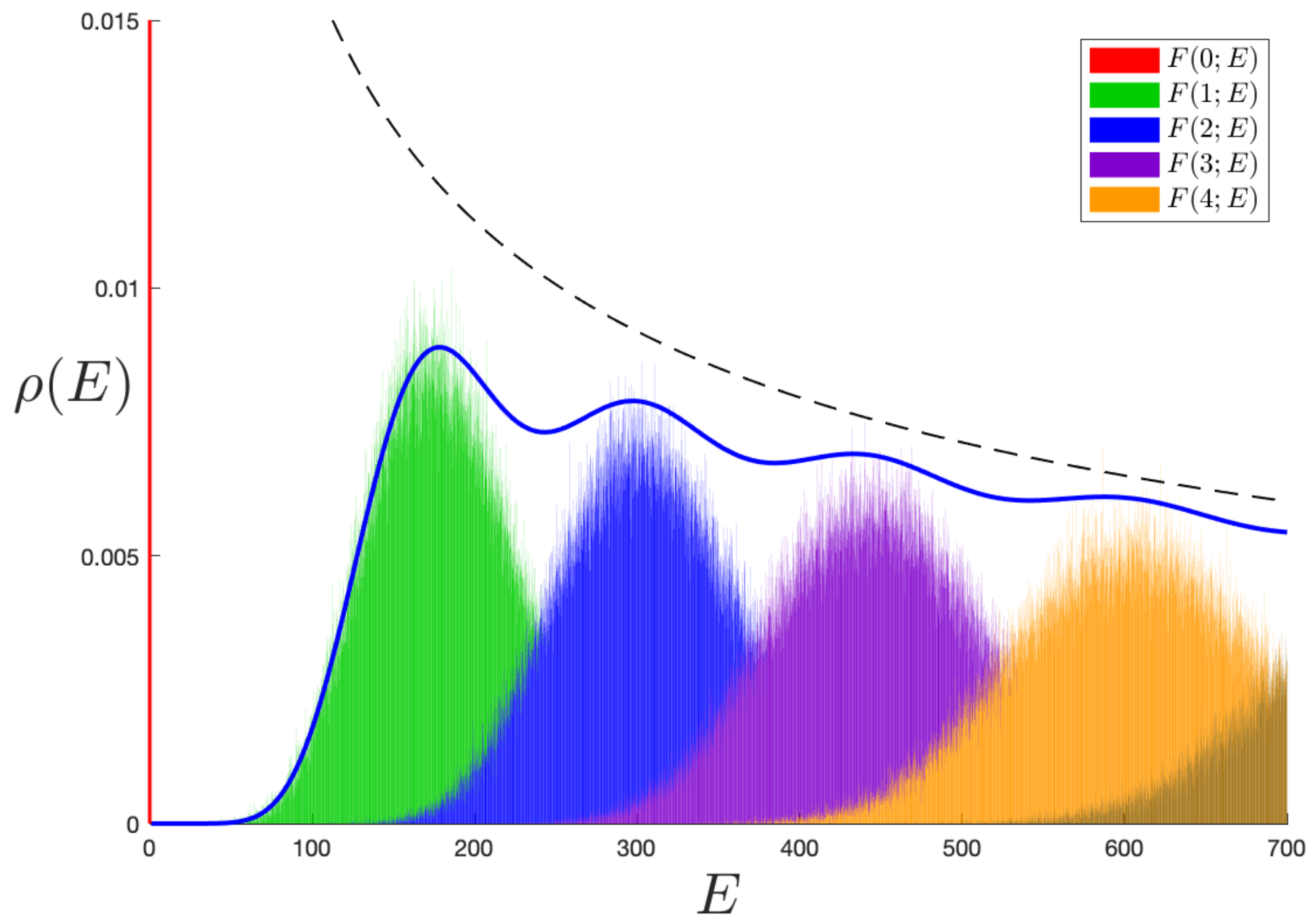}
\caption{\label{fig:toy-densities-6} The spectral density $\rho(E)$ for the $\Gamma{=}-10$ Bessel model (solid line). See figure~\ref{fig:toy-densities-3}'s caption for more details.  $F(0;E)$  is the distribution for the ground state, and here it is a delta function at $E=0$.}
\end{figure}

This latter situation is worth expanding on a little more, as it persists for all negative integer $\Gamma$. In fact, a very interesting situation develops as $|\Gamma|$ increases. The first excited state beyond the degenerate ground state  forms a peak that is increasingly further away from the origin. A gap develops. See figure~\ref{fig:toy-densities-6} for  $\Gamma{=}{-}10$, with a gap  almost 20 times larger than for $\Gamma{=}{-}1$. 

\begin{figure}[t]
\centering
\includegraphics[width=0.48\textwidth]{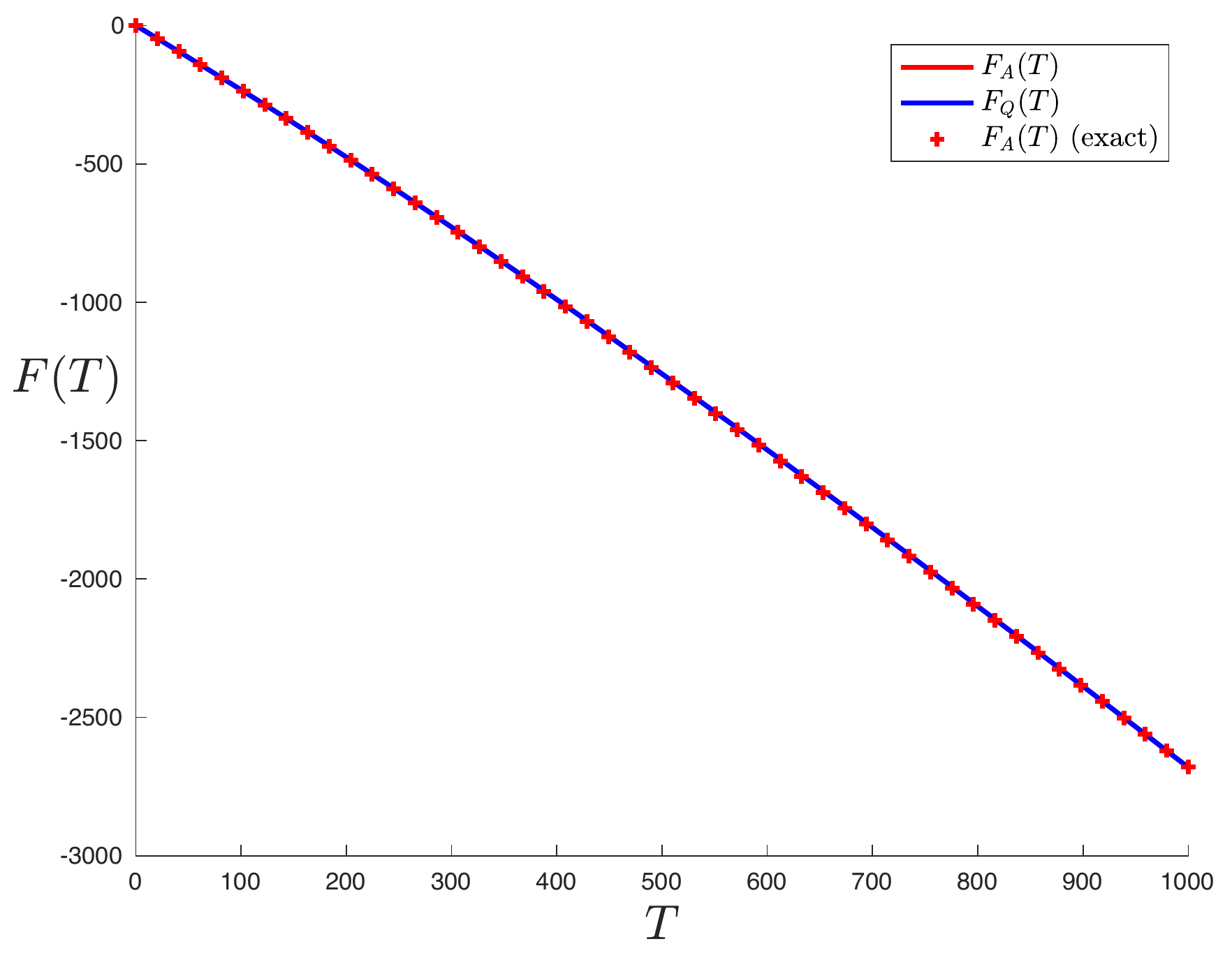}
\caption{\label{fig:bessel-free-4} The directly computed free energy $F_Q$ (upper, blue) for the $\Gamma{=}-10$ Bessel model.   The annealed result,~$F_A$ is in red (lower, completely covered by the $F_Q$ curve) and the cross marks are the result of computing $F_A$ from the $Z(\beta)$ obtained by  Laplace transforming the  spectral density.}
\end{figure}
The result is that at low enough temperatures (falling below the scale of that first excited state), the physics is dominated by the ground state occupancy, and $F_Q(T)$ and $F_A(T)$ become locked together, approaching the origin increasingly linearly (for $\Gamma{<}{-}1)$. It is easy to see why by taking the limit of large $|\Gamma|$ and simply ignoring the other states. Then $Z(\beta){=}|\Gamma|$, with the result 
\be
\label{eq:self-averaged}
F_Q(T)=F_A(T) = -T\log|\Gamma|\ .
\ee
In a sense, the theory crosses over into what might be described as a purely self-averaged phase at low enough temperature scales. The free energy plot in figure~\ref{fig:bessel-free-4} for the case $\Gamma{=}-10$ shows the dramatic effects in action, with the quenched and annealed energies shadowing each other over many decades of temperature. The purely linear behaviour above becomes an increasingly better approximation as the temperature drops. (Even in the case of $\Gamma{=}{-}1$ (figure~\ref{fig:bessel-free-3}) it is remarkable how swiftly the two approach each other.)

Again it should be expected that the leading small $T$ behaviour away from $T{=}0$ is controlled by the gap between the ground state and the first excited state, but this time things are different. The ground state is exactly zero in these models and therefore the function, $p_{\rm gap}(r)$, for the distribution of the gap $r{=}E_1{-}E_0$ is simply given by the first peak away from $E{=}0$, for which the closed form is in equation~(\ref{eq:general-peak}).  The leading power law is what is needed. A quick way of finding the general form is to realize that the leading edge of the peak coincides with  that of the smooth part of the spectral density~(\ref{eq:bessel-density}). So,  setting $E{=}r$ and expanding gives:
\be
p_{\rm gap}(r){=}\frac{r^{|\Gamma|}}{2^{2|\Gamma|+2}|\Gamma|!(|\Gamma|+1)!}+\cdots\ ,
\ee
 where the first few coefficients of $r^{|\Gamma|}$ are $\frac{1}{32}$, $\frac{1}{768}$, $\frac{1}{36864}$, and so on, rapidly decreasing. For the $\Gamma{=}{-}10$  case discussed it is ${1/607545286050447360000}$. This results in:
\be
F_Q(T) = -T\log |\Gamma|-C_\Gamma T^{\Gamma+2}+\cdots
\ee
as the leading small $T$ behaviour for these models, with the coefficient $C_\Gamma$ (computed using the integral in equation~(\ref{eq:leading_FQ})) rapidly becoming small with $\Gamma$ so that an approximately linear behaviour extends out from zero and persists for increasingly large range of $T$. An exception to this is $\Gamma{=}{-}1$, where the linearity disappears and instead the leading form is  $F_Q(T){=}-3\zeta(3) T^3/128$, where $\zeta(z)$ is the Riemann $\zeta$-function. This fits well with the results presented in figure~\ref{fig:bessel-free-3} for small $T$ up to about~$0.16$. This is about the energy range for which the linear approximation $\rho(E){=}\frac{E}{32}+\cdots$ is  reasonable.

All of the examples in this section, for which the quenched free energy has been directly computed and matched to properties of the spectrum, will serve as useful benchmarks for later results obtained other methods. The reason for exploring other methods is primarily one of practicality: The underlying matrix models needed to construct more complete models of JT gravity type do not necessarily lend themselves to the kind of direct construction performed here, so complementary methods are worth exploring. This is the motivation for studying, for example, the new formula of ref.~\cite{Okuyama:2021pkf}, which will be done next.

\section{Free Energy from a New Formula}
\label{sec:the-formula}

The following  expression was proposed in ref.~\cite{Okuyama:2021pkf}  as method for computing $\langle\ln Z(\beta)\rangle$, in terms of the connected correlators for any number of insertions of the partition function, denoted here as $\langle Z(\beta)^n\rangle_c$: 
\begin{eqnarray}
\label{eq:okuyama-formula}
\langle \ln Z(\beta)\rangle  &=&\ln  \langle Z(\beta)\rangle -\int_0^\infty \frac{dx}{x} \left({\rm e}^{-{\cal Z}(\beta,x)} - {\rm e}^{-x \langle Z(\beta)\rangle}\right)\ ,\nonumber\\
&&\hskip-1.85cm\mbox{{\it i.e.,} after multiplying by $ (-\beta^{-1})$}:\,\,\, F_Q=F_A+F_D\ ,
\end{eqnarray}
where  $F_{Q,A}$ are the quenched and annealed free energies, and their difference is $F_D\equiv F_Q-F_A$. In the above, 
\begin{equation}
\label{eq:calZ}
{\cal Z}(\beta,x) = -\sum_{n=1}^\infty \frac{\langle Z(\beta)^n\rangle_{\rm c}}{n!}(-x)^n \ .
\end{equation}

One way of motivating the formula (as ref.~\cite{Okuyama:2021pkf} does, although there are other derivations given there as well) is by starting with the identity:
\begin{equation}
\label{eq:Frullani}
\int_0^\infty \frac{\e^{-ax}-\e^{-bx}}{x} dx = \log\left(\frac{b}{a}\right)\ ,
\end{equation}
and then from there identify $a=Z(\beta)$ and $b=\langle Z(\beta)\rangle$, giving, after rearrangement:
\begin{equation}
 \ln Z(\beta)  =\ln  \langle Z(\beta)\rangle -\int_0^\infty \frac{dx}{x} \left({\rm e}^{-x Z(\beta)} - {\rm e}^{-x\langle Z(\beta)\rangle}\right)\ .
\end{equation}
The average $\langle\cdots\rangle$ is performed term by term on both sides (objects already averaged, being pure numbers, undergo no further change). The final step is to recognize that $\langle \exp(-x Z(\beta))\rangle$ can be written instead in terms of purely connected terms $\langle\cdots\rangle_c$ in a manner that gives precisely expression~(\ref{eq:okuyama-formula}) (with~(\ref{eq:calZ})). (Of course, $\langle Z(\beta)\rangle_c=Z(\beta)$. Sometimes, in an abuse of notation, averaging brackets on a single copy of $Z(\beta)$ will be dropped in later discussions where there can be (hopefully) no confusion.)

The most pertinent question to ask about the formula~(\ref{eq:okuyama-formula}) is whether it is of practical use for extracting key features of $F_Q(T)$. This will be especially useful for models more complicated than the simple ones of the previous section, for which direct evaluation (by sampling) might not be practical. In order to answer this question, some unpacking is warranted, in order to see just how the formula works.

\subsection{Some General Properties}
\label{sec:general-properties}
 Some intuition for how the formula works  comes from how the basic integral~(\ref{eq:Frullani}) operates. It is a form of   integral usually named after Frullani~\cite{Frullani:1828}, and sometimes also Cauchy~\cite{Cauchy:1823,Cauchy:1827} (see {\it e.g.,} the discussion in refs.~\cite{Ostrowski:1949,Ostrowski:1976}).  The functions in the numerator have the same limits (1~and~0) in the case of $x\to0$ and $x\to\infty$, and are well-behaved in between, guaranteeing a finite result. They cancel each other at the dangerous--looking lower limit, but then decay at different rates as $x$ increases, hence generating a finite difference that (if $b>a$) gives a manifestly positive result. The  result is the measure of how much those rates differ. Note that the fact that the answer is $\log(b/a)$ has nothing to do with the integral being built from exponentials. 
Indeed, for later use it is important to know that there are many useful generalizations of the form, with $\e^{-x}$ replaced by a more general function $f(x)$ (with suitable conditions on its behaviour), for which many interesting results have been established. 


An elementary way of directly tackling the integral~(\ref{eq:Frullani}) is to use integration by parts. For the $\e^{-ax}$ term the next step using this method is:
\be
a\!\int_0^\infty\!\! \log x\, \e^{-ax}+\biggl.  \log x\, \e^{-ax}\biggr|_0^\infty\ ,
\ee
with a similar term (with an extra overall minus sign) for the $\e^{-bx}$ term. The boundary term  vanishes at the upper limit, but is divergent as $x\to0$. However the divergences from the $a$ and $b$ sectors cancel each other as $(1-1)\log x $ in the limit. Finally, changing variables to $w=ax$ the bulk term gives
\bea
\int_0^\infty \!\!\log\left(\frac{w}{a}\right) \e^{-w} dw &=& \int_0^\infty\!\! \log w\,\e^{-w}dw - \log a\!\!\int_0^\infty\! \e^{-w}dw\nonumber\\
&=&-\gamma-\log a\ ,
\eea
since the first term is a standard integral yielding the Euler-Mascheroni constant $\gamma\simeq0.57722$. The $b$ term goes a similar way, and the $\log(b/a)$ result follows. There is a reason for recording here this simple set of manipulations in perhaps (for some) too much detail. Some crucial physics will arise from key deviations from these  steps.

Yet  another way of thinking about the integral will turn out to be useful later. The individual  $\log a$ and $\log b$ parts can be thought of as following from taking the $n{\to}0$  limit of Ramanujan's 
``Master Theorem''\footnote{The history of the theorem is interesting. It is related to earlier results by Glaisher~\cite{doi:10.1080/14786447408641072} and O'Kinealy~\cite{o1874new}. See {\it e.g.,} refs.~\cite{9dd4b453c0c54cc79663b13cd563824c,Amdeberhan:2012} for discussion.}),  arising in the study of Mellin transforms of analytic functions: 
\be
\label{eq:ramanujans-master-theorem}
\int_0^\infty f(x) x^{n-1} dx = \overline{\Gamma}(n)\mathbf{a}(-n)\ ,
\ee
 where ${\overline\Gamma}(n)$ is the Gamma-function\footnote{A line is placed over the Gamma-function here because the symbol $\Gamma$ is already in use in the paper to label different Bessel and JT supergravity models.} and  the $\mathbf{a}(k)$ are defined by the expansion $f(x) {=} \sum_{k=0}^\infty (-x)^k \mathbf{a}(k)/k!$. The $\log a$ and $\log b$  terms in the Frullani result each come from expanding both the  Gamma-function and the $\mathbf{a}(-n)$ coefficient in small ({\it i.e., non-integer}) $n{=}\epsilon$ and extracting the finite part upon sending $\epsilon{\to}0$. For example, in the case of the first exponential, $\mathbf{a}(k){=}a^k$, and using ${\overline\Gamma}(\epsilon){=}1/\epsilon-\gamma+O(\epsilon)$ 
 and $a^{-\epsilon}{=}1{-}\epsilon\log a{+}\cdots$ again yields the ${-}\gamma{-}\log a$ result. Working  similarly for the $b$ exponential yields the $\gamma+\log b$ result.    This procedure\footnote{This is all formalised in the  ``method of brackets'', (reviewed in this Frullani context in ref.~\cite{BravoGonzalezKohlMoll+2017+1+12}), which is  a powerful set of techniques used (for example) in computing Feynman diagrams.}  (and to some extent, the previous one) allows for the  treatment of  cases where there are two different functions in the numerator (hence departing considerably from Frullani form~(\ref{eq:Frullani})), which is more akin to the situations to be tackled in this paper.

With those simple remarks made, turn now to the full formula~(\ref{eq:okuyama-formula}).  In a sense, the departure, $F_D$, of $F_Q$ from~$F_A$ is generated (through the $x$-integral) by how much the quantity ${\cal Z}(\beta,x)$ departs from $xZ(\beta)$.  The first thing to notice is that when $\beta$ is small (high temperature) the connected diagrams, as stated earlier, are all subleading compared to the disconnected ones, and so all that is left of ${\cal Z}(\beta,x)$ in the limit is a truncation to $Z(\beta)x$. Hence, the two terms in the integral cancel. So  indeed at high $T$, $F_Q{\to} F_A$, as it should. At lower temperatures is when interesting physics arises, with the connected correlators playing a more significant role. In {\it generic} such cases, ${\cal Z}(\beta,x)$ will differ significantly from $xZ(\beta)$, in ways that will matter a lot in the integral, especially at large~$x$ and large $\beta$.  

There is an important and instructive non-generic case that can occur that is an exception to this picture. Imagine a situation where there are simply $\Gamma$ states at $E=0$. The spectral density is simply $\rho(E)=\Gamma\delta(E)$ in such a case. (In fact, such cases were seen in the previous section, although with other contributions to the spectrum as well, but this simple special case can be recovered by either taking $\Gamma$ large, or working at energies small enough to fall into the gap between $E=0$ and the other states---the examples allow for that quite naturally in fact, since the gap grows with $\Gamma$.) In such an example, the partition function is $Z(\beta){=}\Gamma$. Multi--point correlators are simply $\langle Z(\beta)^n\rangle=\Gamma^n$, {\it i.e.}, the connected contributions are identically zero. In such a case, for any $\beta$, ${\cal Z}(\beta,x)\to xZ(\beta)$, and so the integral in formula~(\ref{eq:okuyama-formula}) vanishes, leaving, of course, $F_Q{=}F_A=-T\log\Gamma$. This fits nicely with the observations  in the directly constructed Bessel examples of non-zero positive $\Gamma$ in Section~\ref{sec:testbeds-B}, where  the curves of $F_Q(T)$ and $F_A(T)$ approach and follow each other to the origin so closely (see figures~\ref{fig:bessel-free-3} and~\ref{fig:bessel-free-4}).

\subsection{A Low Temperature Truncation}
\label{sec:truncation}
A matrix model definition can be used to compute the $\langle Z(\beta)^n\rangle_c$ fully non-perturbatively, although as $n$ increases, the explicit expression for the quantity can get considerably involved. A general pattern emerges~\cite{Banks:1990df,Moore:1991ir,Ginsparg:1993is,Okuyama:2018aij}, and the first four cases are given here for illustration:
\bea
\label{eq:multi_Z}
\langle Z(\beta)\rangle_c &=&{\rm Tr}[ e^{-\beta{\cal H}}{\cal P}] \nonumber\\
\langle Z(\beta)^2\rangle_c &=&{\rm Tr}[ e^{-2\beta{\cal H}}{\cal P}]-{\rm Tr}[ e^{-\beta{\cal H}}{\cal P}e^{-\beta{\cal H}}{\cal P}]\nonumber\\
\langle Z(\beta)^3\rangle_c &=&{\rm Tr}[ e^{-3\beta{\cal H}}{\cal P}]-3{\rm Tr}[ e^{-\beta{\cal H}}{\cal P}e^{-2\beta{\cal H}}{\cal P}]\nonumber\\
&&\hskip1cm +2{\rm Tr}[ e^{-\beta{\cal H}}{\cal P}e^{-\beta{\cal H}}{\cal P}e^{-\beta{\cal H}}{\cal P}] \nonumber\\
\langle Z(\beta)^4\rangle_c  &=&{\rm Tr}[ e^{-4\beta{\cal H}}{\cal P}]-4{\rm Tr}[ e^{-\beta{\cal H}}{\cal P}e^{-3\beta{\cal H}}{\cal P}]\nonumber\\
&&\hskip1.8cm -3{\rm Tr}[ e^{-2\beta{\cal H}}{\cal P}e^{-2\beta{\cal H}}{\cal P}]\nonumber\\
&&\hskip0.7cm +12{\rm Tr}[ e^{-2\beta{\cal H}}{\cal P}e^{-\beta{\cal H}}{\cal P}e^{-\beta{\cal H}}{\cal P}] \nonumber\\
&&\hskip-0.05cm -6{\rm Tr}[ e^{-\beta{\cal H}}{\cal P}e^{-\beta{\cal H}}{\cal P}e^{-\beta{\cal H}}{\cal P}e^{-\beta{\cal H}}{\cal P}] \ .
\eea
Appearing here (although it won't be used here) is the Schrodinger Hamiltonian ${\cal H}=-\hbar^2\partial_x^2+u(x)$ whose potential $u(x)$ satisfies a non-linear ordinary differential equation (``string equation'' in an older language/context), the nature of which depends upon the JT gravity variant in question. ${\cal H}$ has energy eigenstates $E$ and wavefunctions $\psi(x,E)$. The   projection ${\cal P}$ given in the trace is defined as ${\cal P}\equiv\int_{-\infty}^\mu |x\rangle\langle x|$, with a parameter $\mu$ that can be taken as 0 for JT gravity or 1 for JT supergravity (up to scalings). These are all objects that arise naturally in taking the double-scaling limit of various large $N$ matrix models that are equivalent to JT gravity models. The above statements are fully non-perturbative in $\hbar$ and are therefore the natural non-perturbative tools with which to work for JT gravity.

Generically,  $Z(\beta)$ and its correlators have powers of $\beta$ in the denominator (the traces in~(\ref{eq:multi_Z}) involve  integrals over energy with factors of ${\rm e}^{-\beta E}$, amounting to  Laplace transforms).   As can be seen from the pattern of the examples in equation~(\ref{eq:multi_Z}), a key simplification that happens in the matrix model description is that for all $n$  the leading part of $\langle Z(\beta)^n\rangle_c$ in the $\beta$ large (small $T$) limit is simply $\langle Z(n\beta)\rangle_c=Z(n\beta)$. These should be thought of as the dominant part of the key $n$-legged wormholes that can be used to uncover the physics in the low temperature limit (see figure~\ref{fig:n-point-wormhole}). For example, in the case of the $n{=}2$, used in computing the spectral form factor, it is the ${\rm Tr}[ e^{-2\beta{\cal H}}{\cal P}]$ piece which becomes the plateau saturation value at late times, as demonstrated in refs~\cite{Okuyama:2019xbv,Johnson:2020exp}.

 The simplification is extremely convenient, since by Laplace transform
\begin{equation}
\label{eq:n-partition}
Z(n\beta)=\int_0^\infty\rho(E) {\rm e}^{-n\beta E} dE\ .
\end{equation}
 So all that is needed is knowledge of the full non-perturbative spectral density $\rho(E)$. Obtaining it explicitly can be done for a variety of JT gravity  and supergravity models, sometimes in various limits where an analytical form can be written, and often fully (using numerical techniques). This will be the limit in which many of the computations to appear later  are done. Since closed forms for the connected correlators are hard to write down in general,  this truncation  allows the formula~(\ref{eq:okuyama-formula}) to be  used as a practical tool. 
 
It is  important to understand the robustness of  this truncation\footnote{In fact, when the first version of this manuscript appeared, the limitations of this truncation were not appreciated, leading to stronger claims for the results of using the formula in this way than were justified.}. The results computed with  it will be tested in what is to come (for example against the benchmarking models of the previous section), and it will be seen that the accuracy of its performance is mixed. For the truncation to be a useful tool, it is important to establish when it should work. There will be further discussion later when more examples are in hand, but some general observations can be initially recorded here.  

The main point to be made is that the truncation means that the only input about the spectrum that the formula has is the spectral density $\rho(E)$, but as explicitly illustrated in  the previous section, this is a {\it sum} of  peaks representing the statistical distribution of the spectrum. So there is a loss of information in inserting only this into the formula, since no amount of processing of that sum (without additional data or criteria) can reconstruct all the details of the individual peaks it was made from. This is in one-to-one correspondence with the fact that it is the connected correlators of the model that give information about the various statistical variances in the ensemble averaging  that amount to features like the width of the individual peaks, and so forth. Therefore, if parts of the connected correlators are left out,  there is a sort of ``resolution limit'' on this truncation's accuracy meaning that the peak corresponding to the ground state needs to be rather clear and distinct  in order for the formula~(\ref{eq:okuyama-formula}), with this truncation, to reproduce the correct low energy features with good efficacy. A sharp illustration of this is given by  the special case mentioned at the end of the last subsection, or the integer $\Gamma{<}0$ Bessel models of Section~\ref{sec:testbeds-B}. There, $\rho(E){=}|\Gamma|\delta(E)+{\bar\rho}(E)$, where ${\bar\rho}(E)$ is the rest of the spectrum. Using this alone in the formula will give poor results because the $\Gamma$ definite states result (at low enough temperatures) in cancellations between different parts of the contributions to the $\langle Z(\beta)^n\rangle_c$ shown in equation~(\ref{eq:multi_Z}) in order for them to vanish, and so keeping only the leading part won't allow that to happen.

Putting aside these delicate matters for now, it is prudent to explore the mechanics of the truncation. Using equation~(\ref{eq:n-partition}) gives~\cite{Okuyama:2021pkf} a rather nice re-summed form for  ${\cal Z}(\beta,x)$ from equation~(\ref{eq:calZ}):
\begin{eqnarray}
\label{eq:calZ-density}
{\cal Z}(\beta,x) &=& -\int_0^\infty \rho(E) \sum_{n=1}^\infty \frac{e^{-n\beta E}}{n!}(-x)^n dE\nonumber \\
&=&\int_0^\infty \rho(E) \left(1-e^{-xe^{-\beta E}}\right) dE  \ .
\end{eqnarray}
Another useful re-summed form will appear shortly, but it is worthwhile to pause here to see some consequences of this way of writing things. First, note that for small~$x$,
\begin{equation}
\label{eq:small-x-large-beta}
{\cal Z}(\beta,x) \to \int_0^\infty \rho(E) \left(xe^{-\beta E}+\cdots\right) dE \simeq Z(\beta)x+\cdots
\end{equation}
and so the integrand's two exponential pieces cancel and it vanishes at the lower limit, just as with the Frullani prototype. As $x\to\infty$ on the other hand, the exponential can be set to zero, leaving
\begin{equation}
{\cal Z}(\beta,x) \to \int_0^\infty \rho(E)  dE +\cdots
\end{equation}
which is the total energy of the model (divergent since $\rho(E)$ is unbounded). With the minus sign this results in an infinite suppression for first part of the integrand of equation~(\ref{eq:okuyama-formula}). The second part of the numerator also falls to zero exponentially at large~$x$, again as the Frullani prototype dictates.

It is worth taking a second look at large $x$, this time  a bit less hastily. The exponential suppression in expression~(\ref{eq:calZ-density}) can be counteracted by making $\beta$ large enough. So the region of low temperature (large $\beta$) will  be intimately entangled with the behaviour of the upper limit of the integral, which will present some analytic and numerical challenges later on. This will be returned to, but for now it is worth keeping in mind that naive truncations and other estimations of the integral will need to be done with care when interested in large~$\beta$.

In summary, generically the integrand goes to zero as $x\to0$ and as $x\to\infty$.  There are no poorly behaved parts to the integrand away from these limits, and so the integral can be relied upon to yield a well-behaved contribution to $F_Q(\beta)$. Although in the small $\beta$ limit (high temperature) it must vanish (hence $F_Q\to F_A)$, the nature of the theory's partition function $Z(\beta)$ (or alternatively the form of the spectral density $\rho(E)$ at small~$E$) will determine the details of how the integral will behave as $\beta\to\infty$.  Generically, it is clear that (since the difference between ${\cal Z}(\beta,x)$ and $xZ(\beta)$ grows with $\beta$) the integral will also grow with $\beta$ too, and what the dependence is will be determined next, in some key prototype cases. 

\subsection{A  Simplified Model}
\label{sec:simplified-models}

A preliminary attempt at characterising (using matrix models) the low temperature $F_Q$ of JT supergravity and JT gravity was carried out in ref.~\cite{Johnson:2020mwi}. The idea began with a study of  the three Bessel models that appear as a good models of the low energy tail of the spectral density $\rho(E)$ of the $(\boldsymbol{\alpha},\boldsymbol{\beta})=(2\Gamma+1,2)$ JT supergravity models of ref.~\cite{Stanford:2019vob},\footnote{Here, the $(\boldsymbol{\alpha},\boldsymbol{\beta})$ notation refers to the random emsemble classification scheme of Altland and Zirnbauer~\cite{Altland:1997zz}.}  where $\Gamma=0,\pm\frac12$. They capture not just perturbative physics but the key non-perturbative corrections that modify the leading $\rho_0(E)=\frac{\mu}{\hbar\pi\sqrt{E}}$ behaviour from the disc amplitude. (Recall that $\hbar\equiv\e^{-S_0}$. Also, $\mu$ is a parameter that can be set to unity for the purposes of this paper. It is kept in some formulae here mostly for comparison with earlier literature.)

In particular, there was a focus on the case of $\Gamma=0$,  the $(1,2)$ JT supergravity. It is a case without time-reversal symmetry, with non-trivial perturbative and non-perturbative corrections. In fact, a very interesting feature of this model (not just in the Bessel limit but in the full model solved in  ref.~\cite{Johnson:2020mwi}) is that it   has a non-zero spectral density at zero energy, a fully non-perturbatively generated phenomenon. This alone is interesting, but it is also worthy of study since it makes this supergravity theory rather similar to ordinary JT gravity, which also has a non-perturbatively generated~$\rho(0)$.  It clearly  has an impact on the lowest temperature dynamics, and learning  how is  generally instructive. In the exactly solvable Bessel limit, its value is  $\rho(0){=}\frac{\mu^2}{4\hbar^2}$, and in Subsection~\ref{sec:testbeds-B} it was seen to be the peak of the distribution of ground states. Here, the fact that the average ground state energy  is precisely the {\it inverse} of this will emerge naturally from properties of the formula~(\ref{eq:okuyama-formula}). 

As a first step, as done in ref.~\cite{Johnson:2020mwi}, just the leading part of the large $\beta$ expansion of the partition function 
for this  (1,2) model example will be used
\begin{equation}
\label{eq:12-part-fun}
Z(\beta)_{12}=\frac{1}{4}\frac{\mu^2}{\hbar^2}\frac{1}{\beta}+\cdots = \frac{\rho(0)}{\beta}+\cdots\ ,
\end{equation}
which amounts to keeping only the leading tail of the density $\rho(E)=\rho(0)+\cdots$. It is interesting to use this to compute $F_Q(T)$. For a start, it gives:
\begin{equation}
\label{eq:12-ZBX}
{\cal Z}(\beta,x) = -\frac{\rho(0)}{\beta} \sum_{n=1}^\infty \frac{1}{ n n!}(-x)^n = \frac{\rho(0)}{\beta}(\gamma+\ln(x)+{\rm E1}(x)) \ ,
\end{equation}
where 
${\rm E1}(x)=-{\rm Ei}(-x)$  is the exponential integral function, defined as: ${\rm E1}(x)\equiv\int_{x}^\infty t^{-1}e^{-t} dt$. 
This is an alternative to the resummed form given in equation~(\ref{eq:calZ-density}), but the two can be connected in this simple case. Doing the change of variables $t=x{\rm e}^{-\beta E}$ there gives:
\bea
\label{eq:12-ZBX-alt}
{\cal Z}(\beta,x)&=&-\frac{\rho(0)}{\beta}\int_{x}^0\left(1-{\rm e }^{-t}\right) \frac{dt}{t} =\\
&=& -\frac{\rho(0)}{\beta} {\rm Ein}(x)\equiv -\frac{\rho(0)}{\beta} \sum_{n=1}^\infty \frac{1}{ n n!}(-x)^n \ ,\nonumber
\eea
where the latter is indeed the definition of the entire function ${\rm Ein}(x)$  as a formal series.

The next step is to use this form for ${\cal }Z(\beta,x)$ in   the formula~(\ref{eq:okuyama-formula}), the integral performed, and the  compute the free energy.
It is difficult to proceed exactly from this point (but see section~\ref{sec:new-scales}) although some general features can be uncovered.  
First,  take the limit of  large~$\beta$ on the integrand. 
%
%
It turns out that the leading dependence is {\it linear} in $\beta$. When combined with the overall $-\beta^{-1}$ factor, this means that $F_D$ rises and tends to a {\it constant} value as $T$ approaches zero.   To see the leading linear dependence needs a careful approach to taking large~$\beta$ on the integrand. The integration by parts approach perhaps makes things most clear. First, notice that the $\e^{-xZ(\beta)}$ part of the integral is just like the $\e^{-bx}$ part of the usual Frullani form~(\ref{eq:Frullani}) and so  will always contribute a $\log(Z(\beta))$. This will be important in what is to come. In doing it by integration by parts, the divergence of its boundary term at the lower limit will again be cancelled by that of the $\e^{-{\cal Z}(x,\beta)}$ term, since in that limit they (as already established) increasingly resemble each other. While that boundary term vanishes at the upper limit as before, the corresponding part of the boundary contribution from the $\e^{-{\cal Z}(x,\beta)}$  term is:
\be
\label{eq:boundary-term}
\lim_{x\to\infty}\left(\log x\,\e^{-{\cal Z}(x,\beta)}\right)\ .
\ee
As $x$ goes  large, it is tempting to assume that (as happened before)  the right hand factor will go to zero fast enough to overwhelm the logarithmic growth of the left, but this actually depends on what $\beta$ is doing. Large $\beta$ can slow the growth to give a non-zero result. This term must therefore be treated with more care than for the Frullani case. Treating it as a ratio of two diverging terms, the limit can be written (with the help of L'H\^opital)  as:
\bea
\lim_{x\to\infty} \left(\frac{1}{x{\cal Z}(x,\beta)^\prime \e^{{\cal Z}(x,\beta)}}\right)\!&=&\!\lim_{x\to\infty} \left( \frac{\beta}{\rho(0)(1-\e^{-x}) \e^{{\cal Z}(x,\beta)}}\right)\nonumber\\&&\hskip-1.5cm = \lim_{x\to\infty} \left( \frac{\beta}{\rho(0)(1-\e^{-x}) }\right) = \frac{\beta}{\rho(0)}\ , \label{eq:nice-limit}
\eea
where the fact that ${\cal Z}(x,\beta)^\prime {=} \rho(0)(1-\e^{-x})/(x\beta)$ (in this case) was used, and in the penultimate step the expression was simplified by  assuming $\beta$ is  large, so $\e^{{\mathcal Z}(x,\beta)}{\to}1$.\footnote{It is interesting to note here that the result is the inverse of the value of $x\partial_x{\mathcal Z}(x,\beta)$ as $\{x,\beta\}{\to}\infty$, which somewhat resembles a measurement the scaling dimension of an operator. Perhaps there is a useful RG flow framework within which to characterize the large $\beta$ expansion of the integral.}

The above result is the precise linear $\beta$ behaviour that was sought, at large~$\beta$. It combines with the logarithmic piece already established. There are subleading corrections to be found as well, from the boundary term, and from the rest of the integral that is to be done: 
\be
\frac{\rho(0)}{\beta}\int_0^\infty \log x \left(\frac{1-\e^{-x}}{x}\right)\e^{-{\cal Z}(x,\beta)} \ ,
\ee
for which   no ready simplification presents itself.

\begin{figure}[b]
\centering
\includegraphics[width=0.48\textwidth]{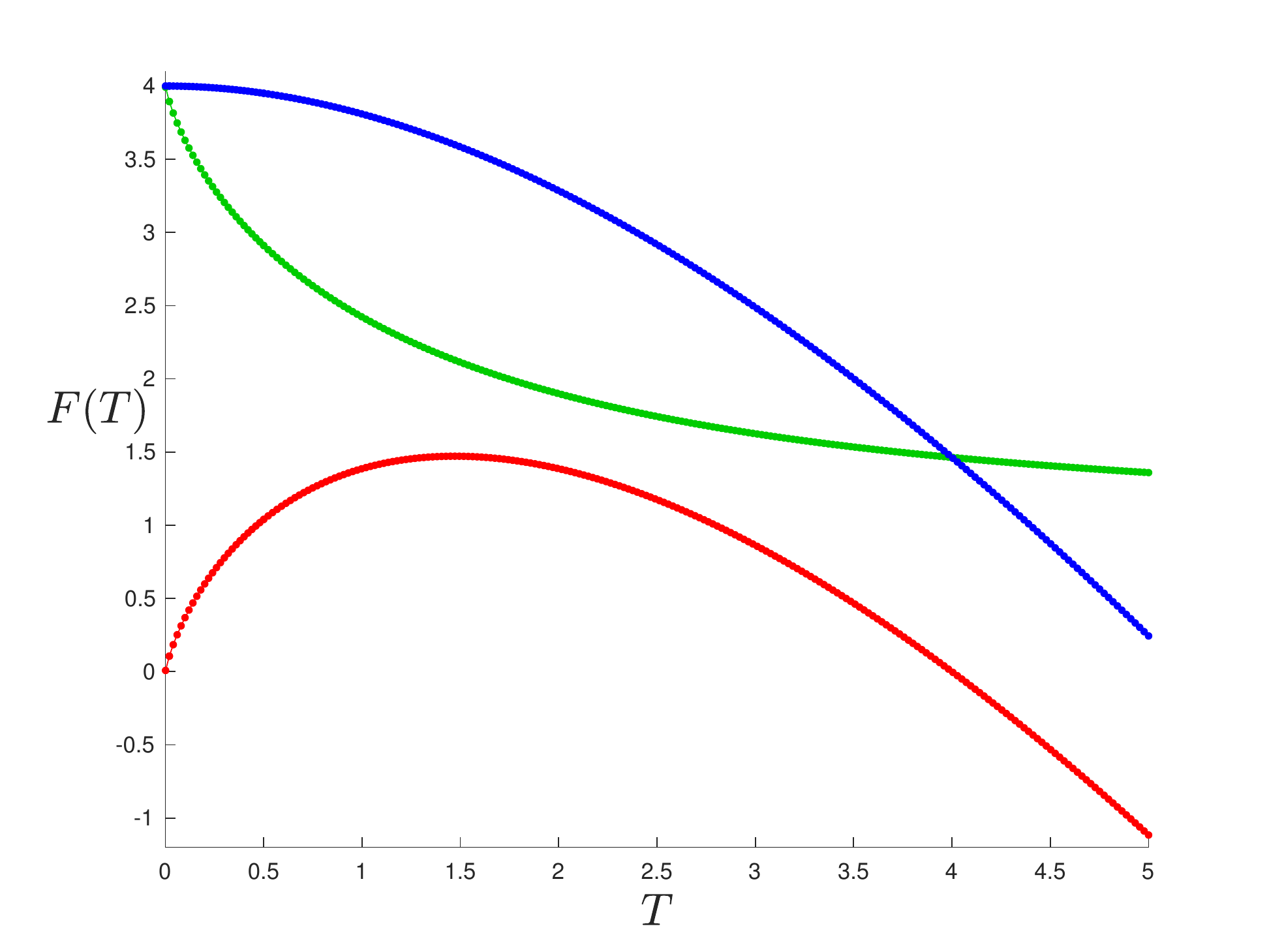}
\caption{\label{fig:12-bessel-model-toy} The free energy $F_Q$ (uppermost, blue) for the leading piece of the (1,2) Bessel  model, with  leading part of partition function given in equation~(\ref{eq:12-part-fun}). The annealed portion $F_A$ is in red (lowermost) and the difference $F_D$ is in green. The region $T<1$ gives the reliable part of the curves.}
\end{figure}

At this point, it is worth checking that all has gone well so far by doing a careful handling of the  integral numerically. Doing the integral~(\ref{eq:okuyama-formula}) with the case~(\ref{eq:12-ZBX}) numerically (delicately handling the large~~$x$ behaviour carefully--it should be clear by now that this is especially important as $\beta$ grows) yields the  large $\beta$ dependence 
\be
\label{eq:leading-with-beta}
I(\beta)= \frac{\beta}{\rho(0)}-\log\left(\frac{\beta}{\rho(0)}\right)-c\frac{\rho(0)}{\beta}+\cdots
\ee where  the first two terms verify the exact analysis so far and $c$ is a pure number so  close to~1 that  it strongly suggests that a simple proof of this can be found.   The dependence at smaller~$\beta$ can of course be extracted numerically too, and   the resulting free energy is displayed in figure~\ref{fig:12-bessel-model-toy}. 

So as anticipated, the free energy $F_Q$ is (minus) quadratic in $T$ in the small $T$ limit, with an additive constant (the coefficient of the linear behaviour in $I(\beta)$):
\be
\label{eq:leading-form-free-12}
F_Q(T)= - \rho(0) T^2 + \frac{1}{\rho(0)}+\cdots
\ ,
\ee
where $c$, although not proven to be unity here, is set to 1 for simplicity of presentation. An argument later on will lend further support for this. Note that in this regime, $F_A{=}-\beta^{-1}\log(\rho(0)/\beta)$'s  contribution has been exactly cancelled by the exact subleading $\log$ dependence shown in equation~(\ref{eq:leading-with-beta}). This will be a general feature, and it confirms the structure observed in ref.\cite{Johnson:2020mwi} that the leading behaviour can all be gleaned from focussing on the fully connected wormhole diagrams.

 As  also anticipated in ref.~\cite{Johnson:2020mwi}, the scale $\rho(0){=}\frac{\mu^2}{4\hbar^2}$ indeed plays a natural role in this simple model: In addition to setting the curvature of the quadratic $T$ behaviour, {\it its  inverse sets the value of $F_Q(T=0)$.} (Notice that $F_Q(0){=}4$ in  figure~\ref{fig:12-bessel-model-toy}, where $\hbar{=}\mu{=}1$ units were chosen, so $\rho(0)=1/4$.) Since $\rho(0)$  is the only scale in the problem, it cannot help but control the behaviour of $F_Q(T)$  in these ways.  
 Looking back to section~\ref{sec:testbeds-B} (see figure~\ref{fig:bessel-free-1}), the value obtained here for $F_Q(0)$ is in fact correct, as it corresponds to $\langle E_0\rangle{=}4{=}\rho(0)^{-1}$. But the leading correction beyond that is quartic instead of quadratic. This is was controlled (see equation~(\ref{eq:leading_FQ})) by the distribution of the difference between the ground state and the first excited peak, given as $p_{\rm gap}{=}\frac12 r^2+\cdots$. There is no such feature here since $\rho$ is constant, which is intuitively equivalent to a constant $p_{\rm gap}$, leading to the quadratic dependence. Once more aspects of the (1,2) model is included, as will be done presently, other scales can enter the problem.

 \subsection{Incorporating Undulations}
\label{sec:new-scales}
The insights gained from last subsection about the formula (with the low-energy truncation) involved only the very leading behaviour of the spectral density. It is now time to learn how well it captures   features that  
appear beyond this regime. 

  Working out how  in complete detail requires better analytic control (than has been acheived so far) over  the integral in equation~(\ref{eq:okuyama-formula}). A different toolbox of  methods mentioned earlier (based on Ramanujan's master theorem) can be brought into play.  The fact that a contribution $-F_A$ always appears (from the second part of the integral) has already  been seen (it is fully analogous to the $\e^{-bx}$ part of the Frullani integral~(\ref{eq:Frullani})).
The focus  will therefore be on applying the master theorem method to the   part of the integral containing $\e^{-{\cal Z}(x,\beta)}$.  

Ramanujan  requires the function in the numerator of the integral to be written as a series expansion of the form $\sum_{n=0}^\infty (-x)^k \mathbf{a}(k)/k!$ Here, the numerator is  an exponential, $\exp(-{\cal Z}(x,\beta))$, but ${\cal Z}(x,\beta)$ is itself a sum, given in equation~(\ref{eq:calZ}). Writing this sum as $\sum_{n=1}^\infty d_nx^n/n! $ where $d_n{=}(-1)^n \langle Z(\beta)^n\rangle$,
 a relation between $d_n$ and the coefficients $\mathbf{a}(k)$ must be found. Actually, this is a standard result in the field of combinatorics: $\mathbf{a}(k) = (-1)^kB_k(d_1,d_2,\cdots d_k)$, where the $B_k$ are the ``complete Bell polynomials''\cite{10.2307/1967979} in the first $k$ of the  $d_n$. For illustration, the first four are:
\bea
\label{eq:bell-polynomials}
B_1 &=& d_1\ , \quad B_2 = d_2+d_1^2\ ,\quad
B_3 = d_3 + 3d_2d_1+d_1^3\ , \nonumber \\
B_4 &=& d_4+4d_3d_1+3d_2^2+6d_2d_1^2+d_1^4\ , 
\eea
and the coefficients simply count the number of ways of partitioning $k$. (For example, the $k=4$ case enumerates the partitions: 4, 3+1, 2+2, 2+1+1, and 1+1+1+1.) 

So the answer to the first term of the integral must come from working out the limit of $\mathbf{a}(-\epsilon)$ as $\epsilon\to0$. The linear (in $\epsilon$) part will cancel against a $1/\epsilon$ from expanding $\overline{\Gamma}(\epsilon)$ to give the result (see the example below equation~(\ref{eq:ramanujans-master-theorem})). It is not entirely clear how best to proceed in general, given the form of the coefficient $\mathbf{a}(k)$ here. A special case already encountered works as follows: If $\Gamma$ states are at some fixed energy $E_0$, then $\rho(E){=}|\Gamma|\delta(E-E_0)$, resulting in $Z(\beta){=}|\Gamma|\e^{-\beta E_0}$, and all $\langle Z(\beta)^n\rangle_c=0$. The $B_k$ just collapse to $(-1)^kd_1^k$, resulting in $\mathbf{a}(k)=a^k=(|\Gamma|\e^{-\beta  E_0})^k$. Following the Ramanujan procedure to the end, the answer for this part of the integral is simply  $-\log a {=} \beta E_0{-}\log|\Gamma|$, and so $F_Q=E_0-T\log|\Gamma|$.

Other examples for which closed (or at least tractable)  forms for $\langle Z(\beta)^n\rangle_c$ can be input are difficult to find, and this is why, in order to render the formula useable,  the low energy truncation $\langle Z(\beta)^n\rangle_c{\simeq}\langle Z(n\beta)\rangle+\cdots$ discussed in the  previous two subsections is used. It has already been noted that this truncation fails for examples such as the special case just discussed, since the vanishing of the connected correlators cannot be reproduced if parts of them have been neglected. Therefore it is of interest to determine how much of the underlying physics the truncated formula can capture.

Now, $\mathbf{a}(k)$ leads as  $(-1)^k d_k {=} Z(k\beta)$ which  is $\int_0^\infty \rho(E) (\e^{-\beta E})^k dE$, plus sums of products of similar pieces (such that the powers add to $k$). An analytic continuation to non-integer $k$ must be found, in order to complete the story, and while it is not clear at present how to do that, an educated guess about the result could go as follows. At large $\beta$, if the result was dominated by one energy scale ${\widetilde E}$, the expansion coefficient would be $a^k{=}(\e^{-\beta {\widetilde E}})^k$. Following the Ramanujan procedure to the end, the answer for the integral would simply be  $-\log a = \beta{\widetilde E}$ reproducing the leading linear $\beta$ dependence seen in the earlier examples. 

In the simple prototype of Section~\ref{sec:simplified-models}, $\rho$ was simply constant, $\rho(0)$, and  the single energy scale that emerged was the inverse ${\widetilde E}=\rho(0)^{-1}$.   Intuitively, in the more general case, the large $\beta$ result for the integral gets its dominant contributions from analogous slowly changing portions of the density $\rho(E)$ as $\beta$ sweeps off to infinity. As will become clear from working with examples, this often comes close what (as shown in Section~\ref{sec:free-energy-direct}) is the  correct answer: $\langle E_0\rangle$, but in general it will not be quite right.

 As discussed in Section~\ref{sec:truncation}, this follows from the fact that neglecting parts of all of the connected correlators throws away crucial information about the {\it individual } peaks in the spectrum that help encode the key features observed to be most important: the average energy of the first peak, and the distribution $p_{\rm gap}(r)$ of the gap $r{=}E_1{-}E_0$ between the first excited state and the ground state.

The latter controlled the coefficient and leading power of the correction $-cT^p$ to $F_Q(T)$ as it grows with $T$. Generically, the integral leads with $\beta{\widetilde E}$ and since the natural dimensionless expansion parameter is $\beta {\widetilde E}$, the 
leading next order  contribution will  be naturally controlled by $1/(\beta {\widetilde E})$ which yields the quadratic fall-off behaviour already seen in the previous section.  

While the above considerations predict that the formula (with the low energy truncation) will not always be a good guide to the precise quantitative features of $F_Q(T)$, it is nevertheless worth exploring how well it does in a variety of examples, especially for more complete theories of JT (super) gravity where other methods are currently not readily available. The following sections will do so first for some of the toy models of section~\ref{sec:free-energy-direct} where the results are known (hence  quantitative comparisons can be done), and then for  a variety of non-perturbative definitions of  JT gravity and supergravity where a spectral density has been exhibited\footnote{Very recently  new  non-perturbative completions of JT gravity were discussed in ref.~\cite{Gao:2021uro}, but the complete spectral densities are not explicitly extracted, and so they cannot be studied here.} The explorations are necessarily numerical, so some remarks on numerical methods  will be made first.

\subsection{Remarks on Numerical Methods}
\label{sec:numerical-tips}
The integral $I(\beta$) in expression~(\ref{eq:okuyama-formula}) will need to be tackled numerically in general (as has already been mentioned).  The key input is the density $\rho(E)$. A straightforward energy integral (Laplace transform) produces the $Z(\beta)$ needed for the second term. However, a more difficult  energy integral~(\ref{eq:calZ-density}) must be  done to produce ${\cal Z}(\beta,x)$, {\it at each $x$}, the result inserted into the first term of the integral and then the whole $x$ integral performed.  As  noted in Sections~\ref{sec:the-formula} and~\ref{sec:simplified-models}, the large $x$ behaviour must be treated with care since this is where the all-important large $\beta$ behaviour emerges. This can  be rather challenging, even if $\rho(E)$ is known in closed form. In such cases, off-the-shelf integration algorithms (such as in {\tt Maple}) can do a good job of the quadrature required to perform the integral, and many accurate data points  obtained relatively swiftly, going down to low enough temperatures to see the trend in the data. A sign that the limits of accuracy of the integration is being reached will be a sudden fall-off of $F_Q(T)$ for low enough $T$, the position of this occurrence being is highly sensitive to the cutoff placed on the numerical $x$ integration, or to the limits of digit accuracy of the computer program. In such cases, the graph of $F_Q(T)$ was simply truncated at some lowest~$T$ once the trend was clear. 

Some careful experimentation can alleviate some of the numerical difficulty. For example, the energy integrals can be safely truncated a high enough energies. This is because as $T$ decreases the physics depends less and less on high energy details. Moreover since   the ansatz used  here for $F_Q(T)$ is  only valid at low $T$,  there is little point in keeping energies that are well above the regime where the non-perturbative oscillations are visible. 

When $\rho(E)$ is not known in analytic form (because it was itself obtained numerically for a full definition of a JT gravity or supergravity such as those in refs.\cite{Johnson:2020exp,Johnson:2020mwi}), the integration procedures mentioned above become additionally more numerically intensive, and integrating up to large values of $x$ to extract accurate large $\beta$ physics can take several orders of magnitude longer to perform carefully. Such computations were done using {\tt MATLAB}, and custom-tailored artisanal integration code was written  to ensure optimum performance, taking care to streamline  steps (such as the $E$-integral) that are performed for every value of $x$. Another useful realization is that the $x$ integration is a local operation, and so a perfect candidate task for which to deploy multiple cores in parallel  (either on a desktop or on large computer clusters).

\section{Tests of the Truncated formula}
\label{sec:FQ-for-Bessel-Airy}

\subsection{The Airy Model}

\begin{figure}[t]
\centering
\includegraphics[width=0.48\textwidth]{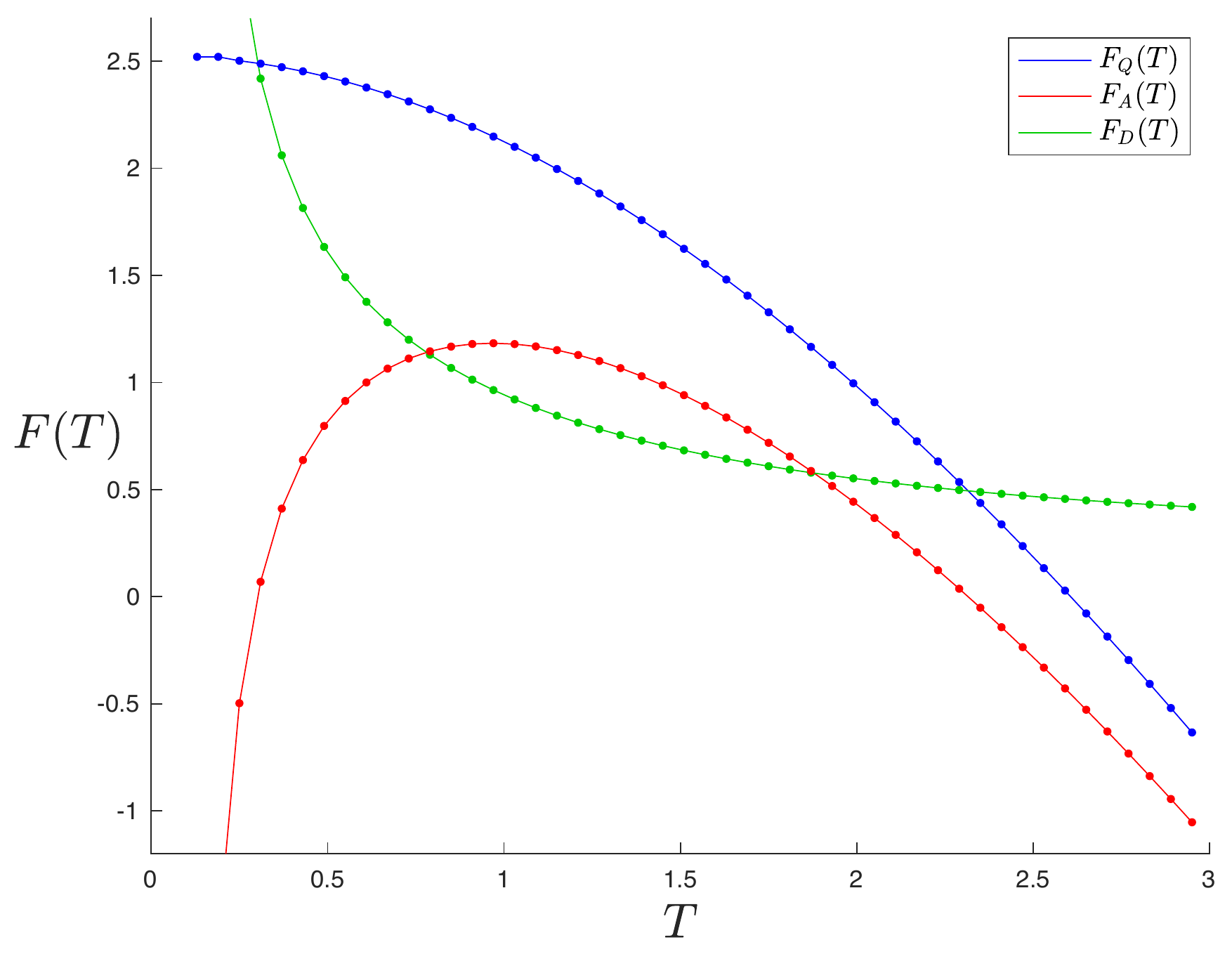}
\caption{\label{fig:free-airy-approx} The result of using the truncation of formula~(\ref{eq:okuyama-formula}) to compute the free energy $F_Q$ (upper, blue) for the Airy model computed using the truncated treatment of formula~\ref{eq:okuyama-formula}.    The annealed result,~$F_A$ is in red (lower) and the difference is in green. It should be compared to the directly computed (benchmark) result in figure~\ref{fig:OMFG}. See text for discussion.}
\end{figure}

The spectral density of the Airy model, given in equation~(\ref{eq:airy-density}), can be inserted into equation~(\ref{eq:calZ-density}) to compute ${\cal Z}(x,\beta)$ (truncated), which in turn is used in the formula~(\ref{eq:okuyama-formula}) to extract the result  for $F_Q(T)$. The numerical methods described in Subsection~\ref{sec:numerical-tips} were used to extract the result, shown in figure~\ref{fig:free-airy-approx}. (Ref.\cite{Okuyama:2021pkf} presents a  similar result.)
This result  should be compared to the benchmark result in figure~\ref{fig:OMFG} (which is {\it not} just a low~$T$ result), which was directly computed from the matrix model by evaluating the ensemble average. The main observation is that the truncated formula's curve for $F_Q(T)$, while qualitatively similar to the directly computed result, fails to capture the two main low temperature features. The value of $F_Q(0)$ is somewhat higher, and is in fact closer to the energy ${\bar E}$ at which the spectral density $\rho(E)$ (shown in figure~\ref{fig:toy-densities-2}) has its first inflection point, {\it i.e.,} between the first and second peaks of the averaged spectrum. The curvature of the curve is also more consistent with a quadratic behaviour $-cT^2$, where $c{\sim}1/{\bar E}$, which follows from  the next generic term in the integral in the large $\beta$ expansion discussed in the previous section. It is clear that this is incorrect,  and the reasons why were identified in the previous section's discussion of the limitations of the low temperature approach to the formula.\footnote{The first version of this manuscript assumed that the truncated formula was more accurate than  it is, and erroneously assumed that these features, which generically appear for other examples too, were an indication of the correct answer.}

\subsection{The Bessel Models}
\label{sec-Bessel}

A trio of  important cases   is considered next. They are the   $(2\Gamma+1,2)$ Bessel models for $\Gamma{=}0,\pm\frac12$. The first case was studied explicitly by sampling ensembles of the randomly generated $H{=}MM^\dagger$ in Section~\ref{sec:testbeds-B} and its spectral density is: 
\be
\label{eq:spectral-12}
\rho(E)=\frac{\mu^2}{4\hbar^2}\left(J_0^2(\mu\sqrt{E}/\hbar)+J_1^2(\mu\sqrt{E}/\hbar)\right)\quad \mbox{\rm for}\,\,\Gamma=0\ ,
\ee
while the two other cases  can be thought of as close cousins of the $\Gamma=\pm1$ cases discussed in Section~\ref{sec:testbeds-B}, but with some special features. They have spectral density:
\be
\label{eq:spectral-22-02}
\rho(E) = \frac{\mu}{2\pi\hbar\sqrt{E}}\pm\frac{\sin(2\mu\sqrt{E}/\hbar)}{4\pi E}\qquad \mbox{\rm for}\,\,\Gamma=\pm\frac12\ ,
\ee
  In the $\Gamma{=}0$ case, there is a leading (disc order) perturbative part $\rhoo(E){=}\mu/(2\pi\hbar\sqrt{E})$, followed by higher order perturbative corrections and non-perturbative pieces. The $\Gamma=\pm\frac12$ cases have the same leading disc behaviour, but the perturbative corrections vanish exactly, leaving only purely non-perturbative physics, which makes them particularly interesting models. Moreover, this structure is mirrored by two full JT supergravity models~\cite{Stanford:2019vob} to be studied later. 
\begin{figure}[t]
\centering
\includegraphics[width=0.42\textwidth]{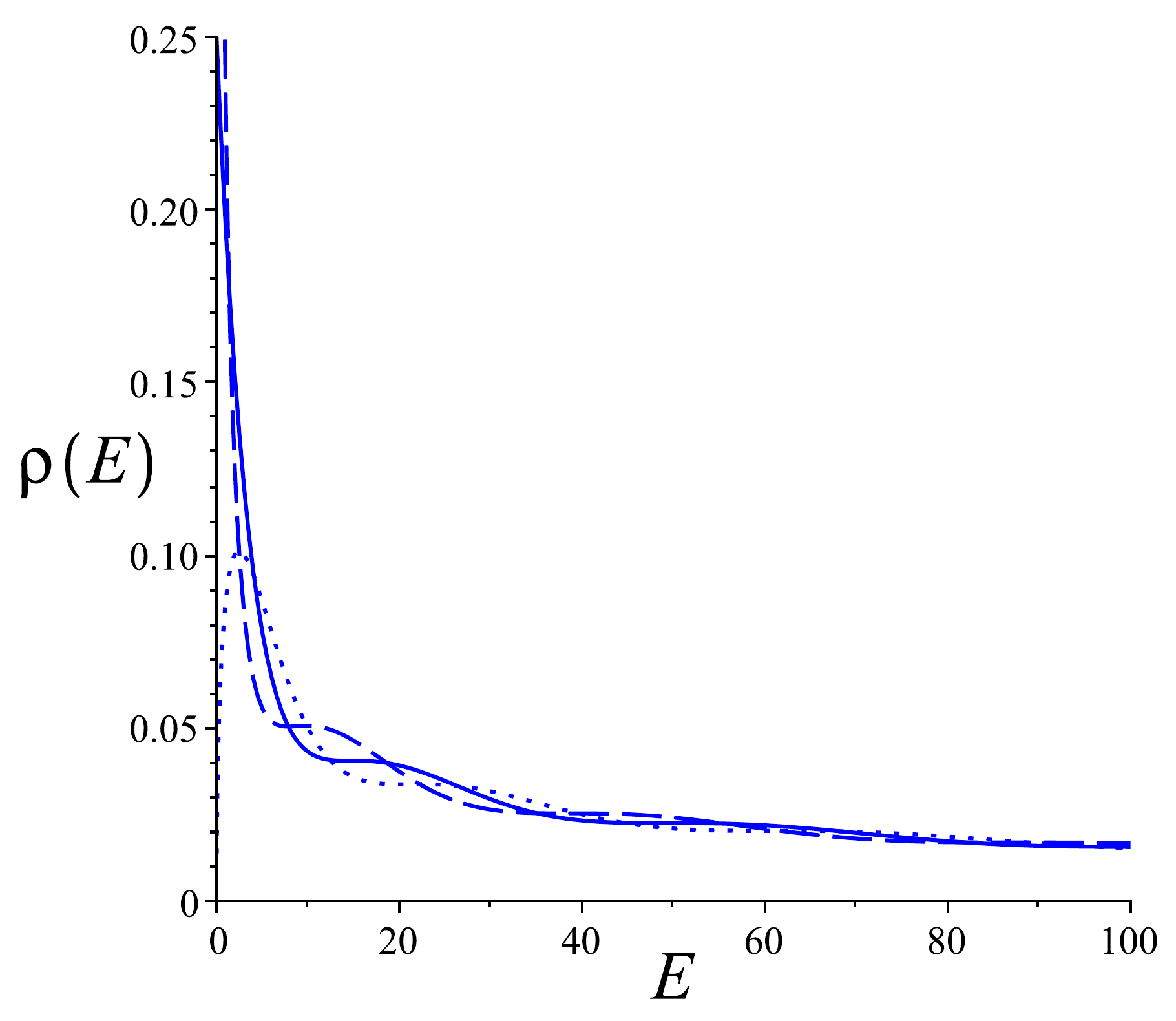}
\caption{\label{fig:toy-densities-1} The spectral densities for various Bessel models. The solid line is the (1,2) case, which goes to a constant (1/4) at $E=0$. The dotted line is the $(2,2)$ case, which starts at  zero at $E=0$, while the dashed line is the (0,2) case that diverges there. The curves' first points of inflection are at ${\bar E}{\simeq} 18, 22, \,{\rm and}\, 12$, respectively.}
\end{figure}
Figure~\ref{fig:toy-densities-1} shows all three models superimposed, and figure~\ref{fig:toy-densities-3} has more details of the individual underlying microscopic spectrum for $\Gamma{=}0$.

\begin{figure}[t]
\centering
\includegraphics[width=0.40\textwidth]{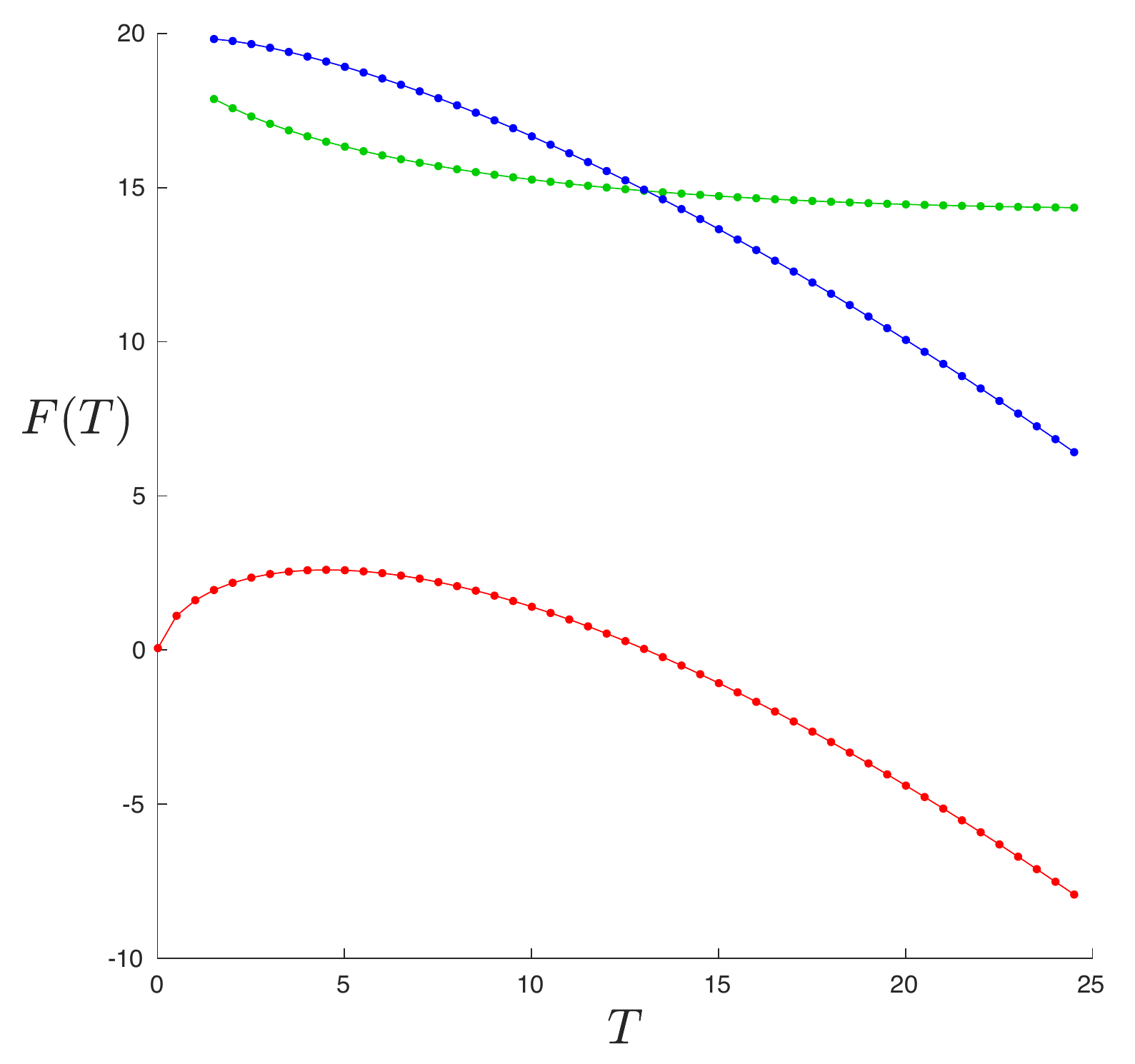}
\caption{\label{fig:12-bessel-model} The result of using the truncation of formula~(\ref{eq:okuyama-formula}) to compute  $F_Q$ (uppermost, blue) for the full (1,2) Bessel   model, with leading part of partition function given in equation~(\ref{eq:12-part-fun}). The annealed portion $F_A$ is in red (lowermost) and the difference $F_D$ is in green. }
\end{figure}

The result of computing $F_Q(T)$ numerically for the $\Gamma{=}0$ Bessel model using the truncation procedure is in   figure~\ref{fig:12-bessel-model}. The previous subsection explains why the integration procedure takes tremendous  care, especially for low $T$ and so the results stop somewhat short of $T{=}0$. However,  a  result similar to the Airy case emerges. The value of the free energy at $T{=}0$ (read off by extrapolating slightly), matches well to the location of the first plateau in the corresponding density curve: $F_Q(0)\simeq 20$. Moreover, the leading part of the curvature of the fall-off from $T{=}0$ is consistent with  quadratic behaviour. 

Comparing this $\Gamma{=}0$ result to the free energy obtained by explicit evaluation over the ensemble shows that the result deviates considerably from the correct result (worse than for Airy). which should be  exactly 4 (see figure~\ref{fig:toy-densities-3}). This is interesting, since when using just the leading piece of the density $\rho(0)=1/4$, in Subsection~\ref{sec:simplified-models}, the formula actually gets this part exactly right. It seems that the inclusion of the rest of the spectral density corrects it away from the result. This again can be traced to the fact that the  large $\beta$ regime of the integral (which controls the $F_Q(0)$ value) is most sensitive to low energy regions where $\rho(E)$ changes slowly--the point of inflection in the curves. Similar comments can be made for the $\Gamma=\pm\frac12$ models. The curves for those cases are similar, with values  of $F_Q(0)$ just above 15 and 25 respectively,  matching the plateau locations in the spectral densities. Even though the individual underlying underlying peaks are not worked out for these cases, it can be seen (by reference to their cousins the $\Gamma{=}\pm1$ cases) the inflection points give too high an energy to correspond to $\langle E_0\rangle$.


These Bessel models describe  the low energy and small~$\hbar{=}\e^{-S_0}$ tail of the full     JT supergravities (the $(2\Gamma+1,2)$ models with $\Gamma=0,\pm1/2$)  for which the non-perturbative spectrum was extracted in refs.~\cite{Johnson:2020exp,Johnson:2020mwi}. The truncated formula's result for the $F_Q(T)$  for those cases will be studied in  Subsection~\ref{sec:JT-supergravity}.

\section{Wider Deployments of the \,\,\,\,\, Truncated Formula}
\label{sec:full-JT-studies}
\subsection{JT gravity}
\label{sec:JT-gravity}
\begin{figure}[t]
\centering
\includegraphics[width=0.48\textwidth]{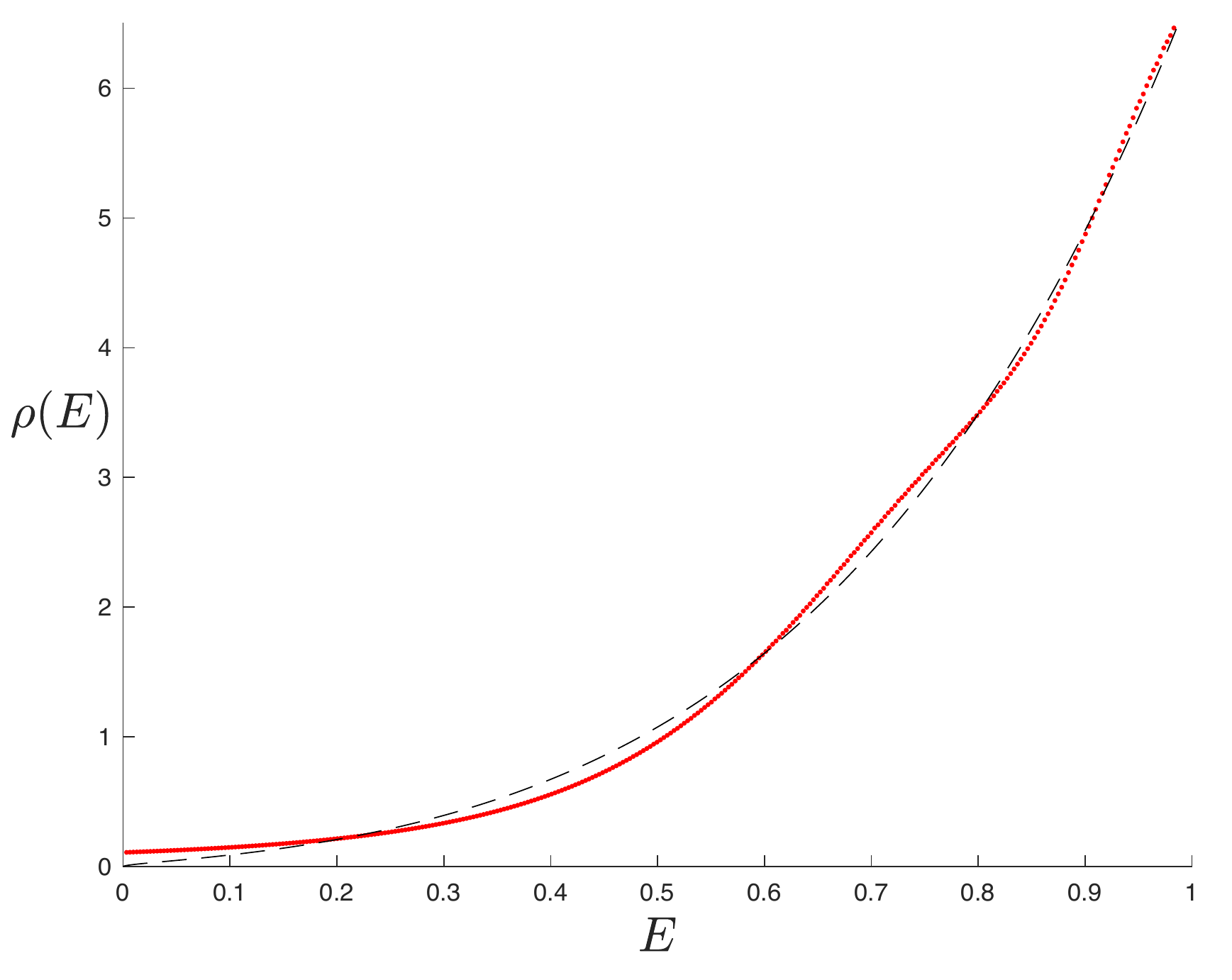}
\caption{\label{fig:JT-gravity-spectrum} The full spectral density $\rho(E)$ for JT gravity, as defined non-perturbatively in refs.~\cite{Johnson:2019eik,Johnson:2020exp}. The dashed curve is the classical result. 
}
\end{figure}

It is possible to construct a non-perturbative definition of JT gravity that has the same perturbative physics as that given by Saad, Shenker and Stanford in ref.~\cite{Saad:2019lba} to all orders in perturbation theory, but which does not possess its non-perturbative instability~\cite{Johnson:2019eik}. The methods used for the definition allow the {\it full} (not just the tail in a special limit) non-perturbative spectral density to be constructed~\cite{Johnson:2020exp}, to any desired accuracy, solving the non-linear equations using numerical techniques. The spectral density obtained is recalled  in figure~\ref{fig:JT-gravity-spectrum}. 
Just as with the toy models of previous sections, there are non-perturbative undulations corresponding to the averaging over the underlying discrete spectrum. In this case, the features are akin to those seen in the $\Gamma{=}0$ Bessel model, with a peak  at $\sim0.7$, but a non-zero $\rho(0)$. 

\begin{figure}[h]
\centering
\includegraphics[width=0.48\textwidth]{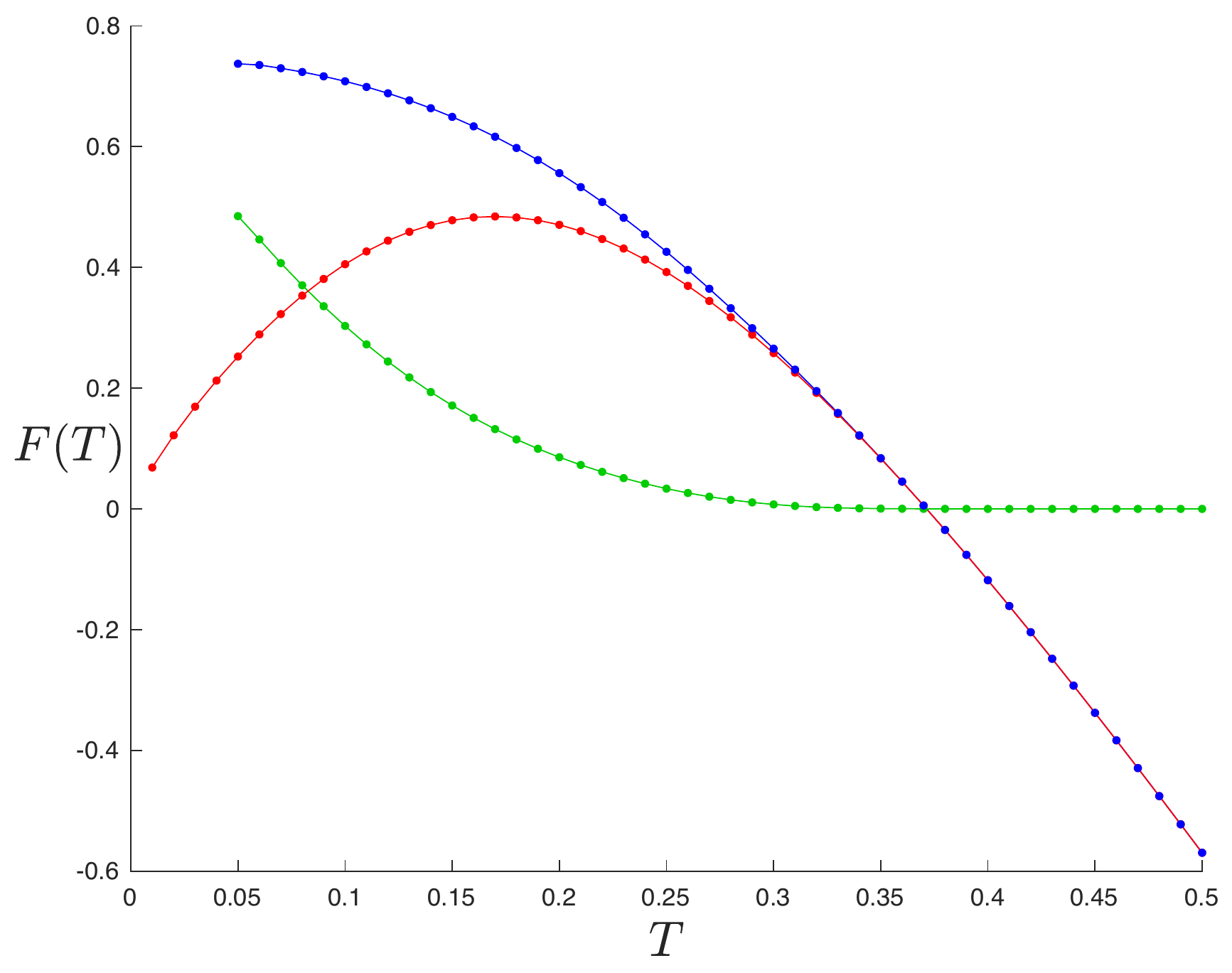}
\caption{\label{fig:JT-gravity-free} The result of using the truncation of formula~(\ref{eq:okuyama-formula}) to compute the free energy $F_Q$ (uppermost, blue) for the non-perturbative definition of JT gravity whose spectrum is given in figure~\ref{fig:JT-gravity-spectrum}. The annealed portion $F_A$ is in red (lowermost) and the difference $F_D$ is in green. }
\end{figure}

It is interesting to see what the low temperature truncation of the $F_Q(T)$ formula~(\ref{eq:okuyama-formula})  gives for this case. The result is given in figure~\ref{fig:JT-gravity-free}. 
As mentioned in Section~\ref{sec:numerical-tips}, since in this case (like others to follow) $\rho(E)$ is only known numerically, it was more labour-intensive numerically to obtain  good points at the lowest temperatures for $F_Q(T)$. As a result, the curves were truncated a little more abruptly than for the cases seen so far, but the clear trend was firmly established. Once again, the  value of $F_Q(0)$ comes out rather higher than it should (slightly above the peak at ${\sim}0.7$). Moreover, there's a quadratic dependence for the fall-off of $F_Q(T)$. (Since this paper's completion, recent new work in Ref.~\cite{Johnson:2021zuo}, using a different approach, has enabled the construction  of the details of the energy spectrum, and a robust computation of $F_Q(T)$, showing that $F_Q(0){=}\langle E_0\rangle{\simeq}0.66$, with a quartic fall-off.)

\subsection{JT supergravity}
\label{sec:JT-supergravity}

In the final cluster of results,  three full supergravity cases are presented. The classification  of these supergravities in terms of random matrix ensembles is given in ref.~\cite{Stanford:2019vob}. The first class are labeled as Altland-Zirnbauer $(\boldsymbol{\alpha},\boldsymbol{\beta})=(2\Gamma+1,2)$ ensembles~\cite{Altland:1997zz}, with~$\Gamma{=}0,\pm\frac12$. The latter two cases are time-reversal invariant, whereas (1,2) is not. The full non-perturbative definition of these was shown to be obtainable as an infinite sum of minimal type~0A string models in ref.~\cite{Johnson:2020heh}, and explicit spectral densities extracted in refs.~\cite{Johnson:2020exp,Johnson:2020mwi}. 
\begin{figure}[t]
\centering
\includegraphics[width=0.48\textwidth]{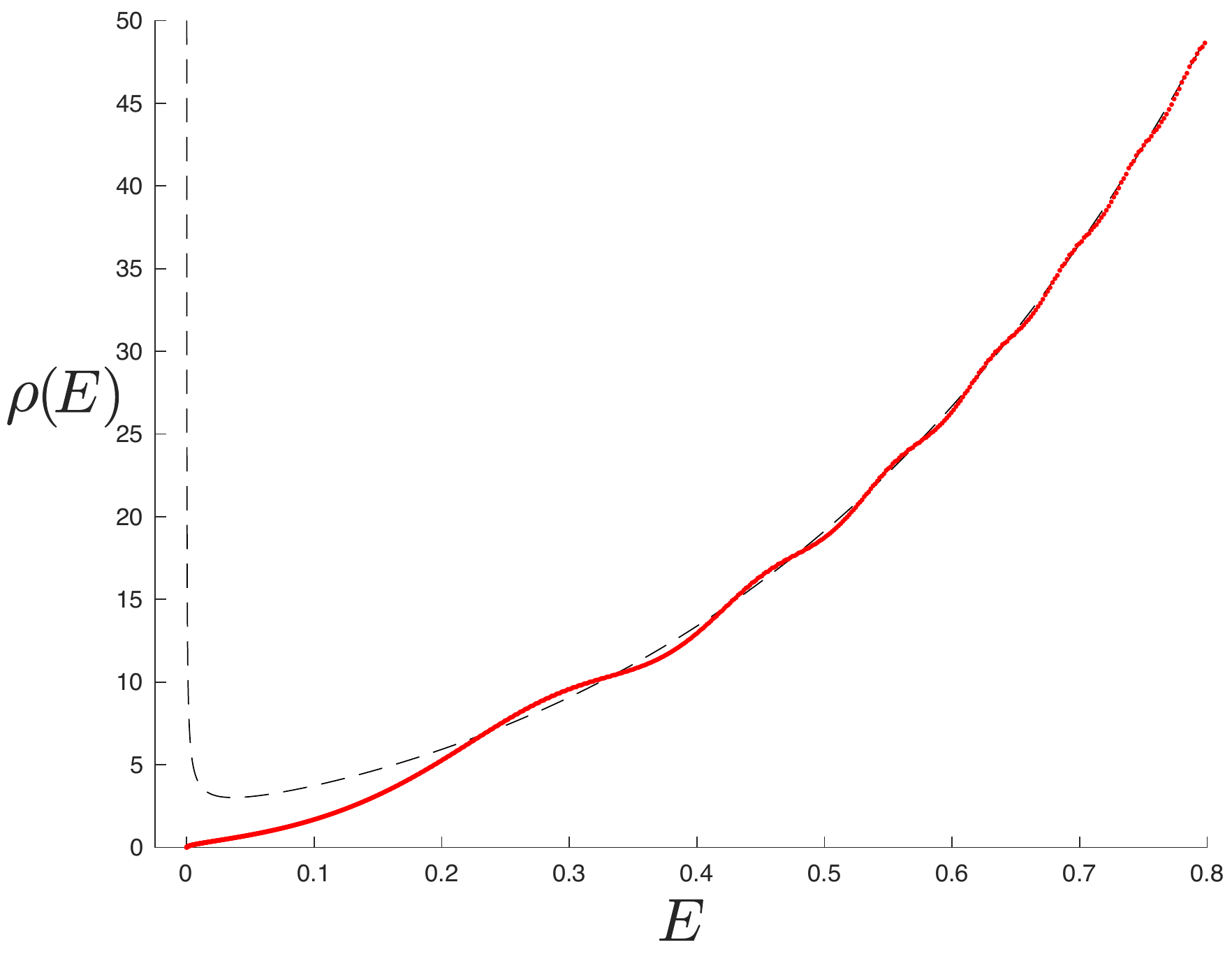}
\caption{\label{fig:22JT-supergravity-spectrum} The full spectral density $\rho(E)$ for (2,2) JT supergravity, from ref.~\cite{Johnson:2020exp}. }
\end{figure}
\begin{figure}[t]
\centering
\includegraphics[width=0.48\textwidth]{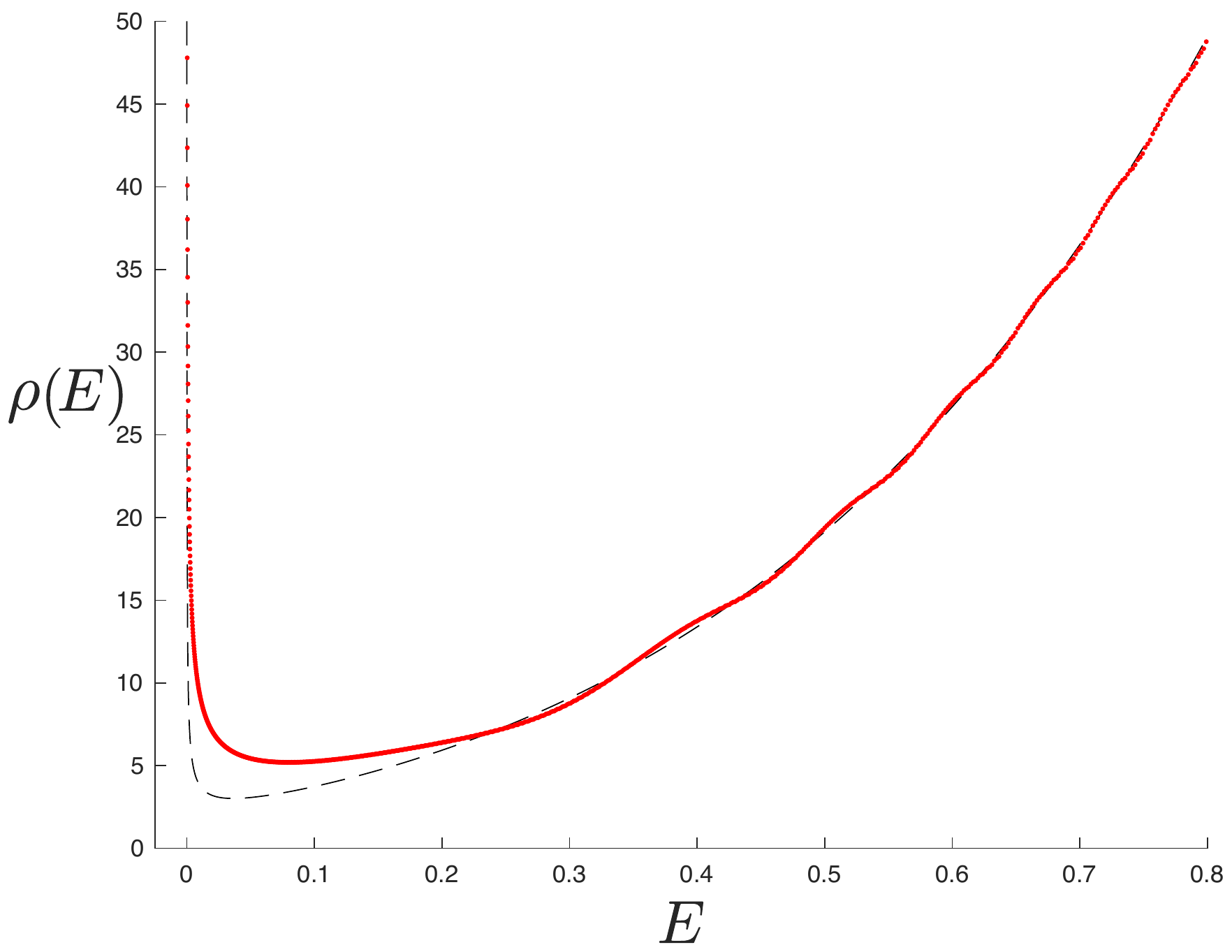}
\caption{\label{fig:02JT-supergravity-spectrum} The full spectral density $\rho(E)$ for (0,2) JT supergravity, from ref.~\cite{Johnson:2020exp}.  }
\end{figure}


For the (2,2) and (0,2) cases the spectral densities are  given in figures~\ref{fig:22JT-supergravity-spectrum} and~\ref{fig:02JT-supergravity-spectrum} respectively, and the quenched free energy provided by the truncated formula can be readily computed from them. 

However, as described in the notes on numerical integration in Section~\ref{sec:numerical-tips}, for improved numerical access to the very lowest temperature points, it  is  helpful (as for the toy models of Section~\ref{sec:new-scales}) if the spectral density is known analytically.  Happily,  for the (2,2) and (0,2) cases Stanford and Witten~\cite{Stanford:2019vob}   proposed an approximate analytic expression for the density: 
\bea
\label{eq:SW-formula}
\rho(E)&\simeq& \rhoo(E)_0\mp\frac{\sin(\pi\int\!\! \rhoo(E^\prime)dE^\prime)}{2\pi E}\ ,\,\,\mbox{for}\,\,\Gamma=\pm\frac12\ ,\nonumber\\
&&{\rm where}\,\,\,\rhoo(0)=\frac{\cosh(2\pi\sqrt{E})}{\pi\hbar\sqrt{E}}\ ,
\eea
expressions that do not include    instanton contributions to the physics. 

\begin{figure}[t]
\centering
\includegraphics[width=0.48\textwidth]{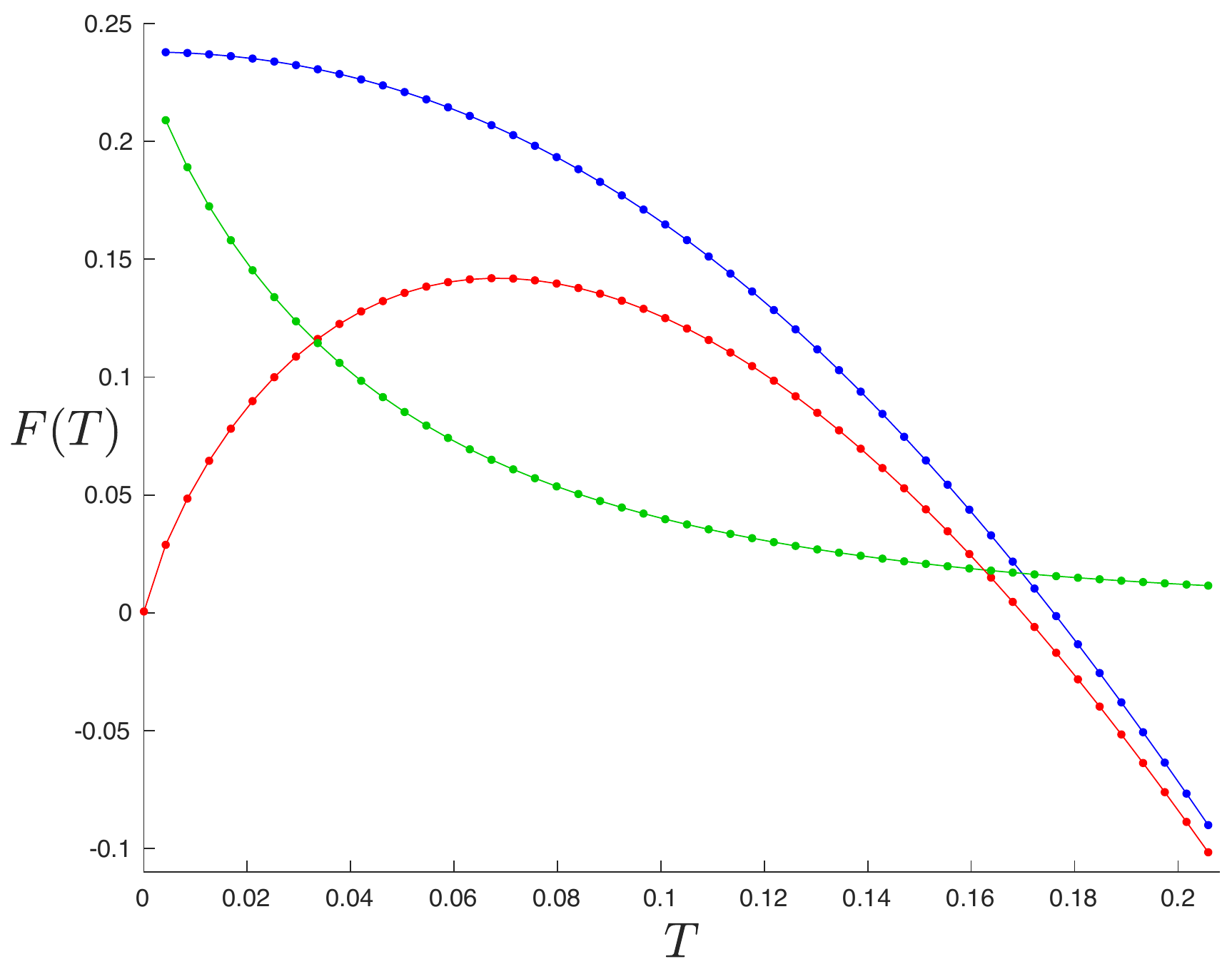}
\caption{\label{fig:22JT-supergravity-free} The result of using the truncation of formula~(\ref{eq:okuyama-formula}) to compute the free energy $F_Q$ (uppermost, blue) for the (2,2)  JT supergravity. The annealed portion $F_A$ is in red (lowermost) and the difference $F_D$ is in green. }
\end{figure}

\begin{figure}[h]
\centering
\includegraphics[width=0.48\textwidth]{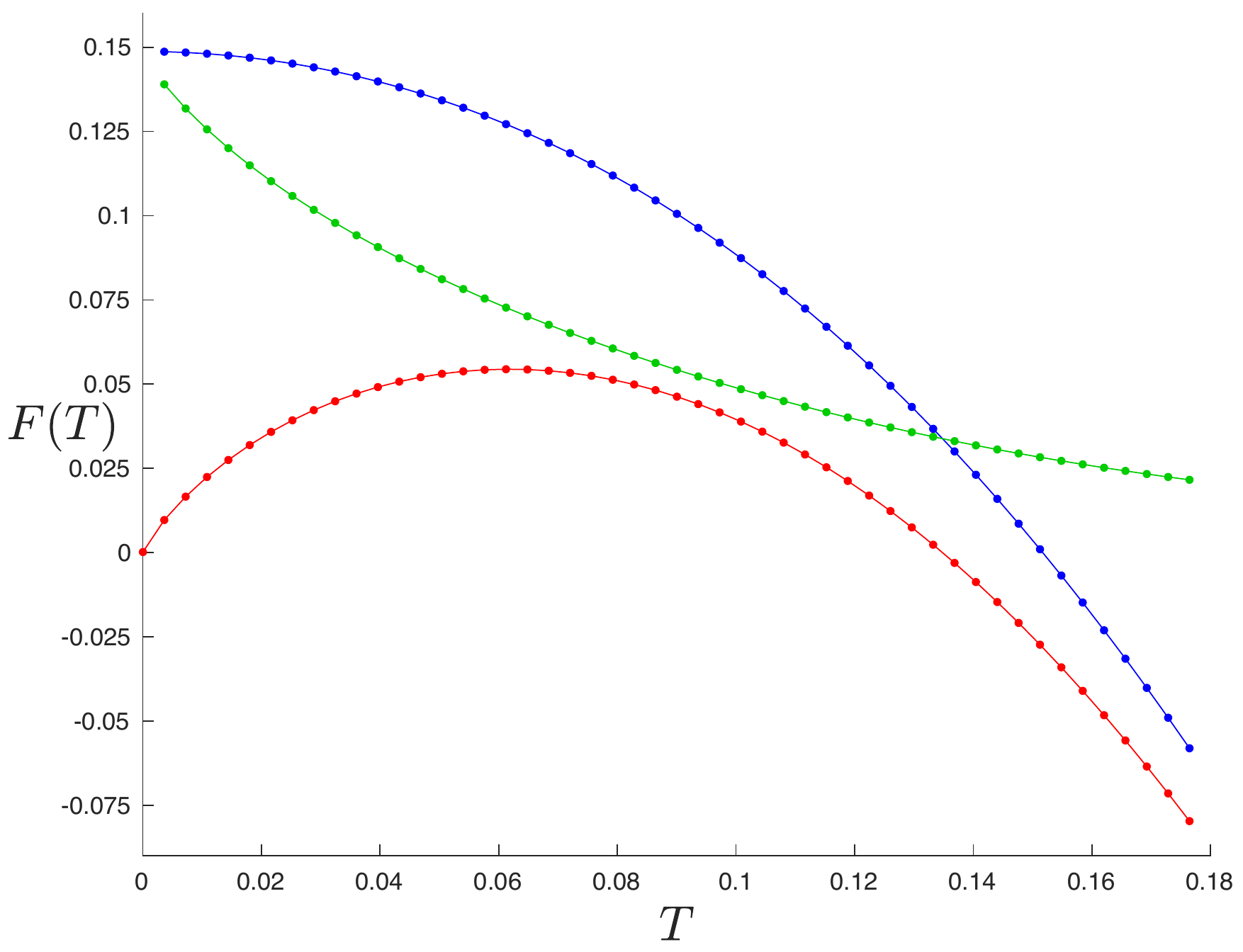}
\caption{\label{fig:02JT-supergravity-free} The result of using the truncation of formula~(\ref{eq:okuyama-formula}) to compute the free energy $F_Q$ (uppermost, blue) for the (0,2)  JT supergravity. The annealed portion $F_A$ is in red (lowermost) and the difference $F_D$ is in green. }
\end{figure}

 These were confirmed using the non-perturbative computations of ref.~\cite{Johnson:2020exp} (including a study of how the neglected instanton effects make their presence felt---in fact, the importance of instantons is reducible by tuning to smaller $\hbar$.).  These expressions can be used for the study of the low temperature truncation approach to the  quenched free energy $F_Q(T)$ here. In fact the results are similar to ones obtained using the full (instanton-rich) numerical $\rho(E)$, with the added advantage of getting access to a few more low temperature points. The results are in figures~\ref{fig:22JT-supergravity-free} and~\ref{fig:02JT-supergravity-free} for (2,2) and (0,2) respectively. 

For the (2,2) case, it is notable that the value of $F_Q(0)$ is close to where  $\langle E_0\rangle$  likely is. This is perhaps attributable to the fact that there is no slow-moving inflection point to confuse matters.   
This can be regarded as a success story for the method, although, as with all cases the generic quadratic fall--off behaviour persists, which is almost certainly incorrect. Again, this follows from the basic input $\rho(E)$ not having the detailed information about the underlying peaks of the averaged spectrum.

For the (0,2) case the quenched free energy computed using the truncated formula  is in figure~\ref{fig:02JT-supergravity-free}. In this special case, there is a very flat minimum appearing somewhat before the first  peak, and the presence of this appears to attract $F_Q(0)$ instead. However, this result is still rather higher than the correct lowest average energy.

\begin{figure}[ht]
\centering
\includegraphics[width=0.48\textwidth]{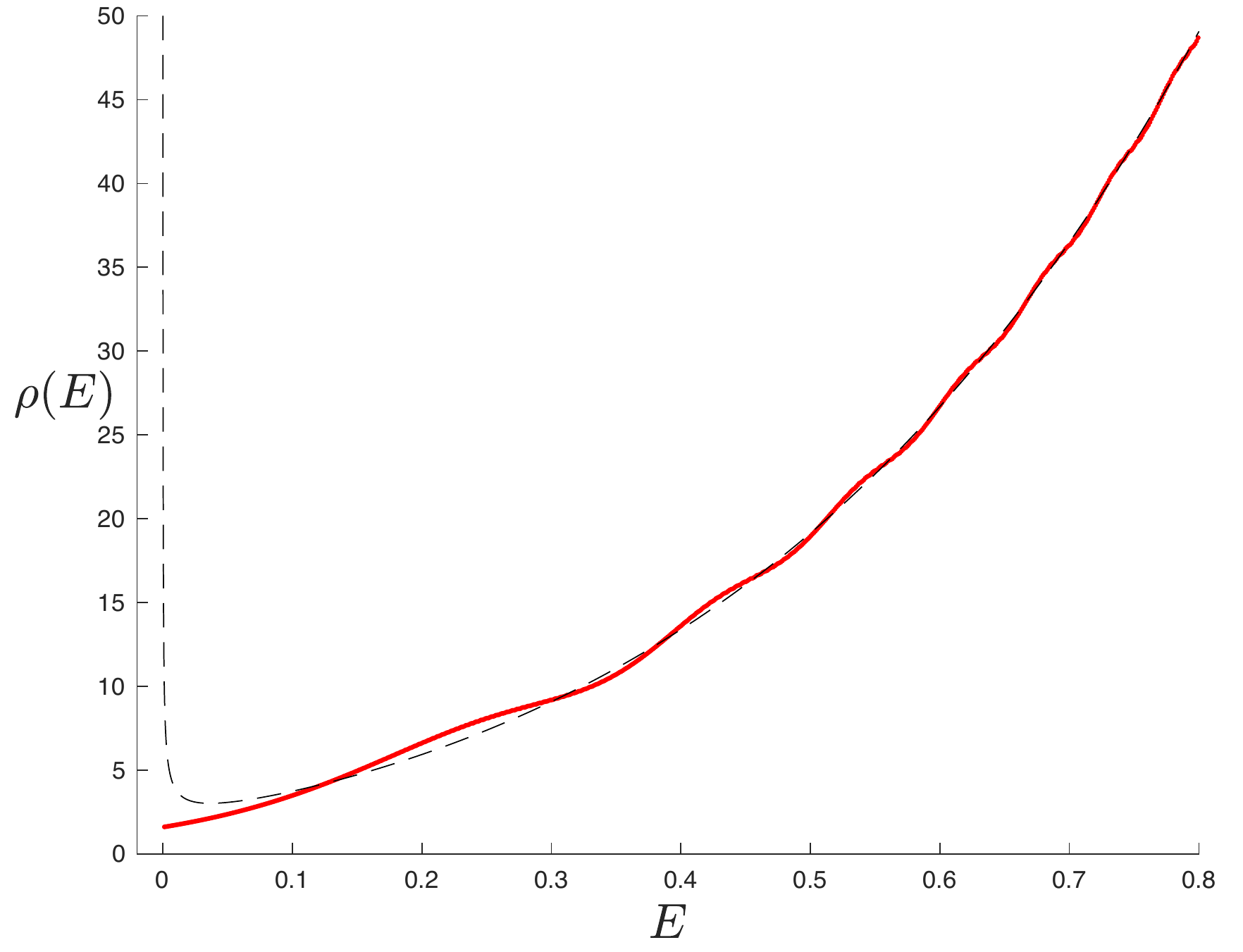}
\caption{\label{fig:12JT-supergravity-spectrum} The full spectral density $\rho(E)$ for (1,2) JT supergravity, from ref.~\cite{Johnson:2020mwi}.  }
\end{figure}

\begin{figure}[ht]
\centering
\includegraphics[width=0.48\textwidth]{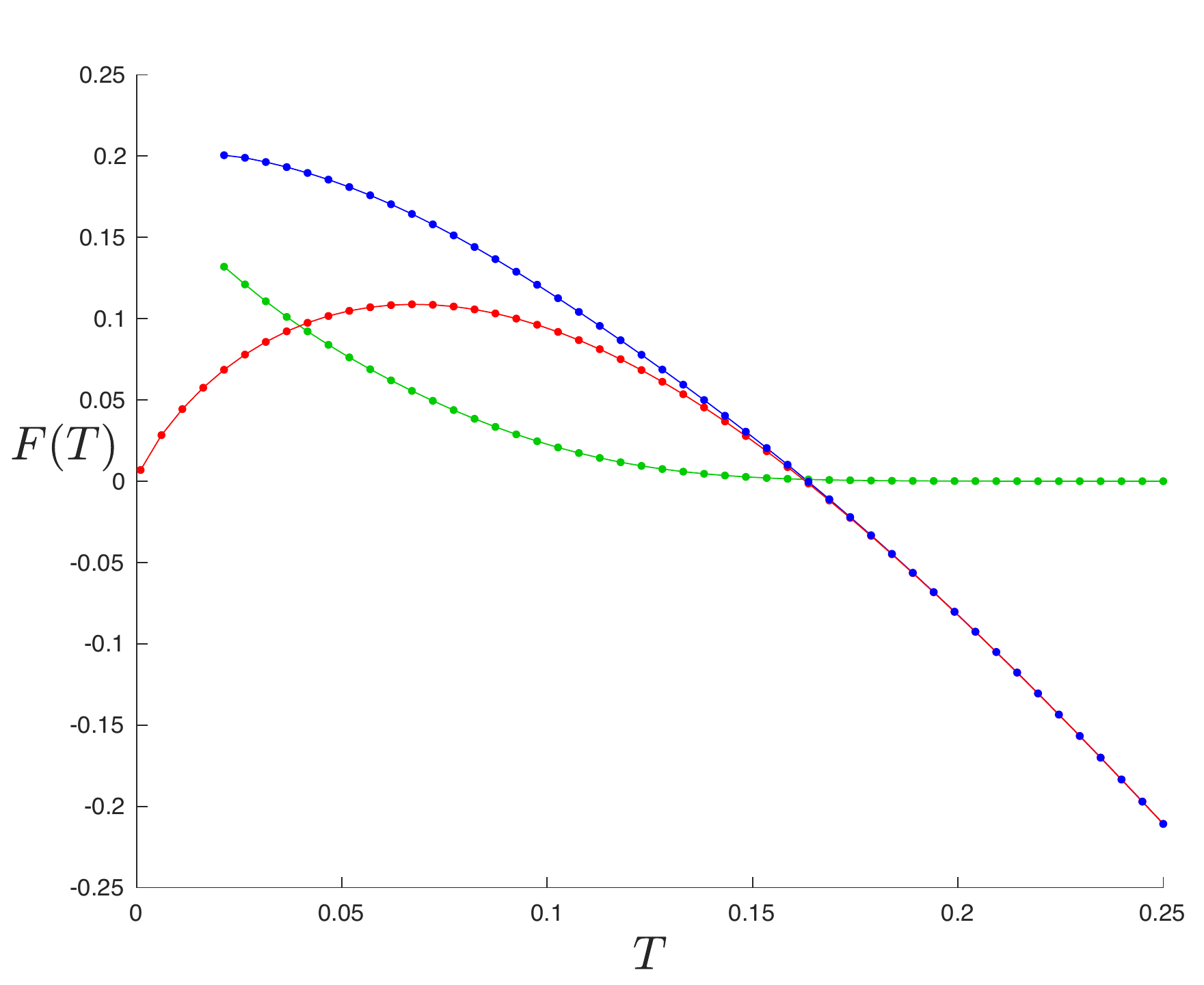}
\caption{\label{fig:12JT-supergravity-free} The result of using the truncation of formula~(\ref{eq:okuyama-formula}) to compute the free energy $F_Q$ (uppermost, blue) for (1,2) JT supergravity. The annealed portion $F_A$ is in red (lowermost) and the difference $F_D$ is in green. }
\end{figure}

For the (1,2) case the density is in figure~\ref{fig:12JT-supergravity-spectrum}, and the truncated formula's result for the quenched free energy in figure~\ref{fig:12JT-supergravity-free}. The outcome is rather similar to the JT gravity case discussed in section~\ref{sec:JT-gravity}. 

\section{Closing Remarks}
\label{sec:closing-remarks}

The matrix model approach to computing the quenched free energy of JT gravity and supergravity (and toy models thereof) is a powerful way of getting access to the essential non-perturbative data that $F_Q(T)$ should be built out of. This includes the physics of the underlying microscopic states, including (crucially) their statistics within the ensemble of matrices. These states make their presence felt as bumps in the spectral density $\rho(E)$, but the full physics requires more than just the bumps, since the data about individual spectral peaks and many other aspects of their statistics must be resolved. This was the lesson learned in testing out the formula~(\ref{eq:okuyama-formula}),  truncated by using the leading low temperature parts of the (wormhole) correlators $\langle Z(\beta)^n\rangle_c$. The truncation amounts to just using $\rho(E)$ to compute, and the resulting behaviour of $F_Q(T)$ using that method did not accurately capture the correct physics, although how close the computed $F_Q(0)$ comes to $\langle E_0\rangle$ was sometimes rather good. The full formula remains very interesting, and deserves further study. It would be key to find a different way of taking a useful low temperature truncation that allows the formula to be applied to non-trivial models while retaining the data equivalent to all of the needed statistics of the lowest energy states. A starting point would be a study of the map between the structure of the connected correlation functions and the properties of the individual peaks that they encode. (See ref.~\cite{Johnson:2021zuo} for very recent progress on precisely this issue, allow a computation of $F_Q(T)$ for JT gravity.)

Directly evaluating properties of the matrix models by sampling the endpoints of the distribution of energies, appropriately double-scaled, is a very fruitful (and surprisingly straightforward) method for getting to the key physics. The full $F_Q(T)$ of the Airy model was computed, as well as that of several Bessel models, and the properties of the leading low $T$ behaviour connected to information about the ground state and first excited states' statistics. There is no reason why this can't be developed further to include information from other states, if so desired. For example it would be interesting develop a dictionary between successive corrections to $F_Q(T)$ at a given $T$ and the nearest energy states (and statistics thereof) at that scale. The connection (which was especially rich for the various Bessel models, indexed by integer $\Gamma$) between the leading properties of $F_Q(T)$ and exact results worked out in the statistical mechanics literature (the work done here was built on foundations started in ref.~\cite{Janssen:2021mek}) was fascinating, and there is almost certainly more to be explored.  There were hints in the results at the existence of a closed form expression for $F_Q(0)=\langle E_0\rangle$ in terms of $\e$ and modified Bessel functions, for example, which would be useful to derive. Moreover, the observation that the ground state peak (whose form is known) of the models at $\Gamma>0$ becomes the peak of the gap $r{=}E_1{-}E_0$ in the $\Gamma<0$ cases may be mirrored by other such exchanges in the spectrum that could be useful to develop.

\begin{figure}[t]
\centering
\includegraphics[width=0.48\textwidth]{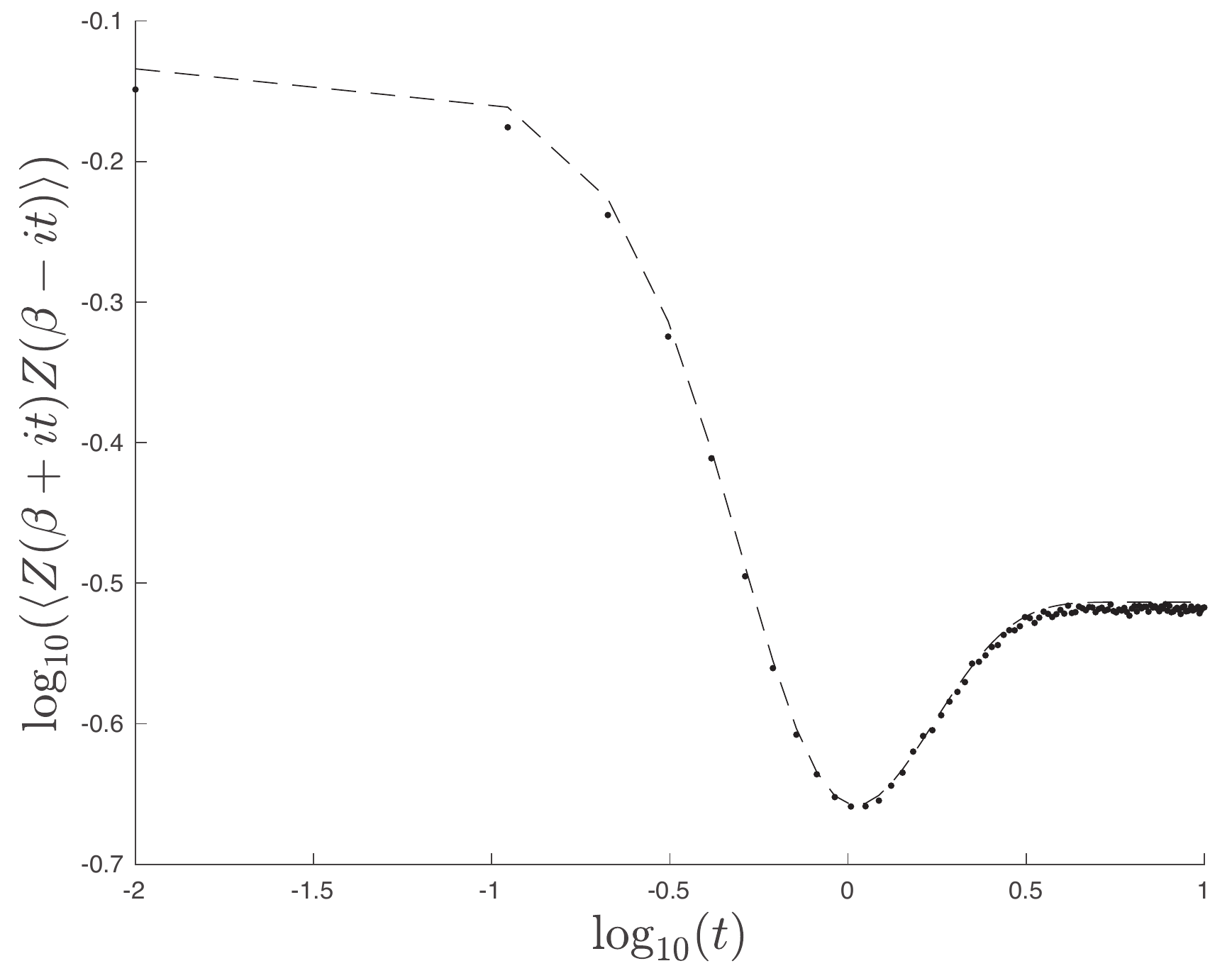}
\caption{\label{fig:airy-sff-direct} The directly computed (black dots) spectral form factor for the scaled $100{\times}100$ random Hermitian matrix model in the normalization that matches the Airy model~(\ref{eq:airy-density}) with $\hbar{=}1$. 100K samples were taken for each point. The dashed line is the exact quantity for Airy.}
\end{figure}

It is clear that the kind of direct evaluation by sampling  the scaled energies done here has more general applications. Other quantities of interest can be readily studied for these same models, and numerous other important cases can be constructed directly. As an example of another easily extracted quantity, the spectral form factor $\langle Z(\beta{+}it)Z(\beta{-}it)\rangle$ of the Airy model can be easily extracted from the same data used to construct $F_Q(T)$ (another few lines of code achieves this) with the result (the black dots) in figure~\ref{fig:airy-sff-direct}. The black dashed line is the known exact result, discussed in this context in {\it e.g.,} refs.~\cite{Okuyama:2019xbv,Johnson:2020exp}.

While the direct methods used here were only used on the ``toy''  Airy and (a variety of) Bessel models, the results were powerful benchmarks for more complicated cases, and for other methods. Moreover, the models themselves are good guides to the low energy physics of the more complete JT gravity and supergravity models. It is the low temperature/energy physics that is most of interest for understanding $F_Q(T)$, and so understanding these simple models represents very good progress.

On the basis of the results of this paper for the JT gravity case, the question of Engelhardt {\it et. al.}~\cite{Engelhardt:2020qpv}  as to whether there is some interesting replica symmetry breaking and/or spin-glass phase of JT gravity can be argued to still be not fully answered. This is because  potentially something unusual could happen between the high~$T$ phase where simple gravity computations (with no non-perturbatative complications) are reliable, and the low $T$ phase where Airy or Bessel-type physics dominates, and where no such exotic phase was evident (the evidence being the computations of this paper). Intuitively, it would seem that the formula~(\ref{eq:okuyama-formula}) of Okuyama (not truncated, the full formula) in principle resolves the replica ambiguity, since the matrix model seems to provide an unambiguous expression for the connected correlators. However, this is not water-tight since even if the individual correlators are well-behaved, it remains to be proven  (especially given the studies of the formula in Section~\ref{sec:the-formula} showing how subtle the emergence of the physics from the core integral can be) whether building the quantity ${\cal Z}(x,\beta)$ out of them can not still produce surprising physical phenomena somewhere at some intermediate temperature.\footnote{The essence  of the issue boils down to the nature of the convergence properties of ${\cal Z}(x,\beta)$, as defined as a sum of correlators in equation~(\ref{eq:calZ}). See ref.~\cite{Janssen:2021stl} for more discussion of related issues. The Author thanks O. Janssen for pointing out this aspect.} On the other hand, new results for the JT spectrum and the statistics of individual energy states, presented in ref.~\cite{Johnson:2021zuo}, have allowed for $F_Q(T)$ to be robustly computed in a manner fully analogous to the direct matrix model enumeration methods of section~\ref{sec:free-energy-direct}. There was  simply no evidence of replica symmetry breaking phase at intermediate temperatures, so this seems to settle the issue.

JT (super)gravity also captures the physics of near-extremal higher dimensional black holes, and so  the physics uncovered here should shed light on issues there too. It would seem that the fully non-perturbative matrix model supplies a very definite  description of the black hole's microscopic degrees of freedom. High temperature or high energy physics sees a smooth density $\rho(E)$, but at scales comparable to $\hbar{=}\e^{-S_0}$, where~$S_0$ is the extremal entropy, the undulations in $\rho(E)$ reveal the presence of the microstate structure. This is natural: Recall that $\hbar$ is the (double scaled) $1/N$ of the large~$N$ matrix model, so scales comparable to $\hbar$ is where the $1/N$ typical spacing between levels/states should become visible, and it does. 

At the very lowest temperatures the effective thermodynamics should depend only upon the properties of the lowest-lying energy states of the black hole, and this is what has emerged here. This matrix model arena could  be the right setting within which to revisit older ideas~\cite{Preskill:1991tb,Maldacena:1998uz,Page:2000dk} (see also recent discussions in the JT gravity content in refs.~\cite{Iliesiu:2020qvm,Heydeman:2020hhw}) about how the thermodynamic description of  near-extremal black holes might break down at the lowest scales and hand over to a different description. Having now a more robust understanding (and underlying statistical picture) of the quenched free energy $F_Q(T)$,  may be helpful in addressing these issues. Interpreting $F_Q(T)$ as the effective description of an ensemble of black holes is one option, but another might be to think of it as describing a single black hole whose microphysics is itself best thought of as an ensemble. Perhaps this is a bridging point to the kind of averaging that was evoked in Mathur's original fuzzball proposal, where  black hole geometries emerge as the net macroscopic effect of some non-black hole underlying description~\cite{Mathur:2005zp}.
 
 \begin{acknowledgments}
CVJ  thanks  Oliver Janssen, Kristan Jensen, John McGreevy,  Felipe Rosso, and Herman Verlinde for questions and comments, the  US Department of Energy for support under grant  \protect{DE-SC} 0011687, and, especially during the pandemic,  Amelia for her support and patience.    
\end{acknowledgments}

%
%
%
%
%
%

\bibliographystyle{apsrev4-1}
\bibliography{NP_super_JT_gravity,LE_super_JT_gravity}

\end{document}